\documentclass[final,3p,authoryear,12pt]{elsarticle}
\usepackage{setspace} % cmr cmss cmtt
\usepackage[round]{natbib}

\usepackage[colorinlistoftodos,textsize=footnotesize]{todonotes}%\usepackage[textsize=tiny]{todonotes}

\usepackage{amsfonts,amsmath,amssymb,lscape,float,graphicx,tabularx,url,times,theorem,color,natbib,afterpage,rotating,hyperref,xcolor,tikz,moresize,multirow,enumerate,colortbl,mathtools,bm,bbm}
\usepackage[utf8]{inputenc}
\usepackage{mathtools}
\usepackage[caption=true]{subfig}
\usepackage[ruled, vlined]{algorithm2e}
\usepackage{enumitem}

\mathtoolsset{showonlyrefs}
\allowdisplaybreaks 

\usepackage{diagbox}

\newcounter{algsubstate}

\definecolor{myblue}{rgb}{0.8,0.8,1}
\definecolor{myred}{rgb}{1,0.8,0.8}
\definecolor{mygreen}{rgb}{0.8,1,0.8}
\definecolor{mygrey}{rgb}{220,220,220}

\usepackage[bottom]{footmisc}
\usepackage[scaled]{helvet}

\DeclareMathAlphabet{\xcal}{OMS}{cmsy}{m}{n}

\definecolor{dbblue}{RGB}{10,65,155}				
\definecolor{dbred}{RGB}{215,0,50}					
\definecolor{blue}{RGB}{0,113.9850,188.9550} 			
\definecolor{red}{RGB}{216.7500,82.8750,24.9900} 		
\definecolor{green}{RGB}{118.8300,171.8700,47.9400} 	
\definecolor{grey}{RGB}{110,110,110}				
\definecolor{lgrey}{RGB}{210,210,210}				
\definecolor{c1}{RGB}{0,113.9850,188.9550}			
\definecolor{c2}{RGB}{216.7500,82.8750,24.9900}		
\definecolor{c3}{RGB}{236.8950,176.9700,31.8750}		
\definecolor{c4}{RGB}{125.9700,46.9200,141.7800}		
\definecolor{c5}{RGB}{118.8300,171.8700,47.9400}		
\definecolor{c6}{RGB}{76.7550,189.9750,237.9150}		
\definecolor{c7}{RGB}{161.9250,19.8900,46.9200}		
\definecolor{c14}{RGB}{0,102,102}					

\hypersetup{colorlinks,urlcolor=dbblue,citebordercolor=dbblue,linkbordercolor=dbblue,filecolor=dbblue,filebordercolor=dbblue,citecolor=dbblue,linkcolor=dbblue,colorlinks=true}

\defcitealias{UK7year:2013}{Griffiths et al., 2013}

\makeatletter
\def\ps@pprintTitle{%
  \let\@oddhead\@empty
  \let\@evenhead\@empty
  \def\@oddfoot{\reset@font\hfil\thepage\hfil}
  \let\@evenfoot\@oddfoot
}
\makeatother

\usepackage{amssymb}
\usepackage{inputenc}
\usepackage{graphicx, animate}
\usepackage{amsmath}
\usepackage{amsfonts}
\usepackage{epstopdf}
\usepackage{dsfont}
\usepackage{bbm}
\usepackage{multirow}
\usepackage{tikz}
\usetikzlibrary{shapes.misc}
\usepackage{pgfplots}

\usetikzlibrary {arrows.meta}
\usepackage{comment}
\usepackage[colorinlistoftodos]{todonotes}
\tikzset{
    cross/.pic = {
    \draw[rotate = 45] (-#1,0) -- (#1,0);
    \draw[rotate = 45] (0,-#1) -- (0, #1);
    }
} 
\usepackage{ctable} % for \specialrule command

\usepackage{lipsum}

\usepackage{bbm}
\usepackage{algorithm2e}
\RestyleAlgo{ruled}

\begin{document} 

\title{\textbf{Deep Attentive Survival Analysis in Limit Order Books:\\ Estimating Fill Probabilities with Convolutional-Transformers}}

%\author[omi]{Álvaro Arroyo$^*\blfootnote{* Denotes equal contibution}$}
\author[omi]{Álvaro Arroyo\corref{contrib}}
\ead{alvaro.arroyo@univ.ox.ac.uk}
% \ead{}
\author[omi,math]{Álvaro Cartea}
%\author[omi,carlos]{Fernando Moreno-Pino$^*$}
\author[omi,carlos]{Fernando Moreno-Pino\corref{contrib}}
\author[omi]{Stefan Zohren}

\cortext[contrib]{Denotes equal contibution}

\address[omi]{Oxford-Man Institute of Quantitative Finance, University of Oxford, UK}
\address[math]{Mathematical Institute, University of Oxford, UK}
\address[carlos]{Signal Processing and Learning Group, Universidad Carlos III de Madrid, Spain}

% \date{April 2022}
\journal{TBC}

\begin{frontmatter}

\begin{abstract}
One of the key decisions in execution strategies is the choice between a passive (liquidity providing) or an aggressive (liquidity taking) order to execute a trade in a limit order book (LOB). Essential to this choice is the fill probability of a passive limit order placed in the LOB. This paper proposes a deep learning method to estimate the filltimes of limit orders posted in different levels of the LOB. We develop a novel model for survival analysis that maps time-varying features of the LOB to the distribution of filltimes of limit orders. Our method is based on a convolutional-Transformer encoder and a monotonic neural network decoder. We use   \textit{proper scoring rules} to compare our method with other approaches in survival analysis, and perform an interpretability analysis to understand the informativeness of features used to compute fill probabilities. Our method significantly outperforms those typically used in survival analysis literature. Finally, we carry out a statistical analysis of the fill probability of orders placed in the order book (e.g., within the bid-ask spread) for assets with different queue dynamics and trading activity. 
\end{abstract}


\end{frontmatter}

\section{Introduction}\label{sec:intro}

Most electronic financial exchanges use limit order books (LOBs) to organise and clear the demand and supply of liquidity in various asset classes. LOBs offer several types of orders, where \textit{limit orders} and \textit{market orders} are the most common types. Market orders cross the bid-ask spread and obtain immediate execution, while limit orders can be placed at various levels of the order book, and, if executed, obtain a better price than that of a market order. The price improvement of the limit order over the market order comes at a tradeoff. Market orders are immediately executed at the best prices available in the book, while limit orders rest in the LOB until they are filled by an incoming market order or they are withdrawn from the LOB; thus, there is no guarantee that a limit order will be executed. The length of time a limit order takes to get filled is known as \textit{time-to-fill}, which can be estimated using different methods. In this work, we use \textit{survival analysis} to calculate the fill probabilities of limit orders placed at different depths of the LOB.  \\

%deep-learning-based method 
We propose a deep learning method to estimate survival functions from longitudinal data. Our approach uses an encoder-decoder neural network architecture based on the Transformer model \citep{vaswani2017attention} and partially monotonic neural networks. The model uses the self-attention mechanism through the encoder to summarise the most recent events over a lookback window before a limit order is placed in the order book, and employs this latent representation of the LOB to estimate the probability of the order being filled after its submission. This self-attention mechanism employs a locally-aware representation of the time series as input, which is obtained through a convolutional network. The attention mechanism and convolutional filters provide a more informative summary of the time series, which improves the estimate of the survival function conditioned on the most recent trades. To evaluate the performance of the estimated survival function we make use of \textit{proper scoring rules} \citep{Gneiting2011, Avati2020, Rindt2022}, which ensure we fit the true survival function and not to an incorrect proxy. This is in contrast with the common approach in the survival analysis literature where performance is typically evaluated using improper scoring rules such as time-dependant concordance, which can lead to an incorrect evaluation of model fit.  \\

% finance application
In this paper, we focus on financial applications and study LOB data from Nasdaq exchange. We use our model to predict the survival functions of orders placed at different levels of the LOB, where matching of orders is determined by price-time priority. Our results show that the proposed architecture significantly outperforms baseline off-the-shelf architectures and standard benchmarks in the survival analysis literature. These results are consistent  throughout several assets with different characteristics, which speaks to the versatility of our model-free approach. We also carry out a comprehensive analysis of the fill probability of limit orders for assets with different queue dynamics and order arrival rates (of both limit and market orders). Furthermore, we provide the first statistical evaluation of the fill probability of limit orders that are placed within the bid-ask spread. Finally, we use Shapley values \citep{lundberg2017unified} to perform an interpretability analysis and show that the model relies heavily on high-frequency information to perform the estimation, and gives less importance to slow-moving features which provide information about seasonal intraday patterns in the fill probability. %\\

\section{Literature Review}
\label{sec:rev}

Time-to-event analysis, also known as survival analysis, is widely used in various fields, including the estimation of the time-to-recovery of a patient \citep{Laurie1989}, clinical trials \citep{Singh2011}, churn prediction \citep{Lariviere2004}, and others \citep{Ziehm2013, Susto2014, Dirick2017}. A fundamental issue in most applications is how to relate the distribution of event times and the features (or \textit{covariates}) describing a system. Simple parametric models, such as the \textit{Cox proportional hazards model}  \citep{Cox1972} and the  \textit{accelerated failure time model} \citep{Wei1992}, are commonly used to make these connections. However, in recent years, several approaches integrate more complex deep learning models into survival analysis, particularly in the medical field. One early example of this is \citet{Faraggi1995}, who use a feed-forward neural network to extend the Cox proportional hazards model. Similar approaches are in \citet{Katzman2018} and \citet{Kvamme2019}, who incorporate techniques such as dropout, which are now common in deep learning. Additionally, some popular models output a discretised survival function without imposing any assumptions on its shape or form  \citep{Lee2018, Lee2019}. Subsequent work aims to improve upon these models \citep{Wang2022, Zhong2021, Hu2021}, while \citet{Rindt2022} highlights the importance of using proper scoring rules in survival analysis and illustrates the shortcomings of previous approaches. Other notable works use Gaussian processes \citep{Fernandez2016, Alaa2017}, random forests \citep{Ishwaran2008}, or adversarial approaches \citep{Chapfuwa2018}.  \\

In quantitative finance, \citet{Cho2000}  assume that the shape of the survival function of filltimes follows a Weibull distribution. The authors track limit orders throughout the trading day, considering cancellations as right-censoring, to determine the fill probability. If a limit order is matched against multiple liquidity taking orders, the total time-to-fill is the weighted average of the individual time-to-fills, where the size of each execution is the weight. The authors treat partial fills as fills of orders that were initially sent with a reduced volume. \citet{Lo2002} use the generalised gamma distribution to model the filltimes and use the accelerated failure time model to incorporate market features. The authors establish separate models for time-to-completion, time-to-first fill, and time-to-cancellation. They discuss hypothetical limit orders, first proposed by \citet{Handa1996}, to study limit-order execution times as the first-passage time to the limit price boundary. \citet{Cartea2015} and \cite{Gueant2016} derive optimal trading strategies with exponential fill rates that do not depend on time and depend on the distance between the price level in the limit order and the midprice in the book. \citet{Maglaras2021} use recurrent neural networks to predict the fill probability of limit orders, which is a widespread approach in the survival analysis literature. To train their model, they use hypothetical limit orders that are placed in the book and kept at a fixed price throughout the trading day, even if the price moves unfavourably over time.
Their work benchmarks its results with the AUC-ROC score, which is an improper scoring rule in survival analysis; hence the assessment of model fit may not be correct. \\

Our work provides a different perspective. We introduce a Transformer-based architecture that outperforms previous benchmarks in terms of proper scoring rules. Furthermore, we evaluate and present several approaches to generate training data, including repositioning hypothetical limit orders and predicting the fill probability directly from the limit orders observed in the LOB.  \\

The remainder of the paper is organised as follows. Section \ref{sec:lob} provides an overview of limit order books. Section \ref{sec:surv_analy} introduces the survival analysis and discusses proper scoring rules. Next, Section \ref{sec:emp_sta_evid}  presents a statistical analysis of the fill probabilities. Section \ref{sec:model}  presents our neural network model, and  Section \ref{sec:experiments}  presents our results and an interpretability analysis.  

\section{Limit Order Books}
\label{sec:lob}

The two most important order types are \textit{market} and \textit{limit} orders. A \textit{market} order is used to buy or sell a given quantity at the best available price. A \textit{limit} order, on the other hand, is an instruction to buy or sell a given quantity at a given price. Market orders guarantee immediate execution, given sufficient liquidity, while limit orders remain in the order book until they are filled or cancelled. The list of pending limit orders is stored in the LOB until they are matched or cancelled. Here, whenever we refer to market orders, we assume that these are either \textit{fill-or-kill} (FoK) market orders, which are to be executed immediately in their entirety (within a predetermined price range) or cancelled entirely, or \textit{immediate-or-cancel} (IoC) market orders, which are to be executed immediately (entirely or partially) at a predetermined price range (i.e., without walking the book). Further, we assume that limit orders are \textit{good for day} (DAY) limit orders, which expire at the end of the trading day if not filled or when they are cancelled. \\

The matching engine of most exchanges follows a price-time priority rule where orders are first ranked according to their price and then ranked according to the time they arrived into the exchange; with earlier orders given priority at the top of the price-level queue. The LOB consists of two sides; the ask side containing all the sell limit orders and the bid side containing all the buy limit orders. 

A snapshot of the LOB at time $t$ is described by the vector
\begin{equation}
   s_{t}=\left\{p_{a}^{l}(t), v_{a}^{l}(t), p_{b}^{l}(t), v_{b}^{l}(t)\right\}_{l=1}^{L} ,  
\end{equation}
where $p_{a}^{l}(t), v_{a}^{l}(t), p_{b}^{l}(t), v_{b}^{l}(t)$ denote the ask price, ask volume, bid price, and bid volume for price level $l \in \{1,...,L\}$ at time $t$, which is typically measured in microseconds after midnight. The matrix $\mathbf{x}_{t} \in \mathbb{R}^{T \times 4 L}$  contains the discrete-time dynamics of the LOB from time $t$ to $t-T$. In the remainder of our work, we use Lobster message data, which provides information on events that update the state of the order book in the Nasdaq stock exchange. For instance, messages to post or cancel orders, to provide the direction (buy or sell), and volume of orders. As an example, Table \ref{table:lob_example} shows the first five messages sent to the book of AAPL on 1 October 2022.

\begin{table}[H]
    \fontsize{11.0}{14.0}\selectfont
    \centering
	\caption{Example of message data from the LOB. Time is measured as seconds from midnight. Price is dollar price times 10000. First five messages on 3 October 2022 for AAPL ticker.}
	\centering % used for centering table
    \begin{tabular}{rrrrrr}
    \specialrule{.1em}{.1em}{.1em}
    \multicolumn{1}{c}{\textbf{Time}} & \multicolumn{1}{c}{\textbf{Event Type}} & \multicolumn{1}{c}{\textbf{Order ID}} & \multicolumn{1}{c}{\textbf{Size}}                & \multicolumn{1}{c}{\textbf{Price}} & \multicolumn{1}{c}{\textbf{Direction}}\\  
    \specialrule{.1em}{.1em}{.1em}
    34200.000841 & 1 & 24974777 & $$100$$ & 1381900 & $$1$$ \\
    34200.000841 & 1 & 24974809 & $$1447$$ & 1383100 & $-1$ \\
    34200.003940 & 1 & 24978469 & $$ 100$$ & 1381900 & $1$ \\
    34200.010366 & 1 & 24986889 & $$100$$ & 1381900 & $1$ \\
    34200.023144 & 1 & 25002805 & $$100$$ & 1381800 & $$1$$ \\\specialrule{.1em}{.1em}{.1em}
    \end{tabular}
    \label{table:lob_example}
\end{table}

\vspace{-0.3cm}

\section{Survival Analysis}
\label{sec:surv_analy}

\vspace{-0.1cm}

The \textit{event time} $T_l\in\mathbb{R}_{\ge 0}$ is a positive valued random variable that describes the filltime of a limit order placed at level $l$ of the order book. Our objective is to predict event times by conditioning on a set $\mathbf{x}\in\mathbb{R}^p$ of market features. All events are subject to right-censoring, i.e. the filltime may not be observed. When a limit order is cancelled or reaches the end of the trading day without being filled, it is considered a censored event. We consider a set of $N$ observations of the triplet $(\mathbf{x}_i, z_i, \delta_i)$, where $\delta_{i}=\mathbbm{1}\{z_i=t_i\}$ is an indicator function that is equal to zero if the event is censored and one otherwise, and $z_i$ and $\mathbf{x}_i$ are the observed event time and the observed market features up to the instant of order submission, respectively. We use the triplet to estimate the \textit{survival function} $S_{T_l}(t\mid\mathbf{x}) = \mathbb{P}\{T_l> t\mid\mathbf{x}\}$.\footnote{This is an instance of \textit{survival regression}, which is equivalent to survival analysis but conditioning on a set of features $\mathbf{x}$.} 

\begin{figure}[H]
    \begin{center}
        \begin{tikzpicture}
            \centering
            % axes
            \draw[thick,->] (0,0) -- (9,0) node[anchor=north west] {$t$};
            \draw[thick,->] (0,0) -- (0,3.5) node[anchor=south east] {};
            \foreach \y in {1,2,3,4}
                \draw (1pt,0.75*\y cm) -- (-1pt,0.75*\y cm) node[anchor=east] {$\footnotesize\text{Order \y}$};
            % 1
            \draw [line width=1.15mm, draw=black, opacity=0.2] (0,0.75) -- (3,0.75);
            \draw [fill=black] (3,0.75) circle (3.5pt) ;
            % 2
            \draw [line width=1.15mm, draw=black, opacity=0.2] (0,1.5) -- (6,1.5);
            \draw [dashed, line width=0.15mm, draw=black, opacity=0.5] (6,0) -- (6,3.5);
            %\draw [dashed, line width=1.15mm, draw=black, opacity=0.2] (6,1.5) -- (8,1.5);
            %\draw [fill=white] (8,1.5) circle (3.5pt) ;
            %\draw (.5,0) node[cross,rotate=10] {};
            \path (6,1.5) pic[red] {cross=5pt};
            % 3
            \draw [line width=1.15mm, draw=black, opacity=0.2] (0,2.25) -- (4,2.25);
            \draw [fill=black] (4,2.25) circle (3.5pt) ;
            % 4
            \draw [line width=1.15mm, draw=black, opacity=0.2] (0,3) -- (2,3);
            \draw [dashed, line width=1.15mm, draw=black, opacity=0.2] (2,3) -- (4,3);
            \draw [fill=white] (4,3) circle (3.5pt) ;
            \path (2,3) pic[red] {cross=5pt};
            %\draw [x=1.4ex,y=1.4ex,line width=.3ex, red]
            %(14,14) -- (14.5,14.5) (14,14.5) -- (14.5,14);
            
            % end of study label
            \draw (7.5 cm , -9pt)
            (7.5 cm, -11pt) node[anchor=east] {$\footnotesize\text{End of trading day}$};
            
            % legend
            \path (9,3) pic[red] {cross=5pt};
            \draw (13.2 cm , 83pt)
            (12.8 cm, 85pt) node[anchor=east] {$\footnotesize\text{cancellation (censoring)}$};

            \draw [fill=white] (9,2.5) circle (3.5pt) ;
            \draw (12.05 cm , 73pt)
            (11.75 cm, 71pt) node[anchor=east] {$\footnotesize\text{fill (unobserved)}$};
            
            \draw [fill=black] (9,2) circle (3.5pt) ;
            \draw (11.675 cm , 59pt)
            (11.4 cm, 56pt) node[anchor=east] {$\footnotesize\text{fill (observed)}$};
            %\matrix [draw,below left] at (current bounding box.north east) {
            %  \draw [fill=white] (4,3) circle (3.5pt) ;
            %};
        \end{tikzpicture}   
    \end{center}    
    \caption{Events that can occur after the submission of a limit order.}
    \label{fig:events_after_limit_order}
    %\todo[inline]{PENDING: to reference this figure}
\end{figure}
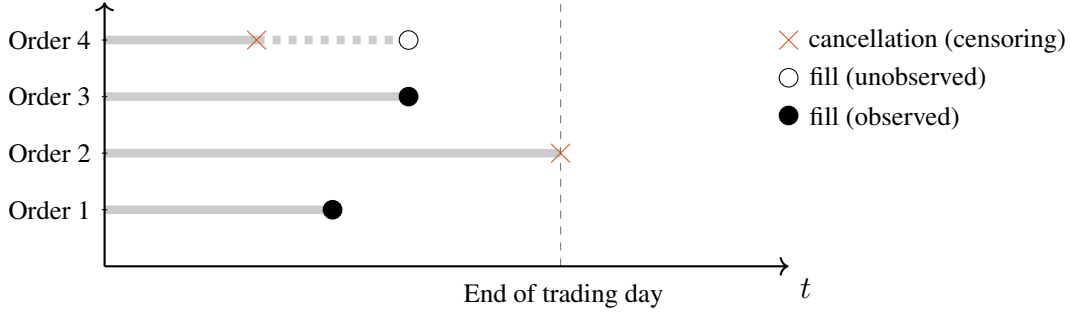

This survival function computes the probability that a limit order posted at level $l$ will not be filled before time $t$. The link between the survival function $S_{T_l}(t|\mathbf{x})$ (which represents the probability of the order not being filled before a particular time) and the cumulative density function (CDF), $F_{T_l}(t|\mathbf{x})$, of the event time is given by $S_{T_l}(t|\mathbf{x}) = 1 - F_{T_l}(t|\mathbf{x})$. Thus, the density function is $f_{T_l}(t|\mathbf{x})=-\frac{d}{dt}S_{T_l}(t|\mathbf{x})$, which describes the probability of an order being filled in a particular amount of time after submission. Further, the \textit{hazard rate}, which indicates the propensity that an order will be filled after time $t$, is given by

\begin{equation}
    h_{T_l}(t|\mathbf{x}) = \frac{f_{T_l}(t|\mathbf{x})}{1-F_{T_l}(t|\mathbf{x})} \ .
\end{equation}
It suffices to obtain one of the previous functions to derive the remaining three:
\begin{equation}
    S_{T_l}(t|\mathbf{x})
    =
    \mathbb{P}\{T_l>t|\mathbf{x}\}
    =
    1 - F_{T_l}(t|\mathbf{x})
    =
    \text{exp}\bigg(-\int_{0}^{t}h_{T_l}(s)\ ds\bigg)
    \ .
\end{equation}
The shape of survival functions is described by a vector of parameters $\boldsymbol{\theta}\in\mathbb{R}$. One may assume that the survival function is given by a tractable distribution (e.g.\ a Weibull distribution) and estimate the relevant parameters with standard methods. However, there is a trade-off between mathematical tractability and goodness of fit to the data. An alternative is to use neural networks to estimate the survival function, which increases the number of parameters substantially but improves the fit to the data. Here, we use the latter approach to model the survival function of limit orders because its shape is expected to have a non-linear relationship with market features. We perform \textit{maximum likelihood estimation}  to obtain the vector of parameters that best fit to the data. Specifically, we train our deep learning model to maximise the \textit{right censored log-likelihood function}

\begin{equation}
    \mathcal{L}(\boldsymbol{\theta}) = \log(L_N(\boldsymbol{\theta})) = \sum_{k=1}^{N}\delta_{k}\log(\hat{f}(z_{k}|\mathbf{x}_{k}, \boldsymbol{\theta})) + (1-\delta_{k})\log(\hat{S}(z_{k}|\mathbf{x}_{k}, \boldsymbol{\theta}))\ ,
    \label{eq:2.4}
\end{equation}
where $\hat{S}$ and $\hat{f}$ are the neural network estimates of the survival function and the density function, respectively.\footnote{Censoring aids in the estimation because it provides information on the order not being filled before it was cancelled.} See \ref{app:b} for a derivation of \eqref{eq:2.4}. Training on right-censored log-likelihood requires a model to output $\hat{S}(z_{k}|\mathbf{x}_{k}, \boldsymbol{\theta})$ at the exact time-instant $z_{k}$. This is challenging in deep learning models that use the softmax activation function to discretise the survival function, because these models require an interpolation scheme to train tractably on \eqref{eq:2.4}. Our model avoids this problem by inputting the observed time $z_{k}$ only to the monotonically restricted decoder, together with the generated latent representation from the LOB time series. This allows us to generate an arbitrarily small time grid over which to evaluate the survival function at no additional parameter cost, as well as respect the monotonicity of the survival function; see Section \ref{sec:model} for more details. \\

To evaluate the quality of model fit to the survival function, we use the concept of \textit{scoring rule}, see \cite{Rindt2022}. A scoring rule $\mathcal{S}$ takes as input a distribution $S$ over a set $\mathcal{Y}$, with an observed sample $y\in\mathcal{Y}$, and returns a score $\mathcal{S}(S,y)$. With positive scoring rules, higher scores indicate an improvement in model fit. In survival regression, a scoring rule $\mathcal{S}$ is \textit{proper} if
\begin{equation}
    \mathbb{E}_{t,c,\mathbf{x}}[\mathcal{S}(S(t|\mathbf{x}),(z,\delta))] \ge \mathbb{E}_{t,c,\mathbf{x}}[\mathcal{S}(\hat{S}(t|\mathbf{x}),(z,\delta))]
\end{equation}
for all survival function estimates $\hat{S}(t|x)$. This means that in expectation, a proper scoring rule will give higher scores to the true survival function over any other estimate. The most commonly used scoring rules in the literature are improper, see \ref{app:c} for more details. On the other hand, \citet{Rindt2022} show that right-censored log-likelihood (RCLL) is a proper scoring rule. Throughout our analysis, we use RCLL as the scoring rule to evaluate the precision with which we fit the true survival function, and evaluate the performance of the proposed model. This guarantees that we evaluate model estimates to fit the true underlying survival function and not an incorrect proxy.

\section{Empirical and Statistical Evidence of Fill Rate Executions}
\label{sec:emp_sta_evid}

We use the messages sent to the LOB to obtain filltime data of limit orders. In this section, we present two approaches to compute this data. The first tracks the outcome of limit orders placed in the LOB, and the second uses hypothetical orders to model limit order repegging. We compare the resulting survival functions associated to different levels of the order book (for the first approach) and the top level for the second approach. We also analyse the changes in the survival functions when orders are placed inside the spread for assets with different queue behaviour and order arrival speed. 

\subsection{Generation of training data}
\label{subsec:synthetically_tracked_orders}

One way to obtain the dataset of triplets $\{(\mathbf{x}_i, z_i, \delta_i)\}_{i=1}^N$ discussed in the previous section is to track all the messages associated to a particular order after its submission. If the final message corresponds to an execution, then the order is recorded as filled, otherwise it is recorded as censored. The time-to-fill is the time elapsed between order submission and the time the final message is observed. 

\begin{figure}[H]
	\includegraphics[height =6.75cm,width=0.49\textwidth]{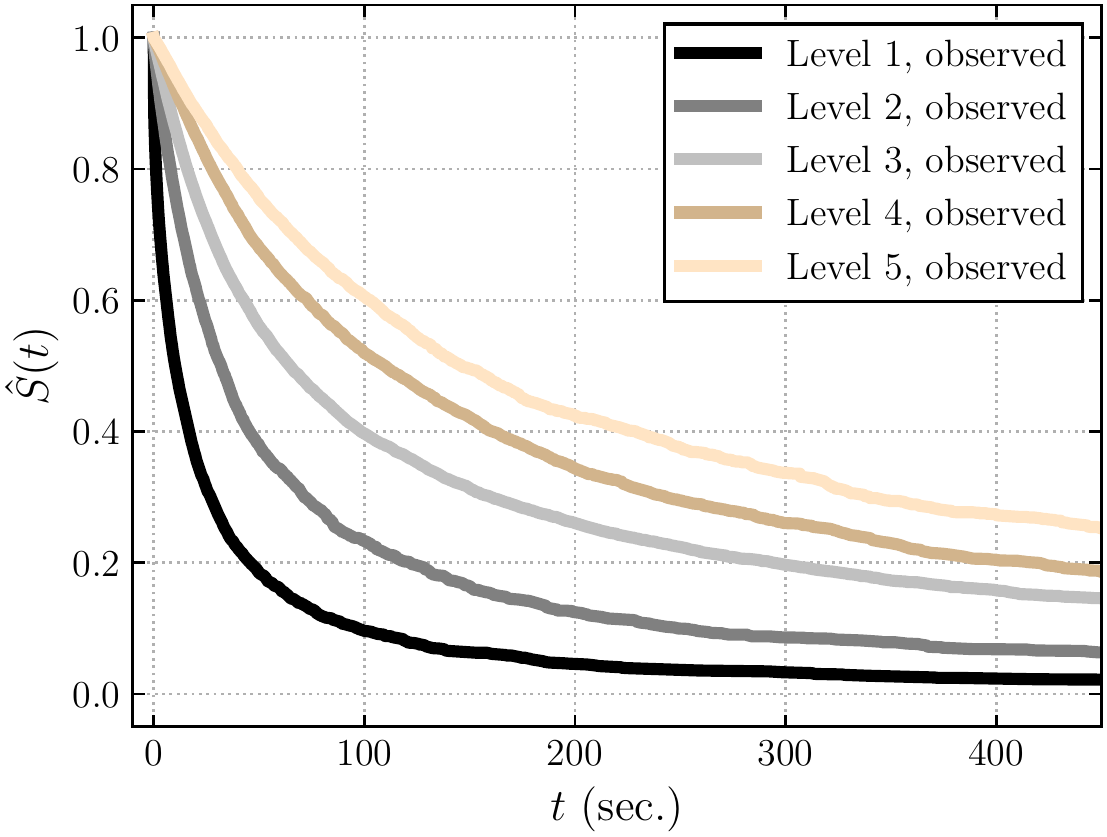}
	\hspace*{\fill}
	\includegraphics[height = 6.75cm,width=0.49\textwidth]{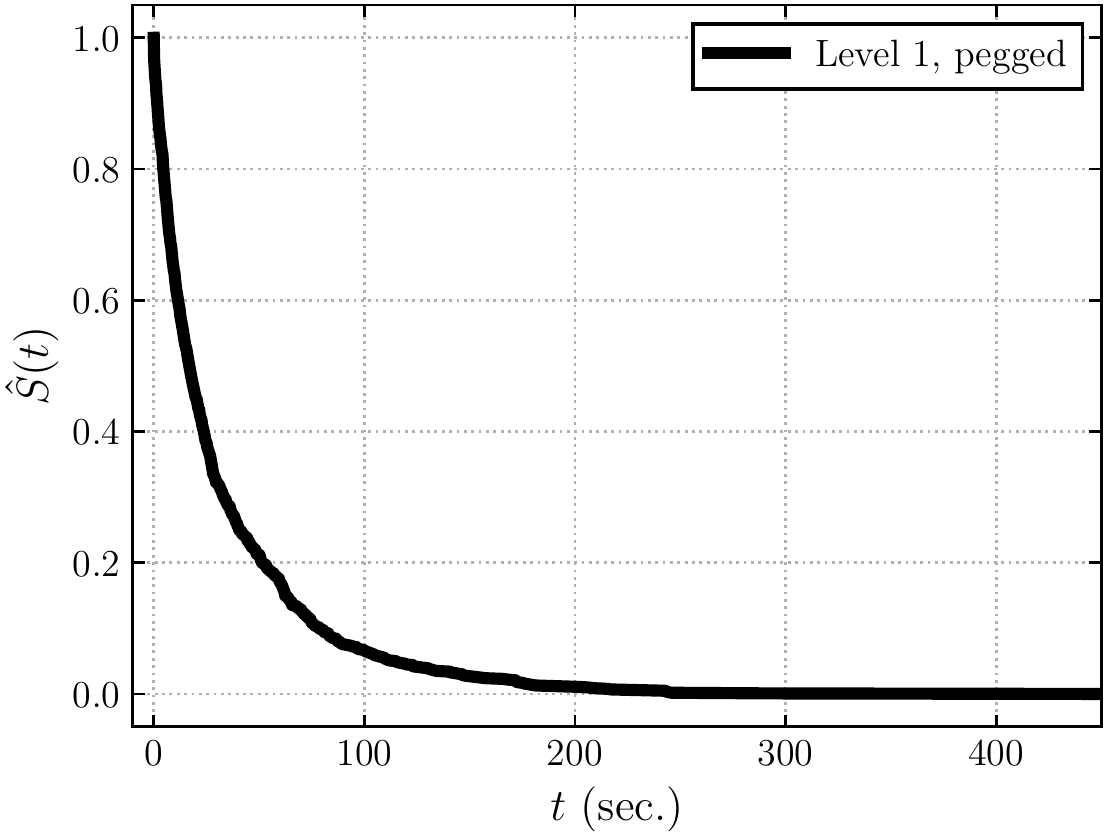}
	\caption{\textbf{Left: }Kaplan--Meier estimates of orders placed at different levels of the AAPL LOB. \textbf{Right: } Kaplan--Meier estimate of hypothetical order pegged to the first level in the LOB of AAPL.}
	\label{fig:kaplan_meier_AAPL}
\end{figure} 

Another approach is to use ``hypothetical" limit orders, which are orders of one share in volume placed last in the queue (i.e. following regular price-time priority) of a level in the order book. In particular, we consider the survival function of the order that follows the price of a particular level. We refer to this as \textit{pegging} the limit order to that price level. Such an approach probes a set of ``fill conditions", see \ref{app:a} for details, every time the order book is updated to determine if an order has been filled. We assume that hypothetical limit orders do not have market impact, which is consistent with the work of \cite{Handa1996} and \cite{Maglaras2021}. With hypothetical limit orders, we model different behaviour to what is observed in the order book data \textit{ex post}. \\

Pegged orders resting outside the best level in the LOB result in censored data because very few aggressive orders walk the LOB beyond the best quotes. Therefore, we focus on estimating the fill probability of orders pegged to the best bid and to the best ask of the LOB.\footnote{For simplicity, we also focus on the best levels with tracked orders, as they also contain a large amount of censoring which results in higher computational cost to obtain data to train the model.} Furthermore, we select the orders to track randomly throughout the trading day. The survival functions are visualised with Kaplan--Meier estimates (see \ref{app:b} for more details on their estimation) in Figure \ref{fig:kaplan_meier_AAPL}. This shows an example of the decrease in fill probability over time for different levels, as well as the increase in fill probability for order repegging. However, this visualisation only provides an averaged view of the survival function and cannot capture the effect of microstructural features on the fill probability of the orders; this motivates our deep learning approach, which we present in the next section.

\subsection{Fill statisics of limit orders placed inside the bid-ask spread}

Here, we compute the fill probabilities of orders placed within the spread of the order book. We consider nine stocks, some of which trade with a small or large tick and also vary in their average trading activity. Table \ref{table:sta_small_large_ticks} shows the average spread, average volume on the best bid and best ask, and average trades per minute over the month of October 2022.\footnote{Large tick stocks have a bid-ask spread that is on average close to one tick in size, while for small tick stocks the spread is several times larger than the tick. There is a significant difference in the dynamics between the two types, see \citet{Eisler2012} for more details. This is a consequence of the ratio between the price and the tick size of the exchange. When this ratio is low, traders tend to be more reluctant to cross the bid-ask spread, which results in the formation of large queues at the best bid and best ask. In our case, we consider a large tick stock to have an average spread of fewer than 1.3 ticks, as detailed in \cite{Binkowski2022}.} \\

\begin{table}[]
\vspace{-1cm}
	\caption{Statistics of small and large tick stocks on October 2022.}
	\fontsize{11.0}{15.0}\selectfont
	\centering % used for centering table
	\begin{tabular}{lrrrrr}
		\specialrule{.1em}{.1em}{.1em}
		&  \textbf{Avg. spread } &  \textbf{Avg. volume} &  \textbf{Avg. volume} &  
        \textbf{Avg. midprice} &
        \textbf{Avg. executed} \\ 
        &  \textbf{(ticks)} &  \textbf{best ask} &  \textbf{best bid} &  
        \textbf{} &
        \textbf{trades/min.}  \\
        
        \specialrule{.1em}{.1em}{.1em}
		\textbf{AAPL} & 1.44 & 487.19 & 442.03 & 144.34 & 405.02   \\
		\textbf{AMZN} & 1.91 &  298.41 & 307.72  & 114.80 & 319.44 \\
		\textbf{BIDU} & 8.75 &  138.98 & 97.24 & 101.89 & 20.86  \\
		\textbf{COST} & 30.89 & 67.80 & 60.89  & 478.21 & 44.27 \\
		\textbf{DELL} & 1.77 &  287.02 & 283.92 & 35.97 & 14.48 \\
            \textbf{GOOG} & 1.66 & 299.26 & 232.08 & 99.59 & 120.84 \\
		\textbf{MSFT} & 2.88 & 159.17 & 159.14 & 236.89 &  204.86 \\
		
		\textbf{CSCO} & 1.15 &  2212.01 & 2193.81 & 41.90 & 74.87 \\
		\textbf{INTC} & 1.13 & 5995.54 & 5533.79 & 26.55 & 110.88  \\\specialrule{.1em}{.1em}{.1em}
	\end{tabular}
	\label{table:sta_small_large_ticks}
\end{table}

\begin{figure}[]
\includegraphics[height =6.75cm,width=0.49\textwidth]{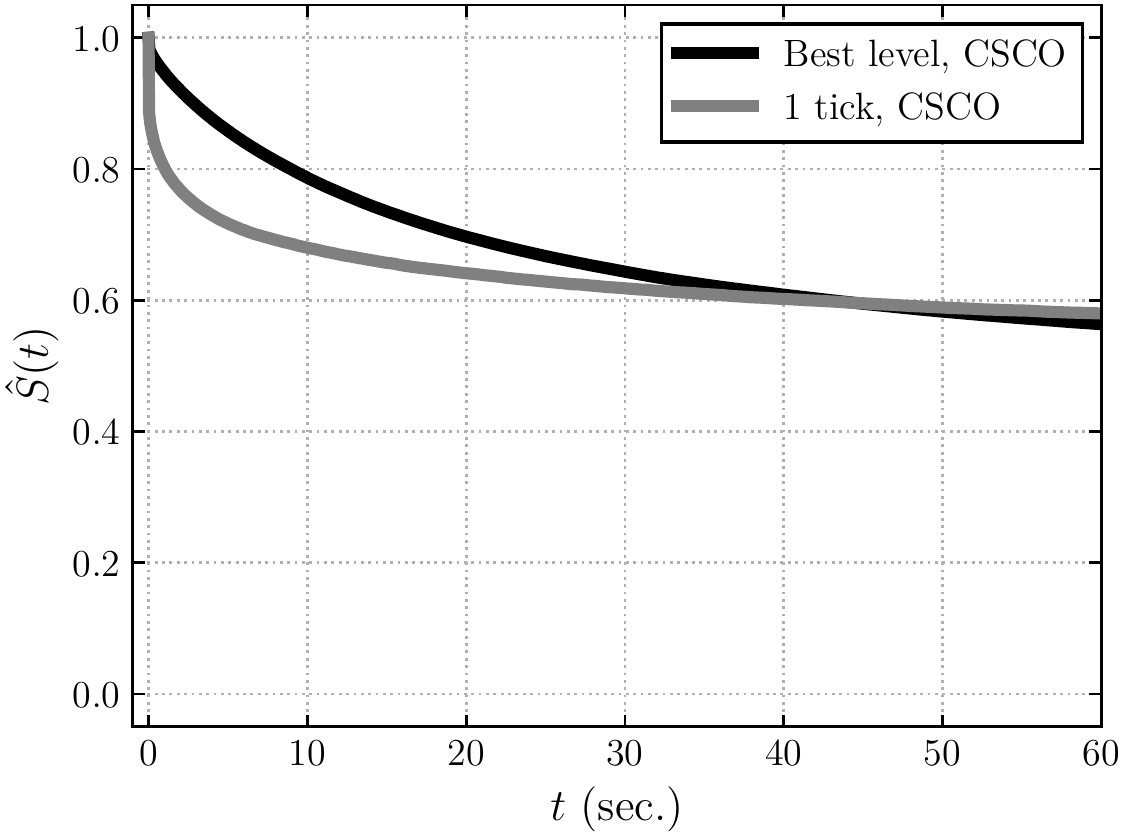}
\hspace*{\fill}
\includegraphics[height = 6.75cm,width=0.49\textwidth]{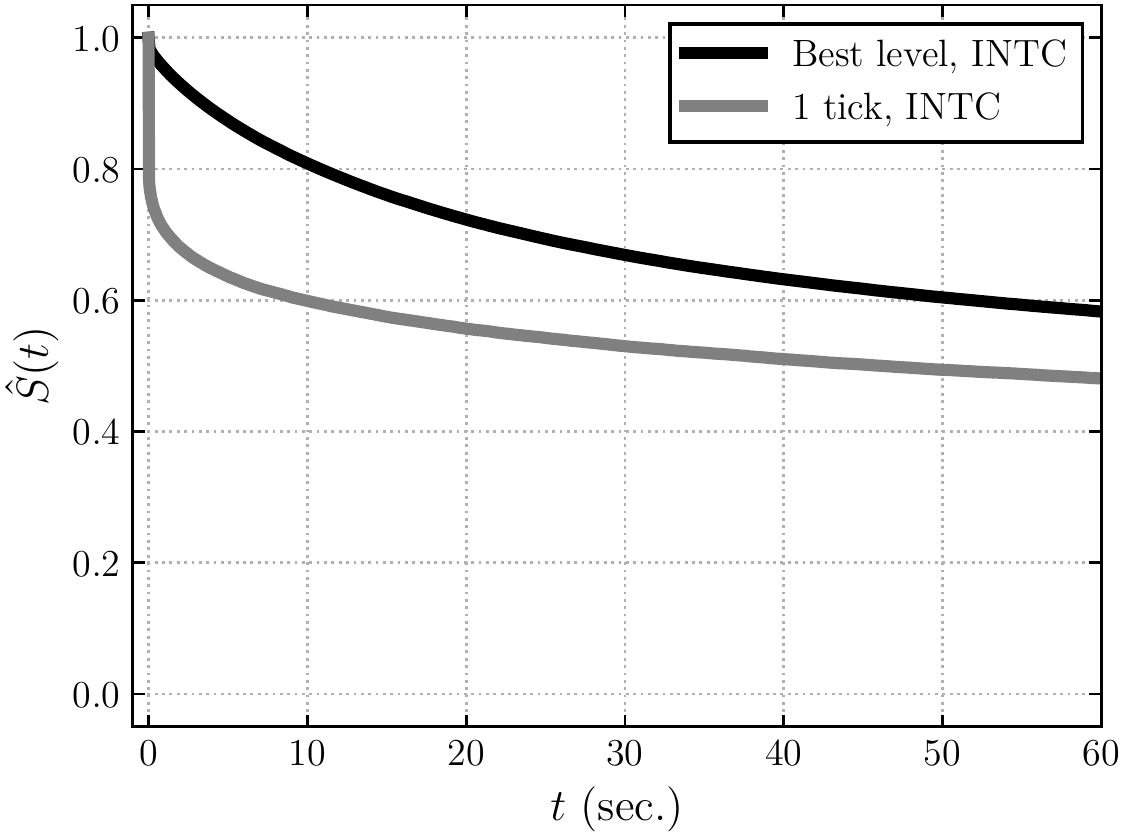}
\vspace*{\fill}
\includegraphics[height =6.75cm,width=0.49\textwidth]{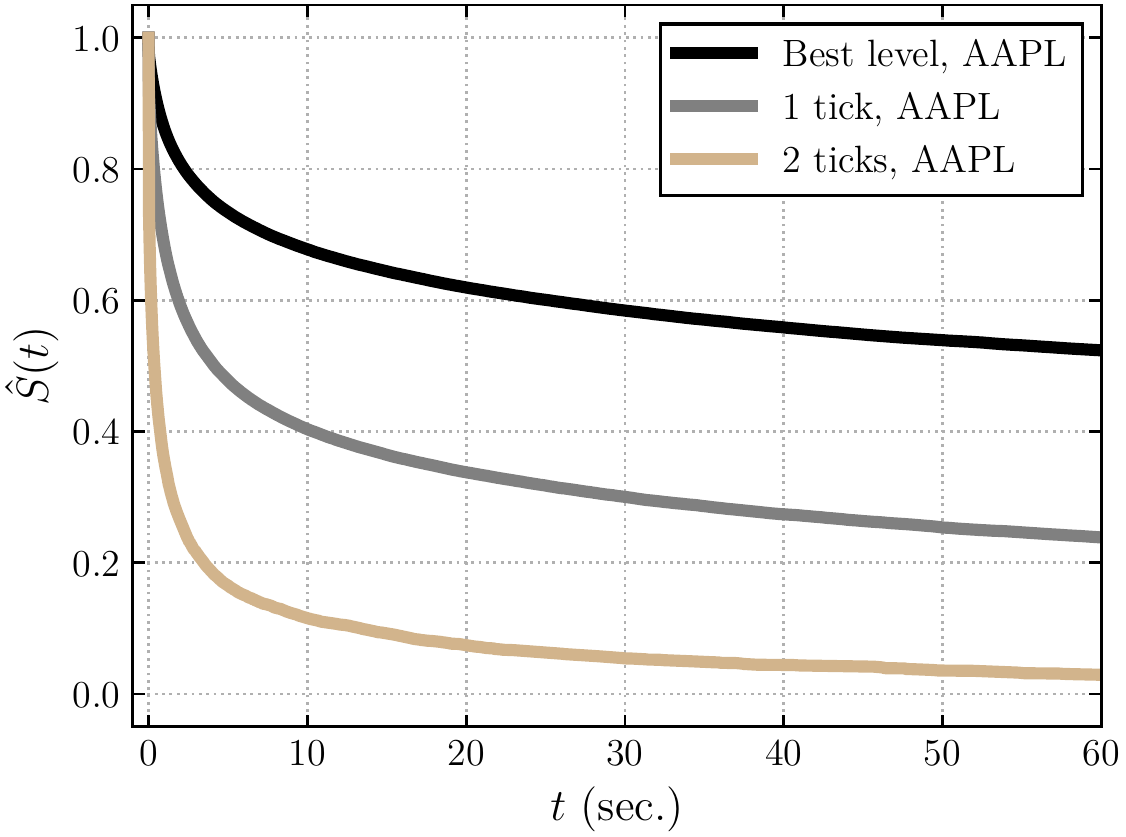}
\hspace*{\fill}
\includegraphics[height = 6.75cm,width=0.49\textwidth]{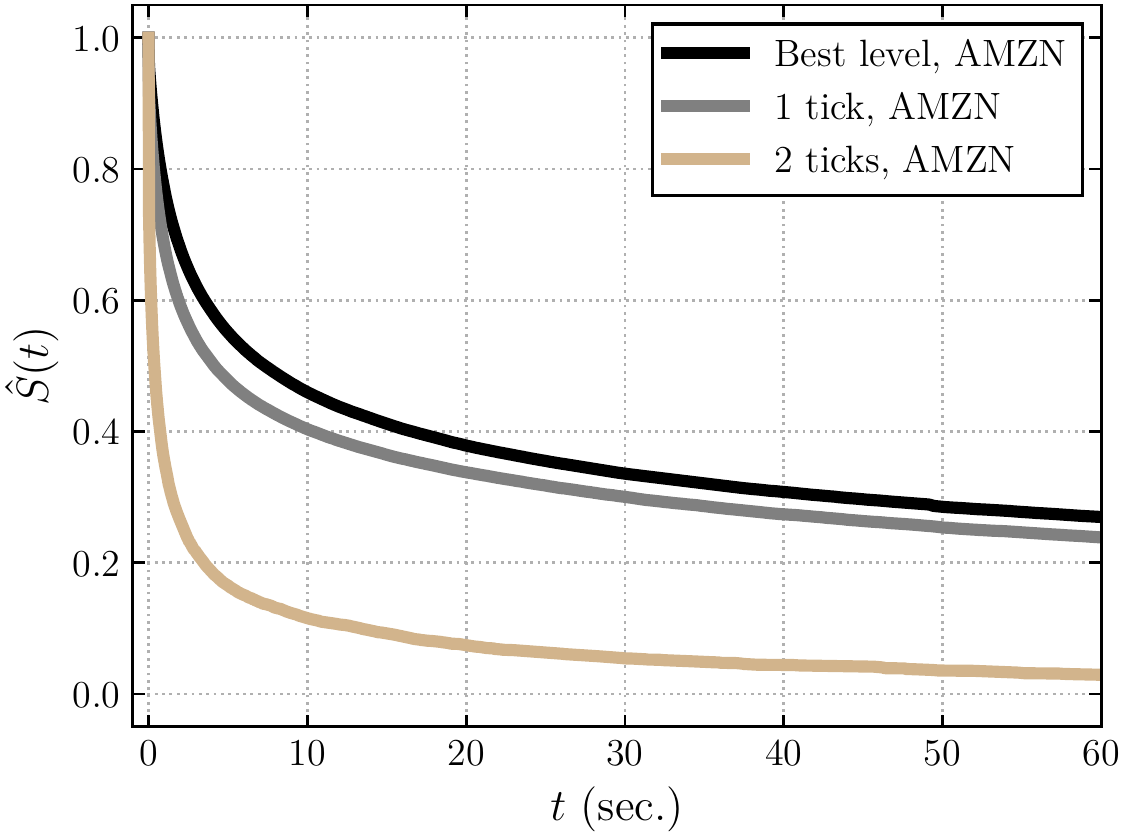}
\vspace*{\fill}
\includegraphics[height =6.75cm,width=0.49\textwidth]{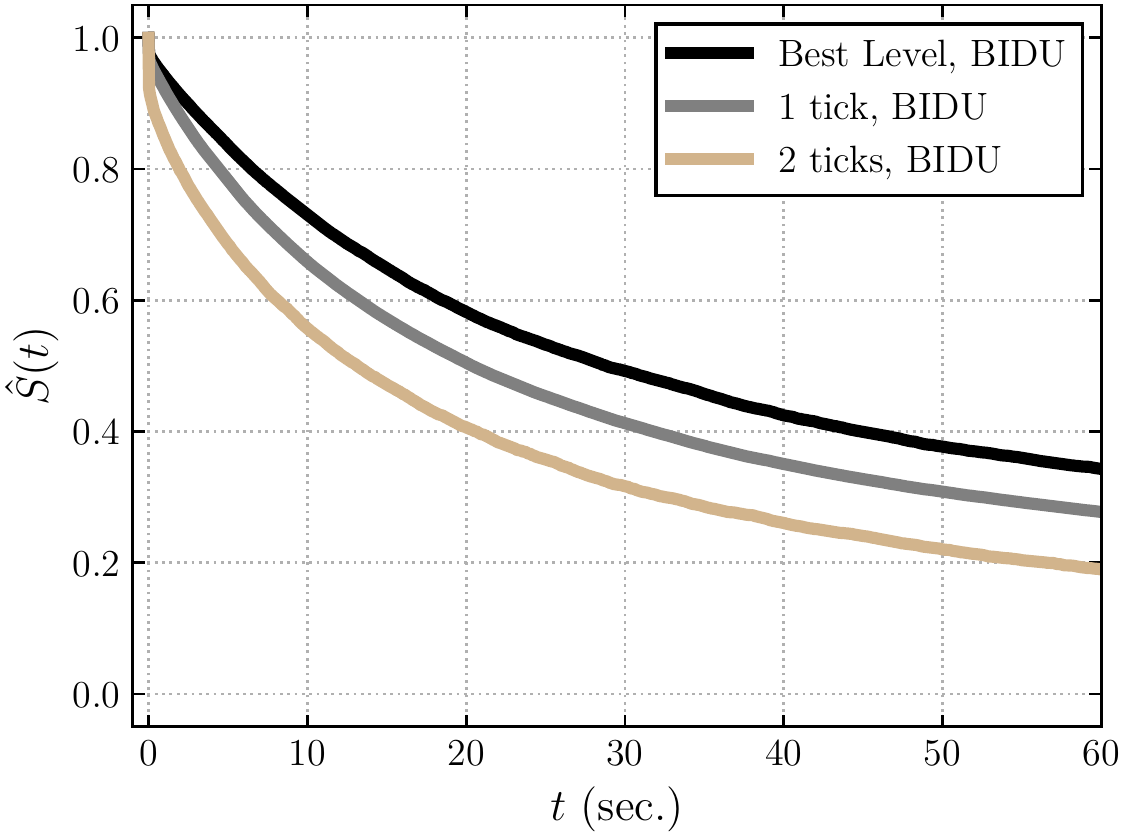}
\hspace*{\fill}
\includegraphics[height = 6.75cm,width=0.49\textwidth]{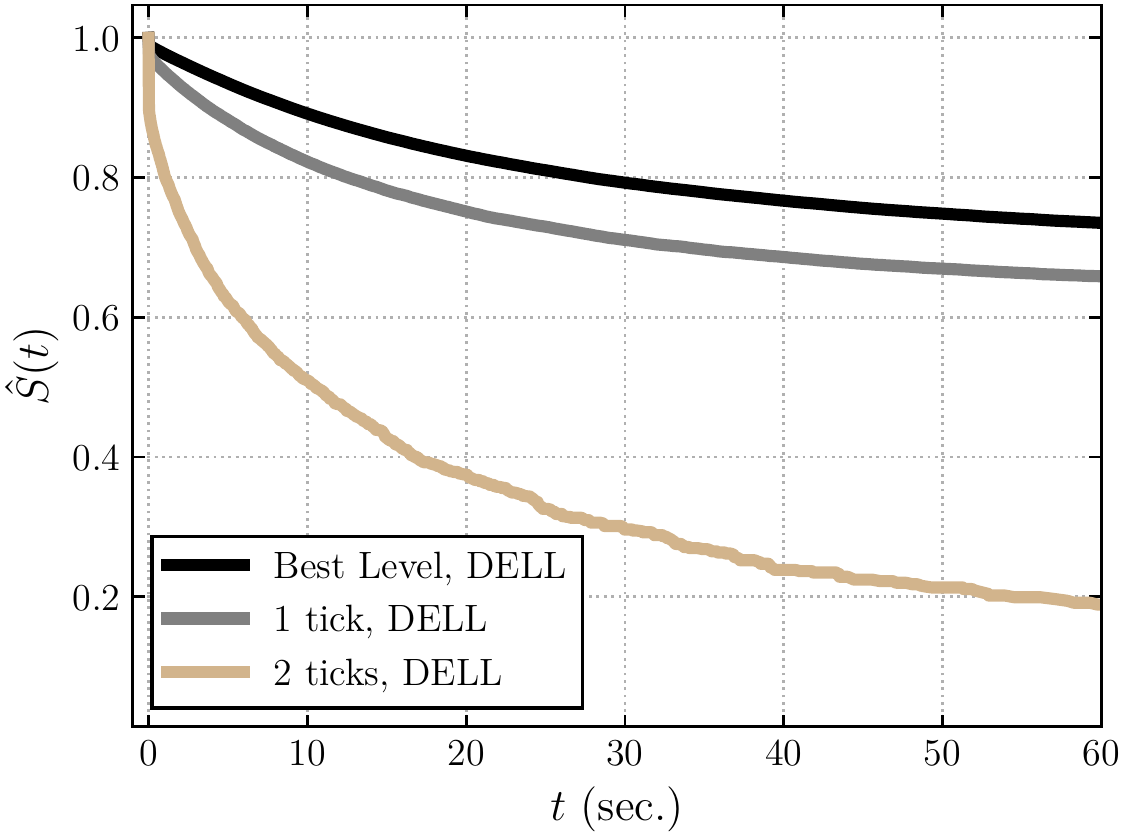}
\caption{Kaplan-Meier estimates of survival functions when placing orders at different depths of the bid-ask spread. \textbf{Top left: }CSCO, \textbf{Top left: }INTC, \textbf{Middle left: }AAPL, \textbf{Middle right: }AMZN, \textbf{Bottom left: }BIDU, \textbf{Bottom right: }DELL.}
\label{fig:surv_func_AAPL_CSCO}
\end{figure} 

Intuitively, agents trading large tick stocks are incentivised to place orders inside the bid-ask spread (whenever this is possible) not to place their LOs at the end of the queue. The cost of executing this trade is almost as much as crossing the spread, and there is no guarantee of immediate execution. However, it is approximately 3 to 6 times more likely that an order is filled when it improves the best quotes and reduces the time to fill by a similar proportion, see Figure \ref{fig:surv_func_AAPL_CSCO} and Table \ref{table:sta_small_large_fpft_ticks}. This is expected because traders who observe orders that improve the best quotes will aim to quickly match them with a liquidity taking order before they are cancelled. Figure \ref{fig:surv_func_AAPL_CSCO} also shows that the fill probability of orders placed in the spread of large tick stocks, when this is possible, rises sharply for small time horizons  before eventually decreasing to levels comparable to those of orders resting at the best level in the asset's limit order book. This suggests that if an order is executed within a few seconds of being placed into the book, the price moved in the favorable direction and there is a fill because the order is at the beginning of the queue. If this is not the case, it is likely that the price moved adversely, making the order rest deeper in its side of the book. Further, the survival function of stocks with larger trading activity exhibit a quicker decay than those of less active stocks.

\begin{table}[H]
	\caption{Fill statistics at the best level and at different depths in the spread between 1 October and 27 December 2022.}
    \vspace{-0.15cm}
	\fontsize{8.5}{15.0}\selectfont
	\centering % used for centering table
	\begin{tabular}{c|cccccc|cccccc}
    		\specialrule{.1em}{.1em}{.1em}
        & \multicolumn{6}{|c}{\textbf{Fill probability}} & \multicolumn{6}{|c}{\textbf{Avg. Filltime (s.)}} \\
        		\specialrule{.1em}{.1em}{.1em}
	\multirow{2}{*}{\textbf{Depth}} & \textbf{Best} &               \textbf{1 Tick} &  \textbf{2 Ticks} &              \textbf{3 Ticks}  &  \textbf{4 Ticks} &               \textbf{5 Ticks}  &  \textbf{Best} &               \textbf{1 Tick} &  \textbf{2 Ticks} &              \textbf{3 Ticks}  &  \textbf{4 Ticks} &               \textbf{5 Ticks}  \\
    & \textbf{} &               \textbf{Inside} &  \textbf{Inside} &              \textbf{Inside}  &  \textbf{Inside} &               \textbf{Inside}  &  \textbf{Inside} &               \textbf{Inside} &  \textbf{Inside} &              \textbf{Inside}  &  \textbf{Inside} &               \textbf{Inside} \\
        		\specialrule{.1em}{.1em}{.1em}
		\textbf{AAPL} & 0.0539 & 0.1317   & 0.3601  & 0.4165  & 0.4382 & 0.4431 &  1.32  & 0.64  & 0.19 &  0.11 & 0.14   & 0.08    \\
		\textbf{AMZN} & 0.0923 & 0.1312  & 0.3360 & 0.4139   &  0.4343 & 0.4485 & 1.07 & 0.70  & 0.38 & 0.25  & 0.11  & 0.21  \\
		\textbf{BIDU} & 0.0948 &  0.0528 & 0.1533 & 0.1692 & 0.1868 & 0.2140 & 4.40 &  3.82  & 3.93 & 3.59 & 3.26  & 3.07   \\
		\textbf{COST} & 0.0458 & 0.0122 & 0.0826  & 0.0898  & 0.1061 & 0.1135 & 5.18  & 4.85 & 6.03 & 5.33 & 4.69  & 4.53  \\
		\textbf{CSCO} & 0.0573 & 0.1753 & 0.3549 & 0.3579 & 0.2484  &  0.4558  & 6.26  & 1.71 & 0.47 & 0.51 & 0.25 &  0.36 \\
            \textbf{DELL} & 0.0392 & 0.0620  & 0.2263 & 0.2762  & 0.2902  & 0.2995 & 6.87 & 4.46 & 3.39 & 2.07 & 2.34 & 1.5 \\
		\textbf{GOOG} & 0.0618 & 0.1088 & 0.3199 & 0.3969   & 0.4450 & 0.5012 & 1.82 & 0.85 & 0.84 & 0.13 & 0.15 &  0.07 \\
		\textbf{INTC} & 0.0570 &  0.2962 & 0.3215 &  0.2463  &  0.2474 & 0.1165 & 7.33 &  1.95 & 0.27  & 0.29 & 0.8 & 0.71   \\
		\textbf{MSFT} & 0.0679 & 0.0734 & 0.2632 & 0.3690  &  0.4292  & 0.4466 & 0.99 & 0.70 &  0.45 & 0.24 & 0.17 & 0.12  \\
        		\specialrule{.1em}{.1em}{.1em}
	\end{tabular}
	\label{table:sta_small_large_fpft_ticks}
\end{table}

The activity inside the spread of small tick stocks is several orders of magnitude bigger than that of large tick stocks, see Table  \ref{table:sta_small_large_ticks_n}. Everything else being equal, this shows how improbable it is for spreads to widen in large tick stocks. Obtaining improvements in the fill probabilities for small tick assets requires placing limit orders closer to the other side of the book, where trading activity is comparable to that of LOs placed one tick closer to the other side in large tick stocks.

\begin{table}[H]
	\caption{Trading activity inside the spread considered between 1 October 2022 and 27 December 2022.}
	\fontsize{11.0}{14.0}\selectfont
	\centering % used for centering table
	\begin{tabular}{rrrrrr}
		
		\specialrule{.1em}{.1em}{.1em}
		\multirow{2}{*}{\textbf{Depth}} &                \textbf{1 Tick} &  \textbf{2 Ticks} &              \textbf{3 Ticks}  &  \textbf{4 Ticks} &               \textbf{5 Ticks}     \\
    &    \textbf{Inside} &  \textbf{Inside} &              \textbf{Inside}  &  \textbf{Inside} &               \textbf{Inside}  \\ 
		
		\specialrule{.1em}{.1em}{.1em}
		\textbf{AAPL} &  51,096  & 17,862  &  11,692  & 4,059 &  2,004          \\
		\textbf{AMZN} &  6,215,869  & 169,024   & 24,456    & 6,469 & 2,470  \\
		\textbf{BIDU} &  1,250,831  & 163,567   & 100,719   & 63,407 & 41,017  \\
		\textbf{COST} & 2,594,091  &  113,709 & 87,820  & 69,689 & 61,457 \\
		\textbf{CSCO} & 466,645   & 2,817 &  841 & 318 & 136 \\
		\textbf{DELL} &   753,603 & 15,584 & 2,548 & 937 & 454 \\
		\textbf{GOOG} &  1,548,080  & 52,565 & 8,344 & 2,274  & 782 \\
		\textbf{INTC} &  317,834  & 992 & 276 & 97 & 103 \\
		\textbf{MSFT} &  9,654,578 &  492,218  & 107,990  & 33,885 & 13,710  \\\specialrule{.1em}{.1em}{.1em}
	\end{tabular}
	\label{table:sta_small_large_ticks_n}
\end{table}

\section{Monotonic Encoder-Decoder Convolutional-Transformer}
\label{sec:model}
\subsection{General Architecture}

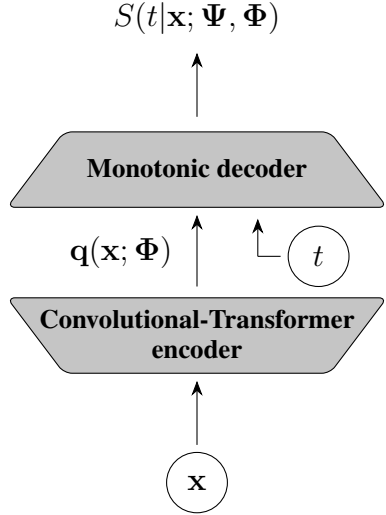
\begin{figure}[H]
	\begin{center}
		\begin{tikzpicture}
			\centering
			%\draw (2cm,2cm) node[anchor=south] {$\mathbf{x}$};
			
			%\draw [-{Stealth[length=2mm]}] (2cm, 2.5cm) -- (2cm, 3cm);
			
			%\draw[fill=grey!40!white,rounded corners=1pt] (0.95cm,3.1cm) -- (3.05cm,3.1cm) -- (3.5cm,3.8cm) -- (0.5cm,3.8cm) -- cycle;
			
			%\draw (2cm,3.3cm) node[anchor=south] {$\text{\footnotesize\textbf{mTAND encoder}}$};
			\draw (2cm,3.3cm) circle [radius=0.4] node {$\mathbf{x}$};
			%\draw (2cm,3.3cm) node[anchor=south] {$\mathbf{x}$};
			
			\draw [-{Stealth[length=2mm]}] (2cm, 3.8cm) -- (2cm, 4.65cm);
			
			\draw[fill=grey!40!white,rounded corners=3pt] (0.25cm,4.75cm) -- (3.75cm,4.75cm) -- (4.5cm,5.75cm) -- (-0.5cm,5.75cm) -- cycle;
			
			\draw (2cm,4.8cm) node[anchor=south] {$\text{\footnotesize\textbf{encoder}}$};
			\draw (2cm,5.2cm) node[anchor=south] {$\text{\footnotesize\textbf{Convolutional-Transformer}}$};
			
			%\draw (1.2cm,3.9cm) node[anchor=south] {$\mathbf{q}(\mathbf{x}; \boldsymbol{\Phi})$};
			
			\draw [-{Stealth[length=2mm]}] (2cm, 5.9cm) -- (2cm, 6.85cm);
			\draw [-{Stealth[length=2mm]}] (2.8cm, 6.3cm) -- (2.8cm, 6.8cm);
			\draw (2.8cm, 6.3cm) -- (3.1cm, 6.3cm);
			
			\draw[fill=grey!40!white,rounded corners=3pt] (-0.5cm,6.95cm) -- (4.5cm,6.95cm) -- (3.75cm,7.95cm) -- (0.25cm,7.95cm) -- cycle;
			
			\draw (2cm,7.2cm) node[anchor=south] {$\text{\footnotesize\textbf{Monotonic decoder}}$};
			
			\draw (1cm,6.0cm) node[anchor=south] {$\mathbf{q}(\mathbf{x}; \boldsymbol{\Phi})$};
			
			\draw (3.6cm,6.3cm) circle [radius=0.4] node {$t$};
			%\draw (3.3cm,5.55cm) node[anchor=south] {$t$};
			
			\draw [-{Stealth[length=2mm]}] (2cm, 8.15cm) -- (2cm, 9.0cm);
			
			\draw (2cm,9.1cm) node[anchor=south] {$S(t|\mathbf{x}; \boldsymbol{\Psi},\boldsymbol{\Phi})$};
		\end{tikzpicture}   
	\end{center} 
	\caption{Encoder-decoder architecture to estimate the survival function. The first block, an encoder with parameter $\boldsymbol{\Phi}\in\mathbb{R}$, uses an attention-based mechanism to project the LOB observations to a latent representation that captures relevant information. The second block, a monotonic decoder takes as input both this latent representation of the time series and the time variable \textit{t}. The weights of the decoder are  positive to enforce a monotonically decreasing survival function.}
	\label{fig:encoder_decoder}
\end{figure}

In this section, we present our encoder-decoder architecture which learns the mapping between states of the LOB and distribution of limit order filltimes. Figure \ref{fig:encoder_decoder} illustrates the two components of our framework. The encoder, parameterised by $\boldsymbol{\Phi} \in \mathbb{R}^{m_{\Phi}}$, processes the LOB data and obtains a latent representation from it, which is  used by the decoder, parameterised by $\boldsymbol{\Psi} \in \mathbb{R}^{m_{\Psi}+}$, to predict the survival function of the limit orders. The decoder comprises a monotonic neural network that guarantees a monotonically decreasing survival function. Further, we use a convolutional-Transformer encoder to model the complex dependencies and interactions within the LOB data and to compress useful information into a lower-dimensional representation, subsequently used by the monotonic-decoder.

\subsection{Convolutional-Transformer Encoder}
\label{subsec:encoder}

\begin{figure}[H]
	%\vspace{0.5cm}
	\begin{center}
		\centerline{\includegraphics[width=0.6\columnwidth]{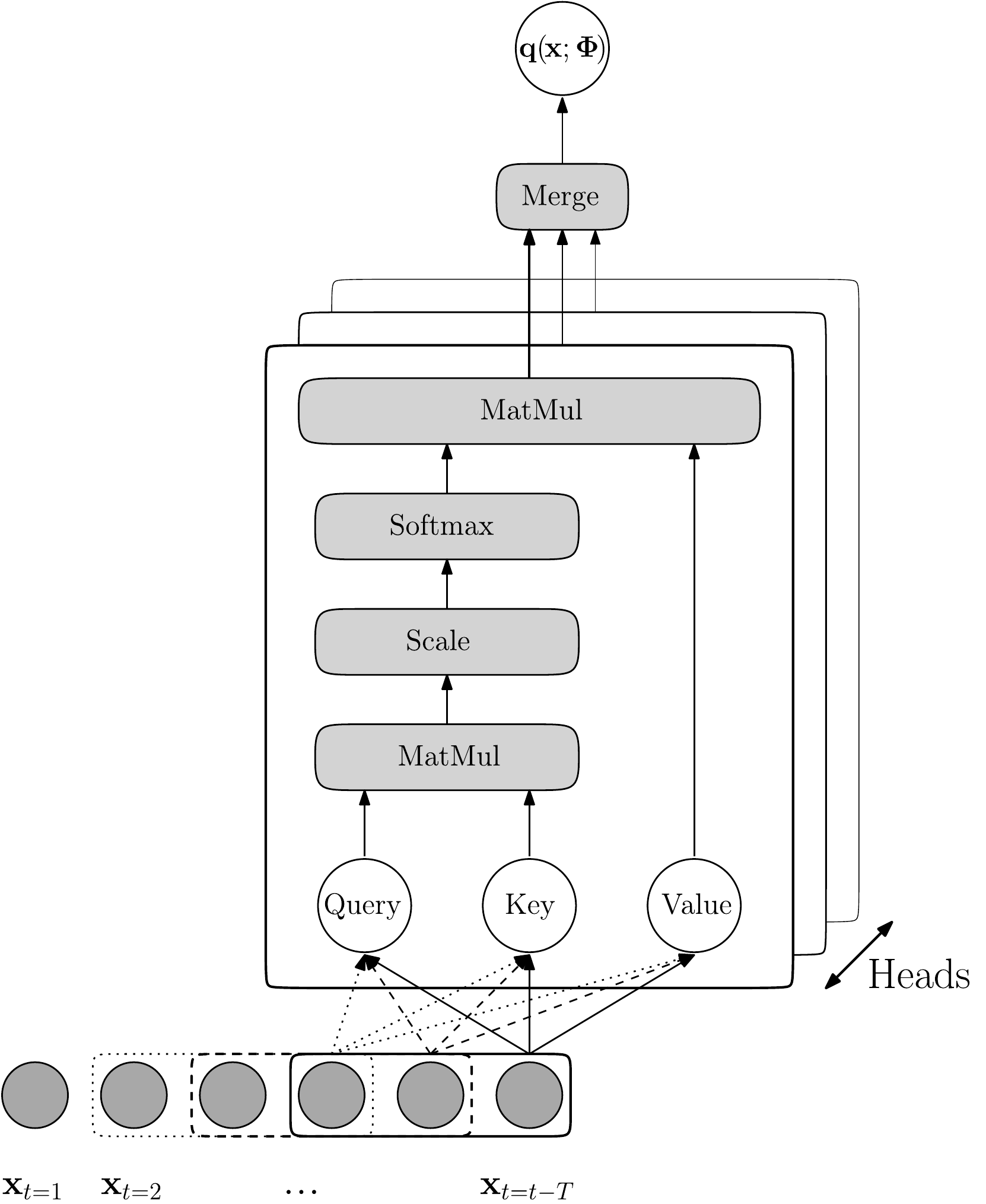}}
		\caption{Structure of the convolutional-Transformer's encoder. In this diagram, the convolutional kernel is of size $s=3$ with dilation factor $p=1$. The locally-aware hidden representation obtained by the CNN is used to feed the Transformer, which uses self-attention over these hidden variables to obtain the latent representation $\mathbf{q}(\mathbf{x};\Phi)$. } 
		\label{fig:encoder_architecture}
	\end{center}
\end{figure}

We propose a convolutional-Transformer encoder to identify patterns in the LOB and to obtain accurate estimates of the fill probabilities of limit orders. The architecture of the encoder, see Figure \ref{fig:encoder_architecture}, processes the LOB time series data and captures its non-Markovian dynamics through  a latent representation of the time series. This representation encapsulates the most relevant information which is used by the decoder to predict the fill probabilities. \\

The encoder consists of two components: a locally-aware convolutional network and a Transformer model. The locally-aware convolutional network consists of three different Dilated Causal Convolutional (DCC) neural networks \citep{oord2016wavenet} that process the LOB data and generate the corresponding queries, keys, and values which serve as inputs to the Transformer model. These DCCs, which are based on Convolutional Neural Networks (CNNs) \citep{lecun1989backpropagation, Zhang2019, zhang2021multi}, use inner product operations based on entries that are a fixed number of steps apart from each other, contrary to CNNs and Causal-CNNs, which operate with consecutive entries, as shown in \eqref{eq:cnn_self_attention}.  Further, causal convolutions ensure that the current position does not use future information. Previously, DCCs have been successfully applied in time series forecasting \citep{borovykh2017conditional, moreno2022deepvol}. 

The queries, keys, and values created by the DCCs are three different representations resulting from the convolution operation on the LOB input features and are collectively used by the Transformer model to perform self-attention and capture dependencies between different parts of the original time series. 
Using convolutional networks to generate the input features to the Transformer  allows our encoder to be more aware of local context, granting the Transformer the ability to discern if observed values are anomalies, part of patterns, etc. This constitutes a clear advantage over using a multi-layer perceptron (MLP) to obtain the queries, keys, and values, which is the most common approach in the literature.  
Therefore, the operation of the DCCs can be understood as a set of data-driven local filters. The projection they perform from the original LOB time series to a hidden representation enhances the depiction of the LOB dynamics. This operation enables the Transformer's self-attention mechanism to capture complex local dependencies between datapoints, because it operates on a locally aware hidden representation, rather than point-wise values that lack local context. Additionally, the convolutional-Transformer optimises the parameters of each of the three DCCs to extract different relevant features from the LOB data. For example, some filters may be optimised to detect trends, while others may identify anomalies or changepoints. Each of the three convolutional neural networks used to obtain the corresponding  query, key, and values, that serve as input to the Transformer model \citep{vaswani2017attention}, consists of only one layer performing a causal convolutional operation between the input sequence, $\mathbf{x}\in\mathbb{R}$, and the corresponding convolutional kernel $\mathbf{k}$ of size $s \in\mathbb{Z}$:
\begin{equation}
\begin{cases}
Q(t) &= \left(\mathbf{x} *_{p} \mathbf{k}^{Q}\right)(t)=\sum\limits_{\tau=0}^{s-1} k_{\tau}^{Q} \cdot x_{t-p\cdot \tau}\,,\\[10pt]
K(t) &= \left(\mathbf{x} *_{p} \mathbf{k}^{K}\right)(t)=\sum\limits_{\tau=0}^{s-1} k_{\tau}^{K} \cdot x_{t-p\cdot \tau}\,,\\[10pt]
V(t) &= \left(\mathbf{x} *_{p} \mathbf{k}^{V}\right)(t)=\sum\limits_{\tau=0}^{s-1} k_{\tau}^{V} \cdot x_{t-p\cdot \tau}\,,\\
\end{cases}
\label{eq:cnn_self_attention}
\end{equation}
where $p$ is the dilation factor.\footnote{ Note that  $p=1$ results in a Causal-CNN.} 

We use the convolutional network solely as a feature extractor that incorporates local context into the Transformer's self-attention mechanism because, in our model, the Transformer model is only responsible for extracting the patterns within the data. Therefore, we restrict the encoder's convolutional network to a single layer, but its complexity can be easily extended to $L$ convolutional layers, see \ref{app:e}. \\ 

After the CNNs extract the relevant features from the LOB data and produce the queries, keys, and values, these are fed to the Transformer model. %, which processes them and obtains a latent representation that encodes the most relevant  information from the selected features. 
Transformer models were initially introduced for Natural Language Processing (NLP), but they have been widely applied in time series-related problems \citep{moreno2023deep}. These models propose a new architecture that leverages the attention mechanism \citep{bahdanau2014neural} to process sequences of data. They have significant advantages over more classical approaches because of their ability to maintain lookback windows with large horizons, which makes them able to detect long-term dependencies in the data. On the other hand, canonical Transformers' point-wise dot-product attention makes them prone to anomalies and optimisation issues because of their space complexity, which grows quadratically with the input length. Furthermore, canonical Transformers are locally-agnostic, because dot-product attention does not allow the model to be aware of the local context while operating with time series data. The convolutional-Transformer alleviates these problems: the integration of convolutional networks makes the model locally-aware and a sparse self-attention mechanism mitigates its space complexity, reducing the cost of computing the attention scores from $\mathcal{O}(L^2)$ to $\mathcal{O}(L(\log(L))^2)$, where $L$ is the input length.

The Transformer-based encoder model grounds its operation on the well-known self-attention mechanism, performed simultaneously by a different number of Transformer's heads $H \in \mathbb{Z}$, resulting in what is known as a \textit{multi-head Transformer}; see Figure \ref{fig:encoder_architecture}.
Each multi-head self-attention sublayer simultaneously applies the scaled dot-product attention over the convolutional network's output. For the $i^{\text{th}}$ head, this scaled dot-product attention is given by
\begin{equation}
    \operatorname{h}_{\mathrm{i}}=\operatorname{Attention}\left({Q}_i, {K}_i, {V}_i\right)=\operatorname{softmax}\left(\frac{{Q}_i {K}_i^T}{\sqrt{d_k}}M \right) {V}_i\ ,\label{eq:self-attention} 
\end{equation}
where $d_k$ is a scaling factor and M is a mask to prevent information leakage.  \\

Each Transformer's head is therefore responsible for learning and modelling attention functions that handle the complex dependencies within the LOB data. With the different heads, the model jointly attends to different temporal subspaces of the original time series. The individual embeddings of each head are then combined to obtain a joint representation
\begin{equation}
\operatorname{MultiHead}(\mathbf{Q}, \mathbf{K}, \mathbf{V})=\operatorname{Concat}\left(\operatorname{h}_1, \operatorname{h}_2, \ldots, \operatorname{h}_i, \ldots \operatorname{h}_{\mathrm{H}}\right)\,, 
\end{equation}
where $\operatorname{h}_i$ represents the output of head $i^{\text{th}}$, and $\operatorname{h}_{\mathrm{i}}=\operatorname{Attention}\left(Q_i, K_i, V_i\right)$.
%The Transformer model is therefore responsible for learning and modelling attention functions able to handle the complex dependencies in the LOB data. Finally, every head's output is merged together through a linear function, obtaining the latent representation $\mathbf{q}(\mathbf{x};\phi$). 
Merging every head's output through a linear function produces the latent representation $\mathbf{q}(\mathbf{x};\boldsymbol{\Phi}$), which encodes the most relevant information from the selected features of the LOB. %This latent representation is used by the decoder to predict the orders' conditional density of survival times in a monotonically-decreasing manner. \\

\subsection{Monotonic Decoder}
\label{subsec:decoder}

\begin{figure}[H]
	\centering
	\scalebox{.9}{\def\layersep{1}
\usetikzlibrary{calc} 
\usetikzlibrary{snakes}
\begin{tikzpicture}[draw=black!50, node distance=\layersep]
            
    \tikzstyle{every pin edge}=[<-,shorten <=1pt]
    \tikzstyle{positive}=[snake=coil,segment aspect=0,segment amplitude=1pt];
    \tikzstyle{neuron}=[circle,fill=gray!30,minimum size=12pt,inner sep=0pt]
    \tikzstyle{input x neuron}=[neuron, fill=gray!30,text width=4mm,align=center];
    \tikzstyle{input y neuron}=[neuron, fill=gray!30,text width=4mm,align=center];
    %\tikzstyle{cdf neuron}=[neuron, fill=gray!30];
    \tikzstyle{pdf neuron}=[neuron, fill=gray!30,text width=10mm,align=center];
    \tikzstyle{output neuron}=[neuron, fill=gray!30,text width=10mm,align=center];
    \tikzstyle{hidden neuron}=[neuron, fill=gray!30,text width=4mm,align=center];
    \tikzstyle{annot} = [text width=4em, text centered]
    \tikzstyle{arrow}=[shorten >=1pt,->]

    %final x trnasformation
    \node[hidden neuron] (xh_out1) at (0,2*\layersep) {$x_1$};
    \node[hidden neuron] (xh_outd) at (\layersep,2*\layersep) {$x_d$};
    \node (xh_out_dots) at ($(xh_out1)!.5!(xh_outd)$) {\ldots};
    
    % dots between x hidden and x final transformation
    %\node at ($(xh_dots)!.5!(xh_out_dots)$) {\ldots};
    
    % y input
    \node[input y neuron] (y) at (2*\layersep,2*\layersep) {$t$};
    
    % 1st layer mixing x and y
    \node[hidden neuron] (xy_1) at (0.5,3*\layersep) {};
    	\node[hidden neuron] (xy_d) at (1.5,3*\layersep) {};
    	\node (xy_dots) at ($(xy_1)!.5!(xy_d)$) {\ldots};
    	
    	% connections between x out, y and 1st xy layer
    \path[arrow] (xh_out1) edge (xy_1);
    \path[arrow] (xh_out1) edge (xy_d);
    \path[arrow] (xh_outd) edge (xy_1);
    \path[arrow] (xh_outd) edge (xy_d);
    \draw[arrow,positive]  (y) -- (xy_1);
    \draw[arrow,positive] (y) -- (xy_d);
    
    %1st hiddent xy layer
    \node[hidden neuron] (xy_h1_1) at (0.5,4*\layersep) {};
    	\node[hidden neuron] (xy_h1_d) at (1.5,4*\layersep) {};
    	\node (xy_h1_dots) at ($(xy_h1_1)!.5!(xy_h1_d)$) {\ldots};
    	
    	% connections between 1st xy layer and 1st hidden xy layer
    \draw[positive,arrow] (xy_1) -- (xy_h1_1);
    \draw[positive,arrow] (xy_1) -- (xy_h1_d);
    \draw[positive,arrow] (xy_d) -- (xy_h1_1);
    \draw[positive,arrow] (xy_d) -- (xy_h1_d);
    	
    	%2nd hiddent xy layer
    \node[hidden neuron] (xy_h2_1) at (0.5,5*\layersep) {};
    	\node[hidden neuron] (xy_h2_d) at (1.5,5*\layersep) {};
    	\node (xy_h2_dots) at ($(xy_h2_1)!.5!(xy_h2_d)$) {\ldots};
     
    % dots between 1st hiddent xy layer and 2nd hiddent xy layer
    \node at ($(xy_h1_dots)!.5!(xy_h2_dots)$) {\ldots};
    
    \node[pdf neuron] (pdf) at (1,6*\layersep) {};
    \node[output neuron] (output) at (1,8*\layersep) {};
    
    %last layer to CDF
    \draw[positive,arrow] (xy_h2_1) -- (pdf);
    \draw[positive,arrow] (xy_h2_d) -- (pdf);
    
    %\node[pdf neuron] (pdf) at (1,7.5*\layersep) {$\frac{\partial{F(y|\mathbf{x})}}{\partial{y}}$};
    \node[pdf neuron, minimum size=1.0cm] (pdf) at (1,6*\layersep) {\vspace*{1.0pt}\footnotesize{$\hat{S}(t | \mathbf{x})$}};
    \node[output neuron, minimum size=1.0cm] (output) at (1,8*\layersep) {\vspace*{1.0pt}\footnotesize{$\hat{f}(t | \mathbf{x})$}};
    
    %cdf 
    \draw[snake=triangles,segment object length=3pt, segment length=3pt] (pdf) -- node[right]{$-\frac{\partial}{\partial t}$} (output);
    
    	 %pdf out edge
    	 %\draw[snake=triangles,segment object length=3pt, segment length=3pt] (pdf) -- node[right]{$f(y|\mathbf{x})$} (1,8.5*\layersep);
    	
    	%legend
    	\path[arrow] (3.5, 3) edge node[at end, label=right:{$w_{ij} \in \mathbb{R}$}]{} (4.5,3);
    	\draw[positive,arrow] (3.5, 2.5) -- node[at end, label=right:{$w_{ij} \in \mathbb{R}^+$}]{} (4.5,2.5);
    	\draw[snake=triangles,segment object length=3pt, segment length=3pt] (3.5,2) -- node[at end, label=right:{No weights}]{} (4.5,2);
    	\node[input x neuron] (input_x_legend) at (3.5,4) {$x_i$} node[right=0pt of input_x_legend] {Input feature};
    	%\node[input y neuron] (input_y_legend) at (2.5,1) {$y$} node[right=0pt of input_y_legend] {response variable}; 
    	%\node[cdf neuron] (cdf_legend) at (2.5,0.5) {$t$} node[right=0pt of cdf_legend] {$\sigma(bias+\sum_i w_{ij}input_i)$};
    	%\node[hidden neuron] (hidden_legend) at (2.5,0) {} node[right=0pt of hidden_legend] {$\tanh(bias+\sum_i w_{ij}input_i)$}; 
        \node[input y neuron] (input_y_legend) at (3.5,3.5) {$t$} node[right=0pt of input_y_legend] {Time}; 
    	
    	%layers annotation
	%\draw[snake=brace]  (-0.3, 3*\layersep) -- node[midway, label=left:{$w_{ij} \in \mathbb{R}^+$}]{} (-0.3, 6*\layersep);
    	
    	%\draw[snake=brace]  (-0.3,2*\layersep) -- node[align=left,midway,anchor=east]{$w_{ij} \in \mathbb{R}$ \\ $w_{ij} \in \mathbb{R}^+$} (-0.3, 3*\layersep);
    	
    	%\draw[snake=brace]  (-0.3,0) -- node[midway, label=left:{$w_{ij} \in \mathbb{R}$}]{} (-0.3, 2*\layersep);

\end{tikzpicture}}
	\caption{Monotonic decoder's structure. The last node represents the operation of differentiating the conditional survival distribution with respect to time, which results in the conditional density function of the survival time. This model guarantees a decreasing survival curve. (Adapted from Figure 1 of \citet{Chilinski2020}.)}
	\label{fig:decoder}
\end{figure}
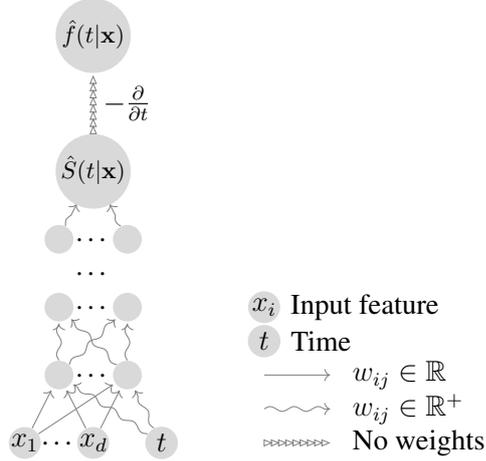

In the context of survival analysis, the survival function needs to be decreasing with respect to time. We encode this inductive bias  into our architecture to avoid the well-known \textit{crossing problem} \citep{Tagasovska2019}. To this end, we use  monotonically restricted neural networks \citep{Chilinski2020, Rindt2022} in our monotonic decoder, see Figure \ref{fig:decoder}. This type of neural network allows us to estimate a cumulative density function (CDF) denoted by $F(t|\mathbf{x})$ with response variable $t$ and conditioned on input features $\mathbf{x}$, which in our case is the latent representation obtained from the LOB time series through the encoder. The output of the decoder's network, $f_{\boldsymbol{\Psi}}(\ \cdot\ )$, is consistent with the properties of a CDF because it satisfies: (i) $\lim_{t\rightarrow -\infty}f_{\boldsymbol{\Psi}}(t,\mathbf{x}) = 0\,,$ (ii) $\lim_{t\rightarrow \infty}f_{\boldsymbol{\Psi}}(t,\mathbf{x}) = 1\, ,$ (iii) $\frac{\partial f_{\boldsymbol{\Psi}}(t,\mathbf{x})}{\partial t}\ge 0\, ,$
\begin{comment}
   \begin{enumerate}[label=(\roman*), leftmargin=*, itemsep=0.4ex, before={\everymath{\displaystyle}}]%
  \item $\lim_{t\rightarrow -\infty}f_{\boldsymbol{\Psi}}(t,\mathbf{x}) = 0$\ , \label{cdf-eq-1}
  \item $\lim_{t\rightarrow \infty}f_{\boldsymbol{\Psi}}(t,\mathbf{x}) = 1$\ , \label{cdf-eq-2}
  \item $\frac{\partial f_{\boldsymbol{\Psi}}(t,\mathbf{x})}{\partial t}\ge 0$\ , \label{cdf-eq-3}
\end{enumerate} 
\end{comment}
where the third condition is the most difficult to guarantee because neural networks are a composition of nonlinear functions, which makes them difficult to interpret or control. See \ref{app:monotonic} for more details. \\

We remark that imposing this monotonicity on the decoder does not hinder other beneficial properties of deep neural networks such as universal function approximation \citep{Cybenko1989, Kidger2020} or convexity \citep{Littwin2020} in the over-parameterised case. The reason behind this is that the restriction is only enforced on the decoder, which can be made arbitrarily small (in terms of parameters) compared to the time series encoder, where the network parameters are not restricted.

\section{Experiments}
\label{sec:experiments}

\subsection{Predictive Features}

\begin{figure}[H]
\includegraphics[height = 6.75cm,width=0.49\textwidth]{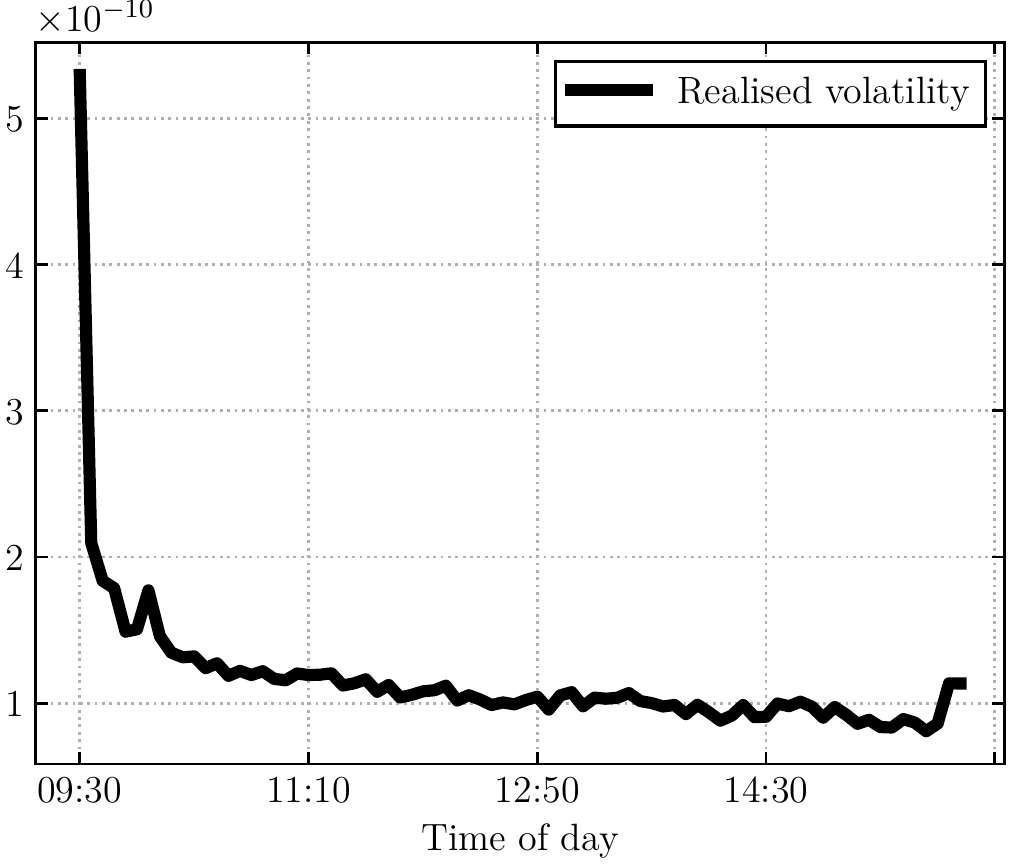}
\hspace*{\fill}
\includegraphics[height = 6.5cm,width=0.49\textwidth]{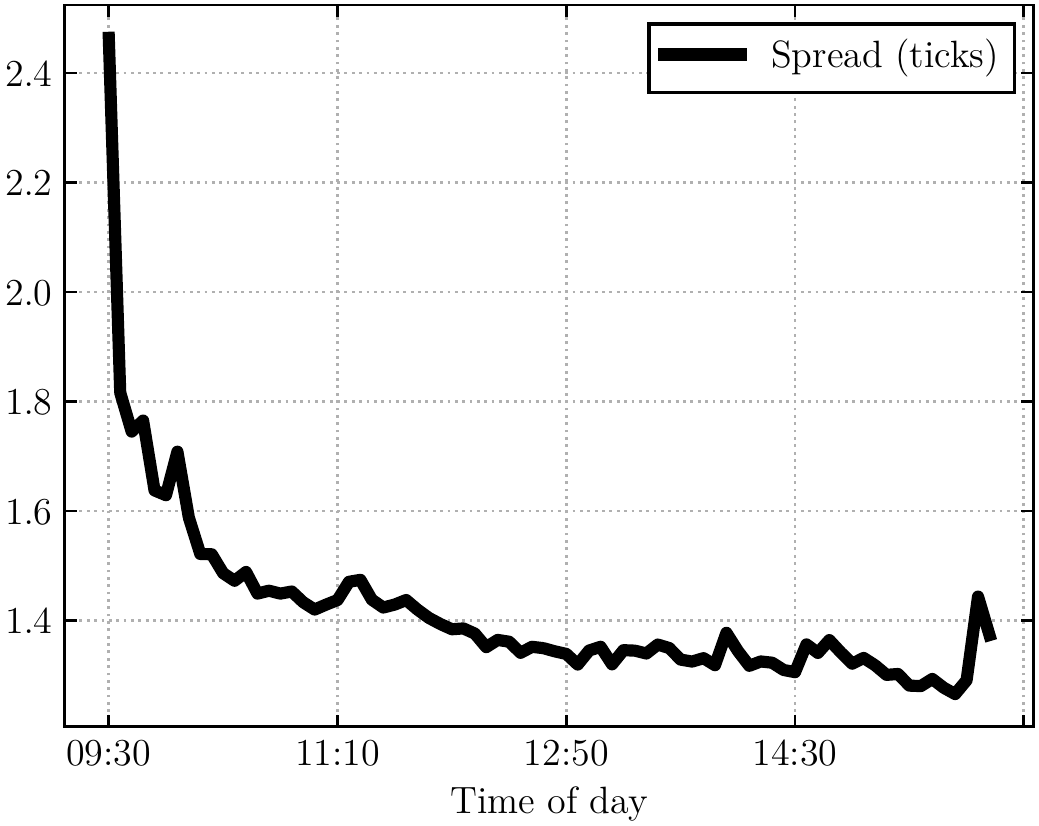}
\caption{Realised volatility and spread of AAPL stock over October 2022. \textbf{Left: } Realised volatility. \textbf{ Right: }Spread.}
\label{fig:realized_vol}
\end{figure}

To estimate the survival function we distinguish between \textit{slow-moving} and \textit{fast-moving} features. Slow moving features provide information about intraday patterns of the trading day, which have predictive power on the probability that a limit order will be filled. One such pattern is the intraday behaviour of the volatility of returns, which is generally high at the beginning of the trading day due to uncertainty and an adjustment to overnight information; see Figure \ref{fig:realized_vol} for an example with AAPL over the month of October 2022.\footnote{The volatility is estimated using a rolling mean of 1000 trades over the squared returns of the midprice time-series.}

\begin{figure}[H]
\includegraphics[height = 6.5cm,width=0.49\textwidth]{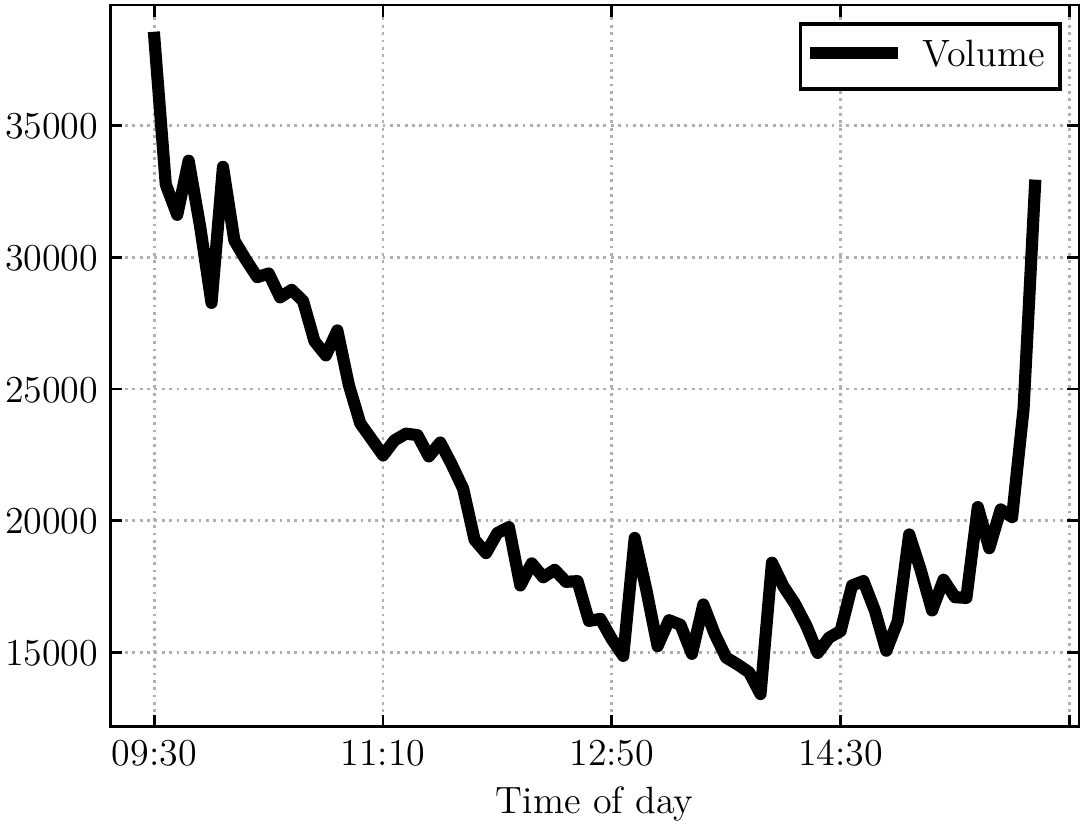}
\hspace*{\fill}
\includegraphics[height = 6.5cm,width=0.49\textwidth]{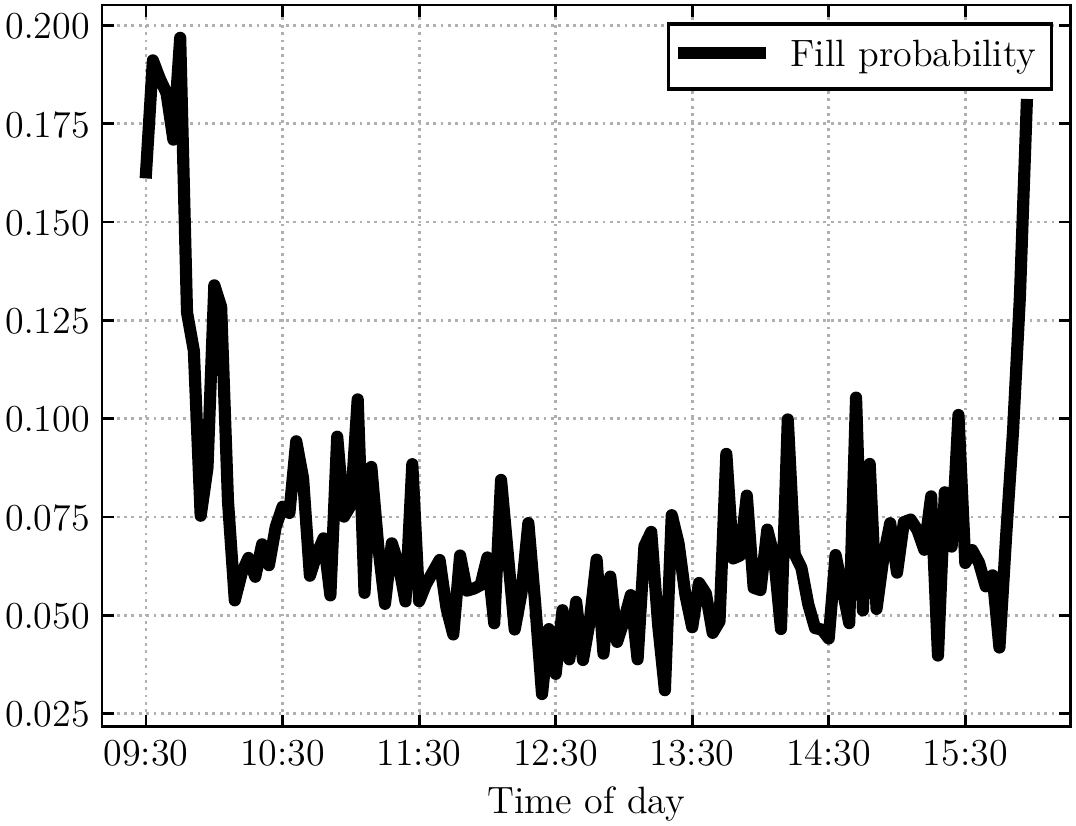}
\caption{5-minute bucket average statistics for AAPL stock over October 2022. \textbf{Left: }Daily traded volume. \textbf{ Right: } Fill probability of limit orders posted at the best level in the book.}
\label{fig:vol_fill}
\end{figure}

A similar effect is visible with daily traded volume. Overnight information generally causes larger trading volume at the beginning of the day, which tends to stabilise at a reduced value during the trading day, and peaks again at the end of the day when traders have more urgency to adjust their positions. This effect is shown in Figure \ref{fig:vol_fill}, along with the evolution of the fill probability (which summarises the asymmetry between supply and demand of liquidity in the order book) during the trading day. Given the persistence of these intraday seasonalities, it is feasible to obtain good estimates of the fill probability based on this information.\footnote{Another possible way of homogenising these effects is to consider the evolution of the day in \textit{transaction time}, see \ref{app:transaction}.}

Aside from seasonal patterns, we want to predict the changes in the fill probability over more granular time-scales by using \textit{fast-moving features}. In particular, we hypothesise that some of the variables that are most important in the estimation of the survival function of limit orders include future evolution of the bid-ask spread (smaller spreads would incentivise traders to place a liquidity taking order on the other side of the book), volatility (due to its correlation with the spread), or order arrival speed (as traders who observe large queues forming will be more incentivised to cross the bid-ask spread). Volatility and the bid-ask spread exhibit significant persistence and cross-correlations, see \cite{Binkowski2022}, which further motivates the use of an attention-based encoder to capture long-ranging dependencies between the different time-series. \\

\begin{figure}[]
\includegraphics[height = 6.75cm,width=0.49\textwidth]{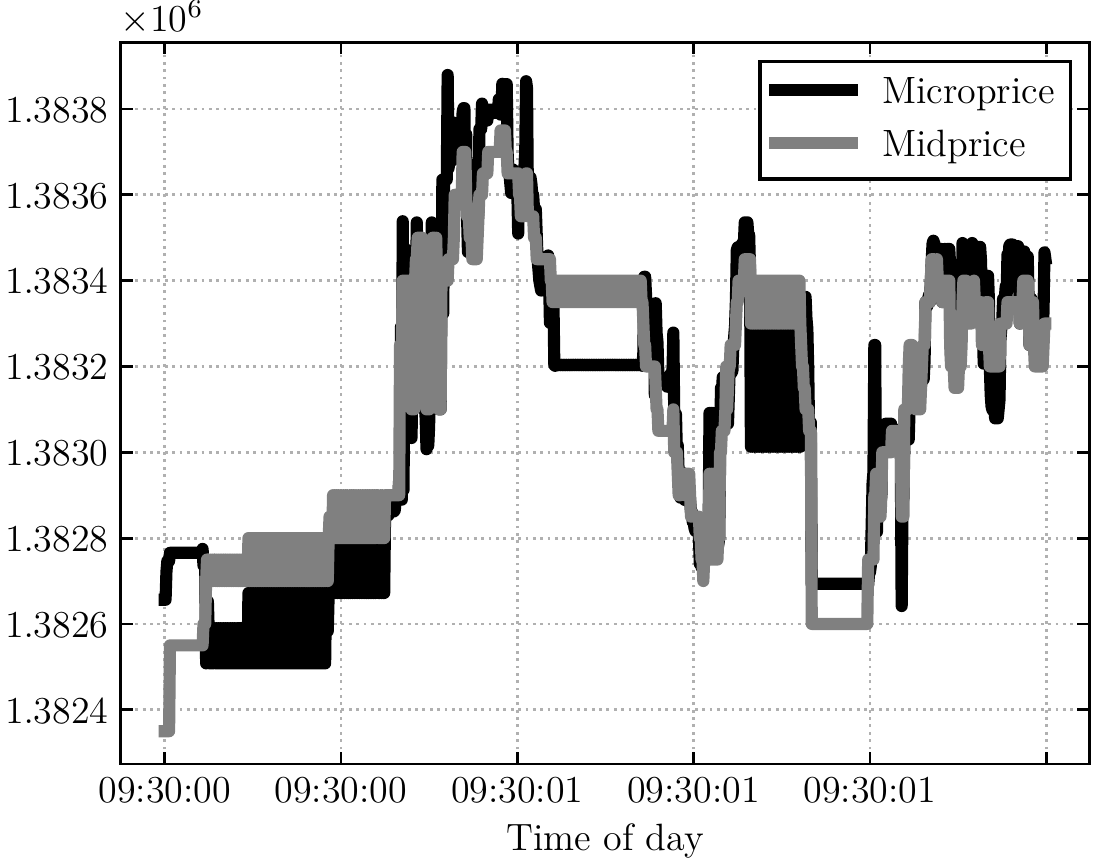}
\hspace*{\fill}
\includegraphics[height = 6.5cm,width=0.49\textwidth]{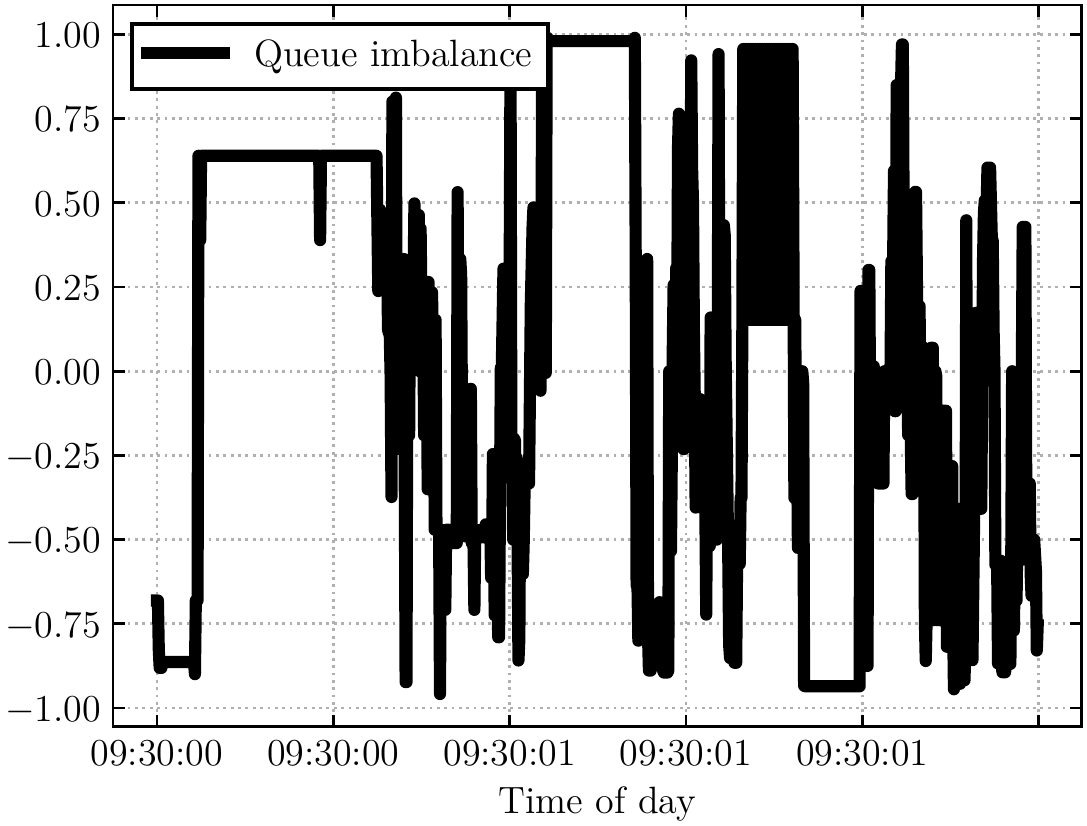}
\caption{Evolution of indicators over the first 1000 trades of 1 October 2022 for the AAPL stock. \textbf{Left:} Midprice and microprice. \textbf{Right:} Queue imbalance.}
\label{fig:mp_queue}
\end{figure} 

We build two signals with fast moving features. These are \textit{volume imbalance}, given by  
\begin{equation}
    \varUpsilon_t = \frac{v_b^1(t)-v_a^1(t)}{v_b^1(t)+v_a^1(t)}
    \in [-1,1]\,,
\end{equation}
and the \textit{microprice}, given by
\begin{equation}
    M_t 
    =
    \frac{v_b^1(t)}{v_b^1(t)+v_a^1(t)}\,p_a^1(t)
    +
    \frac{v_a^1(t)}{v_b^1(t)+v_a^1(t)}\,p_b^1(t)\,.
\end{equation}

Volume imbalance captures the difference between the buy and sell pressures in the order book at time $t$, and it is a predictor of the arrival of aggressive orders, limit orders, and short-term price moves, as detailed in \citet{Cartea2018} and \cite{Cartea2020}. When $\varUpsilon_t $ is close to $1$, there is buy pressure, and when it is close to $-1$, there is sell pressure. Similarly, the microprice reflects the tendency of the price to move toward the bid or the ask side of the book. 
An example evolution of these indicators over a horizon of 1000 trades is shown in Figure \ref{fig:mp_queue}. \\

These two signals are added as inputs to the model given their widespread use and well-known predictive power over short horizons. We also include the raw volumes and prices of the top five levels of the order book to allow our model to find more complex inter-dependencies directly from the data.

\subsection{Model Fit}

In this subsection, we report the results of our model. We compare the performance of our model with classic deep learning benchmarks from the survival analysis literature. In particular, we consider the DeepSurv \citep{Katzman2018} and DeepHit \citep{Lee2018} models. Furthermore, we test the effectiveness of our encoder by replacing it with a Multi-Layer Perceptron (MLP) (which results in the architecture introduced in \citet{Rindt2022}), a CNN, and a long short-term memory (LSTM) network \citep{Hochreiter1997}. The latter two models are the ones expected to be in closest competition with the convolutional transformer, as they also model the temporal dynamics observed in the data.  \\

\begin{figure}[H]
\centering
\includegraphics[height =6.75cm,width=0.49\textwidth]{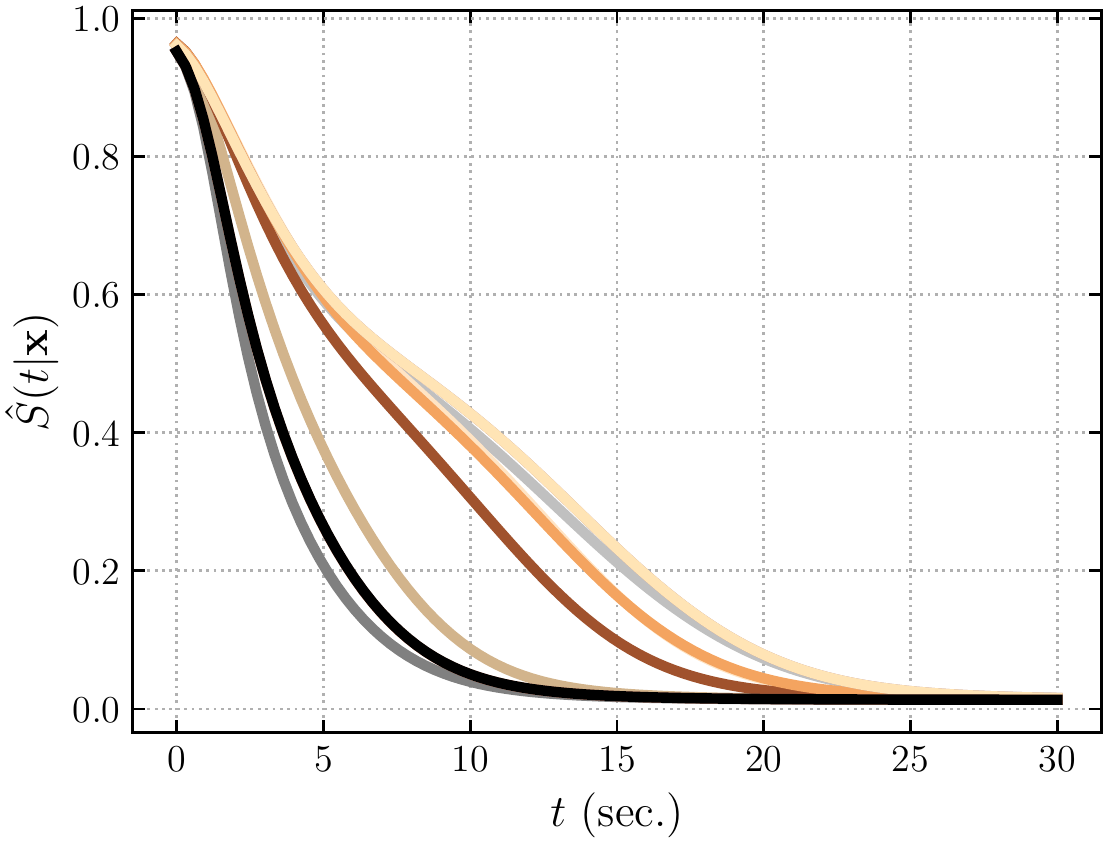}
\includegraphics[height =6.75cm,width=0.49\textwidth]{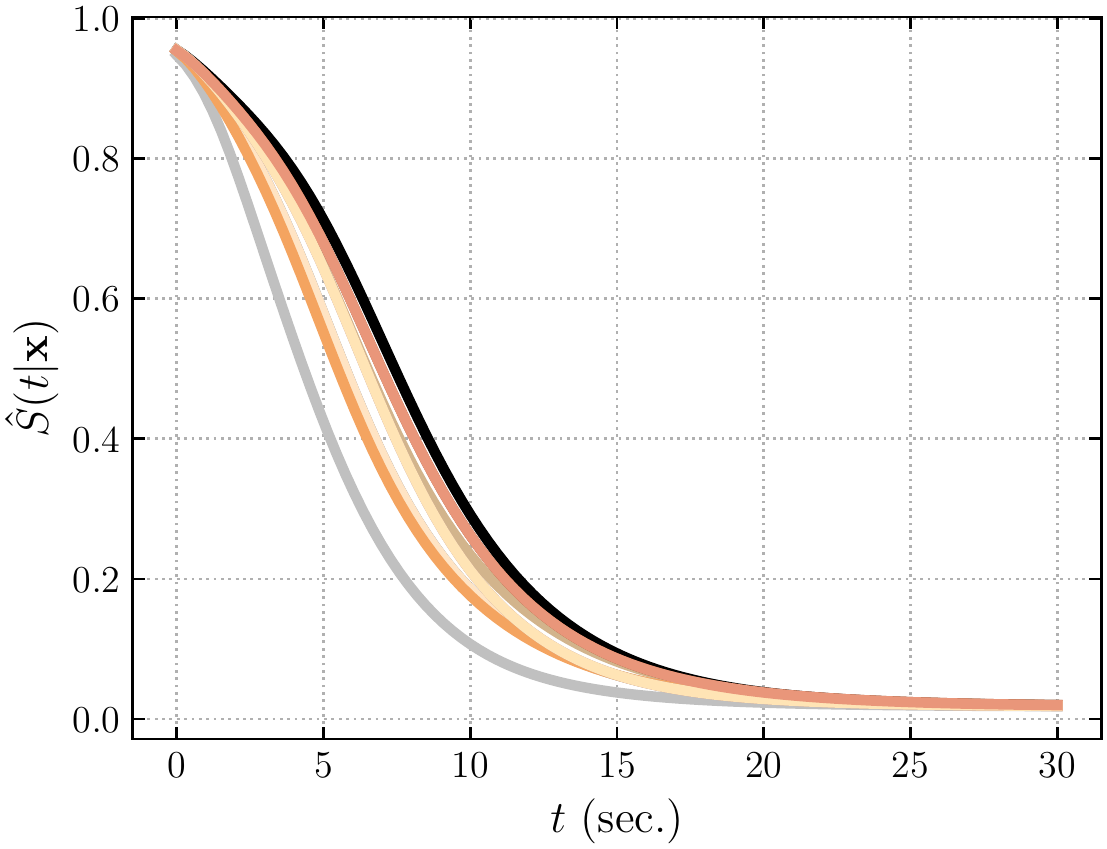}
\caption{Survival functions predicted by the encoder-decoder monotonic convolutional-Transformer model for a batch of different limit orders. \textbf{Left:} AAPL, \textbf{Right:} AMZN.}
\label{fig:convtrans_sf}
\end{figure} 

We train our model with data from 1 September 2022 to 26 December 2022. We use the tickers shown in Table \ref{table:sta_small_large_ticks}, and the features described in the previous subsection (time of day, volatility, volume imbalance, microprice, and prices and volumes of the best five levels). For each trading day, we choose either 100 orders that were placed by market participants into the order book, or we place and track the same number of hypothetical limit orders pegged to the best level of the book.\footnote{We choose random times during the trading day to track both hypothetical and tracked limit orders.} We store the features over different lookback horizons of 50, 500, and 1000 trades to explore whether information in the distant past is informative to predict the fill probability. \\

The performance of the model in terms of negative RCLL is provided in Tables \ref{table:T50-500-1000-RCLL-Tracked} and  \ref{table:T50-500-1000-RCLL} for the tracked and hypothetical limit orders, respectively. Tables \ref{table:T50-500-1000-Improvement-Tracked} and  \ref{table:T50-500-1000-Improvement} show the performance improvement of each model over the one using an MLP as an encoder for both order types. The best result for each ticker is in bold, with our proposed model  outperforming all benchmarks. All models which account for the time-varying dynamics of the LOB achieve significant performance gains, which suggests that high-frequency microstructural information plays a major role in the estimation. Furthermore, models with an LSTM or CNN encoder do not exhibit as significant gains (or even incur in some performance degradation) when considering longer lookback windows, due to the inability to summarise the information of the entire horizon. In contrast, the convolutional-Transformer encoder can weigh the information over the entire horizon by its relevance to the final estimate, allowing it to achieve the best performance for longer time windows. As mentioned, our model achieves this performance with fewer parameters compared to other models such as those using the CNN as an encoder. \\

\begin{table}[]
    \fontsize{10.0}{13}\selectfont
    \centering
    \caption{Model performance for observed orders dataset, in terms of RCLL.}
    \makebox[\textwidth]{
    \begin{tabular}{cccccccccc}
    \specialrule{.1em}{.1em}{.1em} \multicolumn{10}{c}{\textbf{Mean $\boldsymbol{\pm}$ STD Negative RCLL}} \\
    \specialrule{.1em}{.1em}{.1em} 
    \textbf{}              & \textbf{AAPL} & \textbf{AMZN} & \textbf{BIDU} & \textbf{COST} & \textbf{CSCO} & \textbf{DELL} & \textbf{GOOG} & \textbf{INTC} & \textbf{MSFT} \\
    \specialrule{.1em}{.1em}{.1em} 
    \multicolumn{10}{c}{\textit{No recurrence}}                                                                                                                            \\
     \specialrule{.1em}{.1em}{.1em} 
    \multirow{3}{*}{\textbf{DeepSurv}}     & 1.572         & 1.155         & 1.913         & 1.243         & 1.214         & 0.998         & 1.432         & 1.288        & 1.282         \\
    \textbf{}              & $\pm$            & $\pm$            & $\pm$            & $\pm$            & $\pm$            & $\pm$            & $\pm$            & $\pm$            & $\pm$            \\ 
    \textbf{}              & 0.032         & 0.369         & 0.192         & 0.326         & 0.286         & 0.035         & 0.261         & 0.214         & 0.049         \\  \hline
    \multirow{3}{*}{\textbf{DeepHit}}& 1.230         & 1.418         & 1.926         & 1.304         & 1.122         & 0.922         & 1.314         & 1.188         & 1.362         \\
    \textbf{}              & $\pm$            & $\pm$            & $\pm$            & $\pm$            & $\pm$            & $\pm$            & $\pm$            & $\pm$            & $\pm$            \\
    \textbf{}              & 0.054         & 0.098         & 0.093         & 0.243         & 0.089         & 0.096         & 0.116         & .0399         & 0.075         \\  \hline
    \multirow{3}{*}{\textbf{MN-MLP}}         & 1.465         & 1.754         & 1.943         & 1.987         & 1.273         & 1.312         & 1.553         & 1.511         & 1.912         \\
    \textbf{}              & $\pm$            & $\pm$            & $\pm$            & $\pm$            & $\pm$            & $\pm$            & $\pm$            & $\pm$            & $\pm$            \\
    \textbf{}              & 0.364         & 0.295         & 0.972         & 0.261         & 0.410         & 0.349         & 1.405         & 0.352         & 1.058         \\
     \specialrule{.1em}{.1em}{.1em} 
    \multicolumn{10}{c}{\textit{T=50}}                                                                                                                                     \\
     \specialrule{.1em}{.1em}{.1em} 
    \multirow{3}{*}{\textbf{MN-CNN}}       & 0.571         & 0.594         & 0.954         & 0.647         & 1.200         & 0.835         & 0.657         & 0.757         & 0.676         \\
    \textbf{}              & $\pm$            & $\pm$            & $\pm$            & $\pm$            & $\pm$            & $\pm$            & $\pm$            & $\pm$            & $\pm$            \\
    \textbf{}              & 0.095         & 0.071         & 0.042         & 0.037         & 0.165         & 0.057         & 0.006         & 0.174         & 0.109         \\  \hline
    \multirow{3}{*}{\textbf{MN-LSTM}}       & 0.820         & 0.390         & 1.265         & 0.883         & 0.884         & 0.559         & 1.234         & 1.066         & 0.963         \\
    \textbf{}              & $\pm$            & $\pm$            & $\pm$            & $\pm$            & $\pm$            & $\pm$            & $\pm$            & $\pm$            & $\pm$            \\
    \textbf{}              & 0.611         & 0.198         & 0.808         & 0.409         & 0.377         & 0.468         & 0.731         & 0.650         & 0.268         \\  \hline
    \multirow{3}{*}{\textbf{MN-Conv-Trans}} & \textbf{0.142}         & 0.132         & 0.406         & 0.178         & 0.242         & 0.168         & 0.233         & 0.341         & 0.180         \\
    \textbf{}              & $\boldsymbol{\pm}$            & $\pm$            & $\pm$            & $\pm$            & $\pm$            & $\pm$            & $\pm$            & $\pm$            & $\pm$            \\
    \textbf{}              & \textbf{0.185}         & 0.195         & 0.364         & 0.299         & 0.303         & 0.255         & 0.210         & 0.454         & 0.201         \\
     \specialrule{.1em}{.1em}{.1em} 
    \multicolumn{10}{c}{\textit{T=500}}                                                                                                                                    \\
     \specialrule{.1em}{.1em}{.1em} 
    \multirow{3}{*}{\textbf{MN-CNN}}       & 1.043         & 0.629         & 0.714         & 0.362         & 0.805         & 0.754         & 0.854         & 1.188         & 1.456         \\
    \textbf{}              & $\pm$            & $\pm$            & $\pm$            & $\pm$            & $\pm$            & $\pm$            & $\pm$            & $\pm$            & $\pm$            \\
    \textbf{}              & 0.177         & 0.059         & 0.104         & - 0.0         & 0.121         & 0.108         & 0.117         & 0.166         & 0.122         \\  \hline
    \multirow{3}{*}{\textbf{MN-LSTM}}      & 0.977         & 0.414         & 0.966         & 0.078         & 1.145         & 0.449         & 0.620         & 0.959         & 1.322         \\
    \textbf{}              & $\pm$            & $\pm$            & $\pm$            & $\pm$            & $\pm$            & $\pm$            & $\pm$            & $\pm$            & $\pm$            \\
    \textbf{}              & 0.500         & 0.097         & 0.374         & 0.129         & 0.517         & 0.311         & 0.208         & 0.423         & 0.546         \\  \hline
    \multirow{3}{*}{\textbf{MN-Conv-Trans}} & 0.180         & 0.177         & \textbf{0.296}         & \textbf{0.027}         & 0.365         & \textbf{0.167}         & 0.308         & \textbf{0.281}         & 0.188         \\
    \textbf{}              & $\pm$            & $\pm$            & $\boldsymbol{\pm}$            & $\boldsymbol{\pm}$            & $\pm$            & $\boldsymbol{\pm}$            & $\pm$            & $\boldsymbol{\pm}$            & $\pm$            \\
    \textbf{}              & 0.153         & 0.166         & \textbf{0.300}         & \textbf{0.040}         & 0.323         & \textbf{0.250}         & 0.236         & \textbf{0.293}         & 0.119         \\
     \specialrule{.1em}{.1em}{.1em} 
    \multicolumn{10}{c}{\textit{T=1000}}           \\
     \specialrule{.1em}{.1em}{.1em} 
    \multirow{3}{*}{\textbf{MN-CNN}}        & 0.757         & 0.581         & 0.919         & 1.345         & 1.185         & 0.904         & 0.657         & 1.303         & 0.677         \\
    \textbf{}              & $\pm$            & $\pm$            & $\pm$            & $\pm$            & $\pm$            & $\pm$            & $\pm$            & $\pm$            & $\pm$            \\
    \textbf{}              & 0.28          & 0.092         & 0.075         & 0.068         & 0.277         & 0.138         & 0.074         & 0.247         & 0.092         \\  \hline
    \multirow{3}{*}{\textbf{MN-LSTM}}      & 1.013         & 0.402         & 1.633         & 1.053         & 0.889         & 0.871         & 1.240         & 1.405         & 0.858         \\
    \textbf{}              & $\pm$            & $\pm$            & $\pm$            & $\pm$            & $\pm$            & $\pm$            & $\pm$            & $\pm$            & $\pm$            \\
    \textbf{}              & 0.495         & 0.173         & 0.475         & 0.663         & 0.406         & 0.386         & 0.647         & 0.501         & 0.374         \\  \hline
    \multirow{3}{*}{\textbf{MN-Conv-Trans}} & 0.192         & \textbf{0.129}         & 0.405         & 0.241         & \textbf{0.236}         & 0.179         & \textbf{0.208}         & 0.313         & \textbf{0.179}         \\
    \textbf{}              & $\pm$            & $\boldsymbol{\pm}$            & $\pm$            & $\pm$            & $\boldsymbol{\pm}$            & $\pm$            & $\boldsymbol{\pm}$            & $\pm$            & $\boldsymbol{\pm}$            \\
    \textbf{}              & 0.168         & \textbf{0.171}         & 0.320         & 0.236         & \textbf{0.291}         & 0.206         & \textbf{0.197}         & 0.345         & \textbf{0.252}        \\  \specialrule{.1em}{.1em}{.1em} 
    \end{tabular}
    }
    \label{table:T50-500-1000-RCLL-Tracked}
\end{table}

\begin{table}[]
	\fontsize{10.0}{13.0}\selectfont
	\centering
	\caption{Model performance for pegged orders dataset, in terms of RCLL.}
	\makebox[\textwidth]{\begin{tabular}{cccccccccc}
			\specialrule{.1em}{.1em}{.1em} \multicolumn{10}{c}{\textbf{Mean $\boldsymbol{\pm}$ STD Negative RCLL}} \\
			\specialrule{.1em}{.1em}{.1em} 
			\textbf{}                               & \textbf{AAPL} & \textbf{AMZN} & \textbf{BIDU} & \textbf{COST} & \textbf{CSCO} & \textbf{DELL} & \textbf{GOOG} & \textbf{INTC} & \textbf{MSFT} \\ \specialrule{.1em}{.1em}{.1em} 
			\multicolumn{10}{c}{\textit{No recurrence}}                                                                                                                                             \\ \specialrule{.1em}{.1em}{.1em} 
			\multirow{3}{*}{\textbf{DeepSurv}}      & 8.109         & 8.557         & 7.915         & 7.982         & 9.978         & 8.885         & 9.131         & 8.462         & 8.564         \\
			& $\pm$            & $\pm$            & $\pm$            & $\pm$            & $\pm$            & $\pm$            & $\pm$            & $\pm$            & $\pm$            \\ 
			& 0.065         & 0.034         & 0.103         & 0.066         & 0.067         & 0.064         & 0.024         & 0.096         & 0.008         \\ \hline
			\multirow{3}{*}{\textbf{DeepHit}}       & 10.098        & 10.119        & 9.716         & 9.731         & 9.978         & 9.816         & 10.046        & 9.874         & 10.074        \\
			& $\pm$            & $\pm$            & $\pm$            & $\pm$            & $\pm$            & $\pm$            & $\pm$            & $\pm$            & $\pm$            \\
			& 0.065         & 0.028         & 0.146         & 0.165         & 0.021         & 0.081         & 0.065         & 0.022         & 0.055         \\ \hline
			\multirow{3}{*}{\textbf{MN-MLP}}        & 10.554        & 10.709        & 13.300        & 13.603        & 12.364        & 13.151        & 11.330        & 12.110        & 10.784        \\
			& $\pm$            & $\pm$            & $\pm$            & $\pm$            & $\pm$            & $\pm$            & $\pm$            & $\pm$            & $\pm$            \\
			& 0.364         & 0.351         & 0.171         & 0.470         & 0.422         & 0.445         & 0.398         & 0.415         & 0.354         \\ \specialrule{.1em}{.1em}{.1em} 
			\multicolumn{10}{c}{\textbf{$T=50$}}                                                                                                                                                      \\ \specialrule{.1em}{.1em}{.1em} 
			\multirow{3}{*}{\textbf{MN-CNN}}        & 5.075         & 4.795         & 13.743        & 9.474         & 6.569         & 8.778         & 5.535         & 6.588         & 5.351         \\
			& $\pm$            & $\pm$            & $\pm$            & $\pm$            & $\pm$            & $\pm$            & $\pm$            & $\pm$            & $\pm$            \\
			& 0.366         & 0.288         & 0.158         & 0.139         & 0.126         & 0.553         & 0.246         & 0.206         & 0.045         \\ \hline
			\multirow{3}{*}{\textbf{MN-LSTM}}       & 3.896         & 3.302         & 6.452         & 9.539         & 6.065         & 7.056         & 4.385         & 6.954         & 4.077         \\
			& $\pm$            & $\pm$            & $\pm$            & $\pm$            & $\pm$            & $\pm$            & $\pm$            & $\pm$            & $\pm$            \\
			& 0.497         & 0.067         & 0.181         & 1.958         & 0.221         & 0.114         & 0.215         & 0.169         & 0.255         \\ \hline
			\multirow{3}{*}{\textbf{MN-Conv-Trans}} & 3.201         & 2.997         & 5.930         & 5.957         & 5.019         & \textbf{5.522}         & 3.730         & 5.002         & 3.533         \\
			& $\pm$            & $\pm$            & $\pm$            & $\pm$            & $\pm$            & $\boldsymbol{\pm}$            & $\pm$            & $\pm$            & $\pm$            \\
			& 0.827         & 0.675         & 0.863         & 0.808         & 1.149         & \textbf{0.860}          & 0.901         & 1.167         & 1.090          \\ \specialrule{.1em}{.1em}{.1em} 
			\multicolumn{10}{c}{\textbf{$T=500$}}                                                                                                                                                     \\ \specialrule{.1em}{.1em}{.1em} 
			\multirow{3}{*}{\textbf{MN-CNN}}        & 5.056         & 5.401         & 10.801        & 8.910         & 6.699         & 7.422         & 5.536         & 6.768         & 5.284         \\
			& $\pm$            & $\pm$            & $\pm$            & $\pm$            & $\pm$            & $\pm$            & $\pm$            & $\pm$            & $\pm$            \\
			& 0.165         & 0.077         & 0.221         & 0.253         & 0.130         & 0.135         & 0.157         & 0.307         & 0.165         \\ \hline
			\multirow{3}{*}{\textbf{MN-LSTM}}       & 3.701         & 4.135         & 7.658         & 7.681         & 5.636         & 7.479         & 4.338         & 6.545         & 4.369         \\
			& $\pm$            & $\pm$            & $\pm$            & $\pm$            & $\pm$            & $\pm$            & $\pm$            & $\pm$            & $\pm$            \\
			& 0.202         & 0.270         & 0.303         & 0.272         & 0.224         & 0.385         & 0.169         & 0.232         & 0.189         \\ \hline
			\multirow{3}{*}{\textbf{MN-Conv-Trans}} & \textbf{3.171}         & 3.111         & 6.428         & \textbf{5.822}         & 4.997         & 5.814         & 3.729         & 5.077         & \textbf{3.326}         \\
			& $\boldsymbol{\pm}$            & $\pm$            & $\pm$            & $\boldsymbol{\pm}$            & $\pm$            & $\pm$            & $\pm$            & $\pm$            & $\boldsymbol{\pm}$            \\
			& \textbf{0.157}         & 0.249         & 1.231         & \textbf{0.290}         & 0.255         & 0.424         & 0.925         & 0.352         & \textbf{0.309}         \\  \specialrule{.1em}{.1em}{.1em} 
			\multicolumn{10}{c}{\textbf{$T=1000$}}                                                                                                                                                    \\ \specialrule{.1em}{.1em}{.1em} 
			\multirow{3}{*}{\textbf{MN-CNN}}        & 5.925         & 4.796         & 7.485         & 8.052         & 6.581         & 7.171         & 5.536        & 7.676         & 5.347         \\
			& $\pm$            & $\pm$            & $\pm$            & $\pm$            & $\pm$            & $\pm$            & $\pm$            & $\pm$            & $\pm$            \\
			& 0.010         & 0.142         & 0.271         & 0.022         & 0.290         & 0.174         & 0.157      & 0.540          & 0.093         \\ \hline
			\multirow{3}{*}{\textbf{MN-LSTM}}       & 5.404         & 3.932         & 6.945         & 7.199         & 6.043         & 7.202         & 4.338      & 5.52          & 3.904         \\
			& $\pm$            & $\pm$            & $\pm$            & $\pm$            & $\pm$            & $\pm$            & $\pm$            & $\pm$            & $\pm$            \\
			& 0.023         & 0.154         & 0.06          & 0.030         & 0.219         & 0.583         & 0.169      & .241          & 0.141         \\  \hline
			\multirow{3}{*}{\textbf{MN-Conv-Trans}} & 3.724         & \textbf{2.980}         & \textbf{5.887}         & 6.089         & \textbf{4.974}         & 5.625         & \textbf{3.453}      & \textbf{4.951}         & 3.560         \\
			& $\pm$            & $\boldsymbol{\pm}$            & $\boldsymbol{\pm}$            & $\pm$            & $\boldsymbol{\pm}$            & $\pm$            & $\pm$            & $\boldsymbol{\pm}$            & $\pm$            \\
			& 1.226         & \textbf{0.713}         & \textbf{0.918}         & 1.073         & \textbf{0.834}         & 0.919         & \textbf{0.498}      & \textbf{1.053}         & 1.122         \\ \specialrule{.1em}{.1em}{.1em} 
	\end{tabular}}
	\label{table:T50-500-1000-RCLL}
\end{table}

\begin{table}[]
	\fontsize{10.0}{13.5}\selectfont
	\centering
	\caption{Percentage improvement over the MN-MLP model for observed limit orders dataset.}
	\makebox[\textwidth]{
		\begin{tabular}{ccccccccccc}
			\specialrule{.1em}{.1em}{.1em} 
			\multicolumn{11}{c}{\textbf{Improvement over MN-MLP (\%)}}                                                                                                                                                      \\  
			\specialrule{.1em}{.1em}{.1em}
			& \textbf{AAPL}           & \textbf{AMZN}           & \textbf{BIDU}           & \textbf{COST}           & \textbf{CSCO}           & \textbf{DELL}           & \textbf{GOOG}           & \textbf{INTC}           & \textbf{MSFT}           & \textbf{Average}       \\ \hline
			
			\multicolumn{11}{c}{\textit{No recurrence}}                                                                                                                                                      \\  \specialrule{.1em}{.1em}{.1em} 
			
			\textbf{DeepSurv}      & -7.30          & 34.15          & 1.54           & 37.44          & 4.63           & 23.93          & 7.79           & 14.76          & 32.95          & 16.66          \\
			\textbf{DeepHit}       & 16.04          & 19.16          & 0.87           & 34.37          & 11.86          & 29.73          & 15.39          & 21.38          & 28.77          & 19.73          \\
			\specialrule{.1em}{.1em}{.1em}
			\multicolumn{11}{c}{\textit{T=50}}                                                                                                                                                               \\
			\specialrule{.1em}{.1em}{.1em}
			\textbf{MN-CNN}        & 61.02          & 66.13          & 50.90          & 67.44          & 5.73           & 36.36          & 57.69          & 49.90          & 64.64          & 51.09          \\
			\textbf{MN-LSTM}       & 44.03          & 77.77          & 34.89          & 55.56          & 30.56          & 57.39          & 20.54          & 29.45          & 49.63          & 44.42          \\
			\textbf{MN-Conv-Trans} & \textbf{90.31} & 92.47          & 79.10          & 91.04          & 80.99          & 87.20          & 85.00          & 77.43          & \textbf{90.59} & \textbf{86.01} \\
			\specialrule{.1em}{.1em}{.1em}
			\multicolumn{11}{c}{\textit{T=500}}                                                                                                                                                              \\
			\specialrule{.1em}{.1em}{.1em}
			\textbf{MN-CNN}        & 28.81          & 64.14          & 63.25          & 81.78          & 36.76          & 42.53          & 45.01          & 21.38          & 23.85          & 45.28          \\
			\textbf{MN-LSTM}       & 33.31          & 76.40          & 50.28          & 96.07          & 10.05          & 65.78          & 60.08          & 36.53          & 30.86          & 51.04          \\
			\textbf{MN-Conv-Trans} & 87.71          & 89.91          & \textbf{84.77} & \textbf{98.64} & 71.33          & \textbf{87.27} & 80.17          & \textbf{81.40} & 90.17          & 85.71          \\
			\specialrule{.1em}{.1em}{.1em}
			\multicolumn{11}{c}{\textit{T=1000}}                                                                                                                                                             \\
			\specialrule{.1em}{.1em}{.1em}
			\textbf{MN-CNN}        & 60.34          & 66.88          & 13.77          & 32.31          & 32.31          & 31.10          & 57.69          & 13.77          & 13.77          & 35.77          \\
			\textbf{MN-LSTM}       & 30.85          & 77.08          & 7.02           & 47.01          & 47.01          & 33.61          & 20.15          & 7.02           & 7.02           & 30.75          \\
			\textbf{MN-Conv-Trans} & 86.89          & \textbf{93.51} & 79.29          & 87.87          & \textbf{87.87} & 86.36          & \textbf{86.61} & 79.29          & 79.29          & 85.22    \\\specialrule{.1em}{.1em}{.1em}     
		\end{tabular}
	}
	\label{table:T50-500-1000-Improvement-Tracked}
\end{table}

\begin{table}[]
    \fontsize{10.0}{13.75}\selectfont
    \centering
        \caption{Percentage improvement over the MN-MLP model for pegged limit orders dataset.}
\makebox[\textwidth]{\begin{tabular}{ccccccccccc}
\specialrule{.1em}{.1em}{.1em} 
\multicolumn{11}{c}{\textbf{Improvement over MN-MLP (\%)}}                                                                                                                                                         \\ \specialrule{.1em}{.1em}{.1em} 
\textbf{}              & \textbf{AAPL}  & \textbf{AMZN}  & \textbf{BIDU}  & \textbf{COST}  & \textbf{CSCO}  & \textbf{DELL}  & \textbf{GOOG} & \textbf{INTC}  & \textbf{MSFT}  & \textbf{Average} \\ \specialrule{.1em}{.1em}{.1em} 
\multicolumn{11}{c}{\textit{No Recurrence}}                                                                                                                                                       \\ \specialrule{.1em}{.1em}{.1em}
\textbf{DeepSurv}      & 23.17          & 20.10          & 40.49          & 41.32          & 19.30          & 32.44          & 19.41         & 30.12          & 20.59          & 27.44            \\ 
\textbf{DeepHit}       & 4.32           & 5.51           & 26.95          & 28.46          & 19.30          & 25.36          & 11.33         & 18.46          & 6.58           & 16.25            \\ \specialrule{.1em}{.1em}{.1em}
\multicolumn{11}{c}{\textbf{$T=50$}}                                                                                                                                                                \\ \specialrule{.1em}{.1em}{.1em}
\textbf{MN-CNN}        & 51.91          & 55.22          & -3.33          & 30.35          & 46.87          & 33.25          & 51.15         & 45.60          & 50.38          & 40.16            \\
\textbf{MN-LSTM}       & 63.09          & 69.17          & 51.49          & 29.88          & 50.95          & 46.35          & 61.30         & 42.58          & 62.19          & 53.00            \\
\textbf{MN-Conv-Trans} & 69.67          & 72.01          & 55.41          & 56.21          & 59.41          & \textbf{58.01} & 67.08         & 58.70          & 67.24          & 62.64            \\ \specialrule{.1em}{.1em}{.1em}
\multicolumn{11}{c}{\textbf{$T=500$}}                                                                                                                                                               \\ \specialrule{.1em}{.1em}{.1em}
\textbf{MN-CNN}        & 52.09          & 49.57          & 18.79          & 34.50          & 45.82          & 43.56          & 51.14         & 44.11          & 51.00          & 43.40            \\
\textbf{MN-LSTM}       & 64.93          & 61.39          & 42.42          & 43.53          & 54.42          & 43.13          & 61.71         & 45.95          & 59.49          & 53.00            \\
\textbf{MN-Conv-Trans} & \textbf{69.95} & 70.95          & 51.67          & \textbf{57.20} & 59.58          & 55.79          & \textbf{67.09}         & 58.08          & \textbf{69.16} & 62.16            \\ \specialrule{.1em}{.1em}{.1em}
\multicolumn{11}{c}{\textbf{$T=1000$}}                                                                                                                                                              \\ \specialrule{.1em}{.1em}{.1em}
\textbf{MN-CNN}        & 54.56          & 55.22          & 43.72          & 40.81          & 46.77          & 45.47          & 51.15      & 36.61          & 50.42          & 47.19         \\
\textbf{MN-LSTM}       & 48.80          & 63.28          & 47.78          & 47.08          & 51.12          & 45.24          & 61.30      & 54.42          & 63.80          & 53.65         \\
\textbf{MN-Conv-Trans} & 64.71          & \textbf{78.09} & \textbf{55.74} & 55.24          & \textbf{59.77} & 57.23          & 67.08      & \textbf{59.12} & 66.99          & \textbf{62.66}         \\ \specialrule{.1em}{.1em}{.1em} 
\end{tabular}}
\label{table:T50-500-1000-Improvement}
\end{table}

For completeness, we use an order flow representation of the order book to carry out the same analysis and summarise the results in \ref{app:i}. The benefits of such an approach are summarised in \citet{Kolm2021} and \citet{Lucchese2022}, where authors suggest that considering order flow or volume representations of the order book increase predictive performance for step-ahead forecasts, while making the use of more complex models unnecessary. In our case, however, we find that the convolutional transformer still outperforms all other models in this setting. This suggests that some of the representations used to perform directional price forecasts are not as relevant when estimating the survival functions of limit orders. A possible reason for this is that price prediction is a harder task given that it is a directional forecast. Moreover, the prediction of the fill probability is closely linked to the behaviour of the spread, whose prediction is non-directional, and is closely linked to the volatility (which exhibits higher persistence than a regular price time series).  As such, the model might be focusing on properties of the features that aid in the prediction of future volatility, making representations that aid in directional price forecasts redundant.

\subsection{Model Interpretability }

In this section, we focus on the interpretability of our model. To do so, we analyse both the time and feature domains. In particular, we use attention heatmaps to visualise which parts of the past values of the signals are more important to the model to estimate the survival function. Finally, we use Shapley values \citep{hart1989shapley} to  quantify the relative importance of each input feature to the output of the model. 

\subsubsection{Attention Heatmaps}\label{subsub:temporal_interpretability}

The convolutional-Transformer employs \eqref{eq:cnn_self_attention} to obtain, through the convolutional network, the self-attention input features described in Section \ref{subsec:encoder}: the query, key, and value matrices. The model then performs the dot-product computation  between the attention's queries and keys,  see \eqref{eq:self-attention}. This operation results on a matrix of dimensions $\mathbb{R}^{T \times T}$, where  $T \in \mathbb{Z}$ is the lookback window's length. Finally, after the softmax function is applied to this matrix, see \eqref{eq:self-attention}, the output is used to multiply the previously obtained self-attention's values. Therefore, with the matrix resulting from the dot-product operation, one visualises which regions of the lookback window (more precisely, of its non-linear projection), are given the highest weighting by the model when estimating the survival function. These are commonly known as \textit{attention heatmaps}.\\

\begin{figure}[H] 
	%\vskip -0.2in
		%\centering{\includegraphics[width=0.6\columnwidth]{Figures/features.pdf}}
	%	\caption{Features used to predict the survival function for a specific order, using $T=500$.} 
	%	\label{fig:order}
        %\vspace{1.0cm}
		\centerline{\includegraphics[width=1.0\columnwidth]{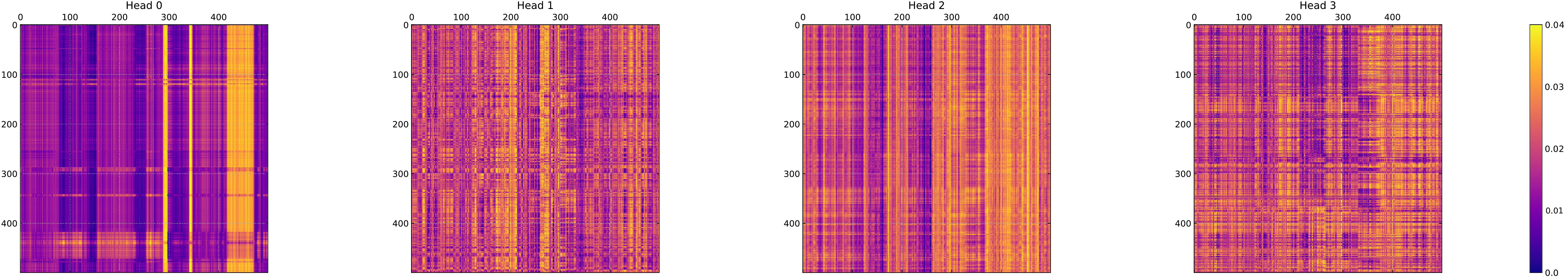}}
		\caption{Attention heatmaps obtained for an example order.} 
		\label{fig:heatmap}
\end{figure}

Figure \ref{fig:heatmap} shows the four attention heatmaps of our model, one per head, for a single estimate. Further, \ref{app:g} shows the corresponding evolution in time of the features. The attention heatmaps display the self-attention weights and provide information on the weight given for each time-step by the model.  As shown in the plots, head 0 focuses on samples of 400 trades ago when there was a significant reduction in volatility and the size of the spread. The remaining of the heads have a sparse attention pattern, showing that the models accounts for both short and long-term information to make the forecast.  

\subsubsection{Shapley Values}\label{subsub:features_interpretability}

Shapley values provide a method to attribute the predictions of the model to individual features of the input. In this case, this helps to understand which of the market features are most important and how they contribute to the model's overall prediction. To calculate the Shapley values, we follow the DeepSHAP approach in  \citet{lundberg2017unified}. 
To measure the importance per feature, we define $\mathbf{S} \subseteq \mathbf{F}$ as all possible feature subsets, where $\mathbf{F}$ is the set of all features. First, integrate over many background samples to obtain the expected model output $\mathbb{E}[\hat{f}_{\mathbf{S}}(t|\mathbf{x})]$, where $\hat{f}_{\mathbf{S}}(t|\mathbf{x})$ denotes the model's prediction while using all the available features. Next, we approximate Shapley values such that they sum up to the difference between the expected output of the model and the predicted values, i.e., $\hat{f}_{\mathbf{S}}(t|\mathbf{x}) - \mathbb{E}[\hat{f}(t|\mathbf{x})]$.
The contribution of the $i^{\text{th}}$ feature is
\begin{equation}
    C_{i} = \hat{f}_{\mathbf{S}}(t|\mathbf{x}) - \hat{f}_{\mathbf{S} \backslash \{i\}}(t|\mathbf{x}),
\end{equation}
where $\hat{f}_{\mathbf{S} \backslash \{i\}} (x)$ is the model's prediction without using the $i^{\text{th}}$ feature, whose Shapley value is given by
\begin{equation}
\partial_{i} = \frac{1}{n!} \sum_{p=0}^{n} p!\ (n - p)!\ C_{i}\ ,
\end{equation}
where $n$ is the total number of features and $p$  is the number of features present in the input sample. \\%, which can be any subset over the total number of features depending on the specific input being evaluated. 

\begin{figure}[H] 
	%\vskip -0.2in
	\begin{center}
		\centerline{\includegraphics[height=16cm, width=0.65\columnwidth]{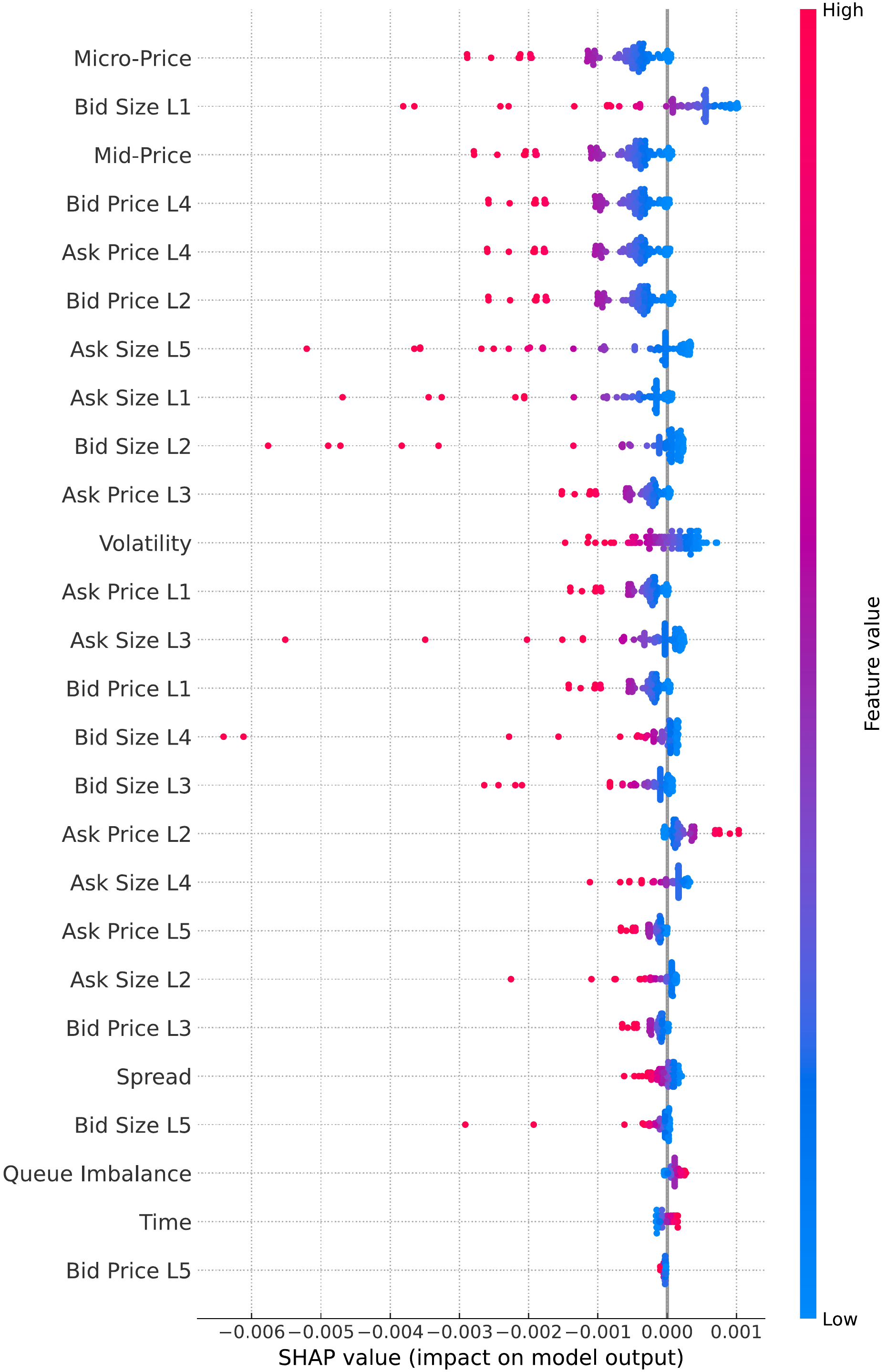}}
		\caption{Shapley values for $100$ predictions the MN-MLP model.} 
		\label{fig:shap}
	\end{center}
\end{figure}

Figure \ref{fig:shap} depicts a Beeswarm plot, which shows the relative importance of each feature and its relationship with the predicted outcome. Each of the datapoints is represented with a single dot, where colours ranging from blue to red indicate lower to higher values per feature. The vertical axis is ordered in accordance with the importance of each feature for the convolutional-Transformer on average, while the horizontal axis displays the associated Shapley value. This analysis is associated with the more basic MN-MLP model, because these quantities cannot be computed easily for models which explicitly model the time evolution of the features. This means that the information we extract from this cannot be directly translated to analyse our proposed model, but can give a good indication of which features are most influential. \\

The figure shows that the model gives the most importance to fast-moving features. For example, the microprice is the most important feature because it provides a relatively good estimate of the perceived fundamental value of the asset, which has an effect on where liquidity taking orders are placed. The volatility of the asset also plays a major role in the prediction, which could be attributed to the fact that it is a good proxy for the volume being traded. If more volume is being traded in the market, there is a higher chance that a liquidity taking order will cross the spread and match an outstanding order. Time of day, as well as other features, shows little to no importance, which is likely due to the fact that time of day shows a very persistent pattern it is an intraday pattern, meaning that small perturbations in its value should not affect the output of the model significantly.  

\section{Conclusion}
This paper presented a novel approach to estimating the fill probabilities of limit orders posted in the LOB. The proposed data-driven approach integrates a novel convolutional-Transformer model to raise local-awareness from LOB data, and a monotonic neural network to guarantee the monotonicity of the survival function of limit orders. To train and evaluate the model, we use right-censored log-likelihood, which is a proper scoring rule, unlike other scoring rules commonly used in the literature. To demonstrate the effectiveness of the proposed method, we conducted a set of experiments on real LOB data. These experiments showed that the monotonic encoder-decoder convolutional-Transformer significantly outperforms state-of-the-art benchmarks, and provides a new general framework with which to perform survival analysis from time-series observations. Finally, we provided an interpretability analysis based on Shapley values and attention heatmaps, which provides insight on which predictive features are the most influential. \\

In this paper we have focused on the financial applications of survival analysis. However, future work could make use of the same architecture for alternative application domains like healthcare, where the use of survival analysis is prevalent. From a financial standpoint, it would be interesting to further explore if the fill probability results in improved LOB modelling when used in a realistic market simulator. Furthermore, we considered a setup in which we summarised the time-series representation up to a point in time and did not use any information after order submission. To this end, an interesting follow-up could leverage or extend recent ideas of temporally-consistent survival analysis, see \cite{Maystre2022}, to obtain improved estimates of the survival function of limit orders. Finally, it would be interesting to explore a multi-asset framework, see \cite{bergault2022multi, drissi2022solvability, drissi2023models}, to understand how the fill probability is affected by correlated instruments, which opens up the possibility of using graphs, see \cite{Arroyo2022} and \cite{Borde2023}, as a modelling technique.

\section{Acknowledgements}

We are grateful to Fayçal Drissi and Leandro Sánchez-Betancourt for insightful comments and suggestions. We also thank the participants of the Victoria seminar for interesting discussions. We thank the participants of the 67$^{\text{th}}$ EWGCFM Meeting for comments. Álvaro Arroyo acknowledges support from the Rafael del Pino Foundation. Fernando Moreno-Pino acknowledges support from Spanish Ministerio de Ciencia, Innovación y Universidades (FPU18/00470), Ministerio de Ciencia e Innovación jointly with the European Commission (ERDF) (PID2021-123182OB-I00, PID2021-125159NB-I00), by Comunidad de Madrid (IntCARE-CM and IND2020/TIC-17372), and by the European Union (FEDER) and the European Research Council (ERC) through the European Union’s Horizon 2020 research and innovation program under Grant 714161.

\newpage

\appendix

\section{Conditions for a hypothetical order fill}\label{app:a}

Following a similar approach to that of \cite{Maglaras2021}, we use the following fill conditions for hypothetical limit orders:

\begin{enumerate}
	\item A new hypothetical limit order is filled if:
	\begin{itemize}
	   \item A new buy/sell order arrives at a higher/lower price than the hypothetical sell/buy limit order; 
	   \item An order in front of the hypothetical order in the execution queue is filled/partially filled, then we consider the hypothetical order to be filled.
	\end{itemize}
	\item A new market order comes in at the same price than the hypothetical order:
	\begin{itemize}
		\item If the market order executes against any of the orders being tracked, then our hypothetical order is considered filled.
	\end{itemize}
\end{enumerate}
From an implementation perspective, we do not require maintaining a full log of all the orders in the queue. It suffices to track the messages associated to orders in front of the hypothetical limit order in the queue and market orders arriving above/below the limit price of the hypothetical order, depending on the side it was placed in the book.

\section{Derivation of right-censored log-likelihood}\label{app:b}

Assume we have a dataset of observations $\mathcal{D} = \{(\mathbf{x}_k, z_k, \delta_k)\}_{k=1}^{N}$, where $z_k = \min\{T_l, C_l\}$, where the random variables $T_l$ and $C_l$ denote the random fill and cancellation (censoring) times, respectively. For clarity, in the remaining derivations of the appendix, we drop the subscript $l$. The likelihood function is
\begin{equation}
    \begin{aligned}[b]
        L &= f(z_1, \delta_1, \hdots, z_N, \delta_N) \\
        &= \prod_{k=1}^{N} f(z_k, \delta_k), \\
    \end{aligned}
    \label{eq:B.1}
\end{equation}
where we assume that pairs are independent. To re-write this equation depending on the values of the indicator variable $\delta_k$, we first consider the case in which the order is filled, and the event is therefore observed ($\delta_k=1$)
\begin{equation}
\begin{aligned}[b]
    f(z_k, \delta_k) &=
    \mathbb{P}\{Z=z_k, \delta_k=1\} \\
    &= \mathbb{P}\{T=z_k, T \le C\}  \\
    &=  \mathbb{P}\{T=z_k, z_k \le C\} \\
    &=  \mathbb{P}\{T=z_k\} \mathbb{P}\{z_k \le C\}  \quad \text{(assuming $T\text{ is independent of } C$)}\\
    &= f_{T}(z_k)S_C(z_k). \\
\end{aligned}
\label{eq:B.2}
\end{equation}
Moreover, if the observation is censored, we have that $\delta_k=0$ and
\begin{equation}
\begin{aligned}[b]
    f(z_k, \delta_k) &=
    \mathbb{P}\{Z=z_k, \delta_k=0\} \\
    &= \mathbb{P}\{C=z_k, T > C\} \\
    &=  \mathbb{P}\{C=z_k, z_k < T\}  \\
    &=  \mathbb{P}\{C=z_k\} \mathbb{P}\{z_k < T\} \quad \text{(assuming $T\text{ is independent of } C$)} \\
    &= f_{C}(z_k)S_T(z_k),
\end{aligned}
\end{equation}
and re-write \eqref{eq:B.1} as
\begin{equation}
\begin{aligned}[b]
    L &= \prod_{k=1}^{N} \big[f_{T}(z_k)S_C(z_k)\big]^{\delta_k}\big[f_{C}(z_k)S_T(z_k)\big]^{(1-\delta_k)} \\
    &= \prod_{k=1}^{N} \big[f_{T}(z_k)^{\delta_k} S_T(z_k)^{(1-\delta_k)} \big]\big[f_{C}(z_k)^{(1-\delta_k)} S_C(z_k)^{\delta_k}\big].
\end{aligned}
\label{eq:B.4}
\end{equation}
We are  concerned with the estimation of $S_T(t)$ and not $S_C(t)$, because $S_C(t)$ contains information related to censoring mechanism. Only $S_T(t)$ contains
information about the filltimes, which is the variable of interest. Thus, the terms that do not involve $T$ are considered constants, and the log-likelihood is 
\begin{equation}
    \mathcal{L} =
    \log(L) = \sum_{k=1}^{N}\delta_{k}\log(\hat{f}(z_{k})) + (1-\delta_{k})\log(\hat{S}(z_{k})).
    \label{eq:B.}
\end{equation}

\section{Typical survival models and scoring rules}\label{app:c}

\subsection{Survival Models}

An initial approach is to use a Kaplan--Meier estimate \citep{Kaplan1958} to estimate the survival function i.e., 
\begin{equation}
    \hat{S}(t) = \prod_{i:t_{i}\le t}\bigg(1-\frac{k_i}{n_i} \bigg),
\end{equation}
where $t_i$ is the time when at least one event occurred, $k_i$ is the number of events that occurred at time $t_i$, and $n_i$ denotes the limit orders known to have survived up to time $t_i$. A follow-up model to the  Kaplan--Meier estimate which conditions on a feature vector is the Cox proportional hazards model \citep{Cox1972}, where the hazard rate follows the definition
\begin{equation}
    h(t|\mathbf{x}) = h_0(t)\text{exp}(\boldsymbol{\beta}^{
    T}\mathbf{x}),
\end{equation}
where $\boldsymbol{\beta}$ are coefficients for the feature vector $\mathbf{x}$, and $h_0(t)$  is a baseline hazard directly estimated from the data. Given $N$ observations, the regression coefficients which maximize

\begin{equation}
    \mathcal{L}(\boldsymbol{\beta})
    =
    \prod_{\delta_i=1}\frac{\text{exp}(\boldsymbol{\beta}^{
    T}\mathbf{x})}{\sum_{j:t_j\ge t_i} \text{exp}(\boldsymbol{\beta}^{
    T}\mathbf{x})}.
\end{equation}

Another popular model used in survival analysis is the accelerated failure time model \citep{Wei1992}, in which the hazard rates are determined by

\begin{equation}
    h(t|\mathbf{x}) = \phi(\boldsymbol{x})h_0(\phi(\boldsymbol{x})t),
\end{equation}
where $\phi(\boldsymbol{x})$ models the effect of the covariates, usually through the relationship $\phi(\boldsymbol{x})=\text{exp}(\boldsymbol{\beta}^{T}\mathbf{x})$. In practice, the assumption of linear interaction between features and the proportional hazards assumption are often violated. This motivated the extension of the Cox model with deep learning to capture non-linearities between features of interest \citep{Katzman2018, Kvamme2019}
\begin{equation}
    h(t|\mathbf{x}) = h_0(t)\text{exp}(f_{\boldsymbol{\theta}}(\mathbf{x},t)).
\end{equation}
More recent work \citep{Lee2018,Lee2019, Rindt2022} focuses on directly learning the survival function conditioning on input features
\begin{equation}
    S(t|\mathbf{x}) = f_{\boldsymbol{\theta}}(\mathbf{x}, t).
\end{equation}

\subsection{Scoring Rules}

Commonly used scores for survival functions include time-dependant concordance \citep{Antolini2015}, given by
\begin{equation}
    \begin{split}
        C_{td} &= \mathbb{P}[\hat{S}(z_{i}|\mathbf{x}_{i})<\hat{S}(z_{i}|\mathbf{x}_{j})|z_{i}<z_{j}, \delta_i=1]  \\
        &\approx \frac{\sum_{i=1}^{N}\sum_{j=1; i\neq j}^{N}\mathds{1}[\hat{S}(z_{i}|\mathbf{x}_{i}) < \hat{S}(z_{j}|\mathbf{x}_{j})]\pi_{ij}}{\sum_{i=1}^{N}\sum_{j=1; i\neq j}^{N}\pi_{ij}},
    \end{split}
\end{equation}
where $\pi_{ij}$ is an indicator of the pair $(i,j)$ being amenable for comparison, i.e., if the pair is ``concordant". The intuition behind time-varying concordance \citep{Harrell1982} is based on the idea that the predicted survival probability for an order $i$ evaluated at time $z_i$ and conditioned on market features $\mathbf{x}_i$ should be lower than that of order $j$ evaluated at the same time and conditioned on market features $\mathbf{x}_j$ if order $i$ was filled faster than order $j$.  In addition to time-varying concordance, another frequently used scoring is the Brier score for right-censored data \citep{Graf1999}, defined as
\begin{equation}
    \beta = \frac{\hat{S}(t|x)^{2}\mathbbm{1}\{z\le t, \delta=1\}}{\hat{G}(z)}
    + \frac{(1-\hat{S}(t|x))^{2}\mathbbm{1}\{z> t\}}{\hat{G}(z)},
\end{equation}
where $\hat{G}$ is the Kaplan--Meier estimate of the censoring distribution. \\

Time-varying concordance and Brier score are the two most common scores to evaluate models in the survival analysis literature \citep{Lee2018, Lee2019, Zhong2021}. However, recent work \citep{Rindt2022} shows that both scoring rules (as well as a number of others) are  improper, meaning that they can potentially give higher scores to wrongly fitted distributions.\footnote{Brier score is a proper scoring rule under the assumption of independence between censoring and covariates, as well as a perfect estimate of the censoring distribution. These assumptions do not hold in the context of limit order executions, given that orders of larger size are prone to cancellations (which are interpreted as censoring in this work) and there is not a tractable way of obtaining a perfect estimate of the distribution of cancelled orders.} Right-censored log-likelihood is a proper scoring rule, which is the key result in \citep{Rindt2022}.

\section{Extending the Encoder's Dilated Causal Convolutional Neural Network to $L$ layers}
\label{app:e}

To extend the original the DCC in the encoder, which is responsible for obtaining the corresponding queries, keys, and keys in the proposed convolutional-Transformer, the  causal convolutional network should be modified as follows. The first layer would perform the same operation, convolving the input sequences $x$ and the kernel $k$:
\begin{equation}
F^{(l=1)}(t)=\left(x *_{p} k^{(l=1)}\right)(t)=\sum_{\tau=0}^{s-1} k^{(l=1)}_{\tau} \cdot x_{t-p\cdot \tau},
\end{equation}
 where $p$ is the dilation factor and $k$ the convolutional filter with size $s \in\mathbb{Z}$.
%\todo{Needs some more explanation. $*_{d}$ is the dilated convolution operation as defined on the RHS. $k$ is not the size of the filter but the filter (weight vector) itself, it seems to me.}
For each of the rest $l$ layers, we define the convolution operation as
\begin{equation}
  F^{(l)}(t)=\left(F^{(l-1)} *_{d} k^{(l)}\right)(t)=\sum_{\tau=0}^{s-1} k^{(l)}_{\tau} \cdot F_{t-p\cdot\tau}^{l-1}(t).
\end{equation}
Each of the layers in this hierarchical structure defines the kernel operation as an affine function acting between layers
\begin{equation}
    k^{(l)}: \mathbb{R}^{N_{l}} \longrightarrow \mathbb{R}^{N_{l+1}},\ 1 \leq l \leq L.
\end{equation}
The previous equation shows how, through the use of residual connections, firstly proposed in \citet{he2016deep}, the encoder's convolutional network could connect $l^{\text{th}}$ layer's output to $(l+1)^{\text{th}}$ layer's input, enabling the usage of deeper models with larger receptive fields to generate the hidden representation that is used by the Transformer self-attention's inputs.

\section{Monotonicity of the Decoder}\label{app:monotonic}

This condition is satisfied, because we consider a set-up where all the intermediate layers $h$ of the network, with $1<h<H$ (not to be confused with $h_i$ in Section \ref{subsec:encoder} to accredit the Transformer model's head $h_i \in H$), have the following input-output relationship for each node $j$ and input $\mathbf{x}^h\in\mathbb{R}^{M^h}$ (ignoring the bias terms)
\begin{equation}
    \mathbf{y}^{h}_{j} = \text{tanh}\bigg(\sum_{i=1}^{M^h}w_{ij}^{h} \, x_i^h\bigg)\ ,
\end{equation}
where the final output is given by
\begin{equation}
    f_{\boldsymbol{\Psi}}(t,\mathbf{x})
    =
    P(T\le t | \mathbf{X}=\mathbf{x}) = \sigma\bigg(\sum_{i=1}^{M^H }w_{i1}^{H} \, x_i^H\bigg)\,.
\end{equation}
Here, $w_{ij}^h$ and $b_{ij}^h$ are the individual weight and bias terms associated to node $j$ and layer $h$, respectively, and $\text{tanh}(\,\cdot\,)$ and $\sigma(\,\cdot\,)$ are the hyperbolic tangent and sigmoid functions, respectively. To enforce monotonicity of the output with respect to the response variable $t$, we impose $w_{ij}^{h}\ge 0 \quad \forall h\in\{1,\hdots,H\}$, because the derivative of the output of each node in the first layer is
\begin{equation}
    \frac{\partial\mathbf{y}^{1}_{j}}{\partial t} = \text{tanh}'\bigg(w_{1 M^1}^{1}t + \sum_{i=2}^{M^1}w_{ij}^{1} \, x_i^1\bigg)w_{1 M^1}^{1}\,,
\end{equation}
which requires positivity of the $w_{1 M^1}^{1}$ to guarantee the monotonicity condition. From the chain rule, a similar argument holds for all nodes in subsequent layers of the network. Finally, the remaining two conditions are satisfied empirically given the likelihood-based training method. \\

\newpage

\section{Order Features}
\label{app:g}

\begin{figure}[H] 
	%\vskip -0.2in
		\centering{\includegraphics[width=0.9\columnwidth]{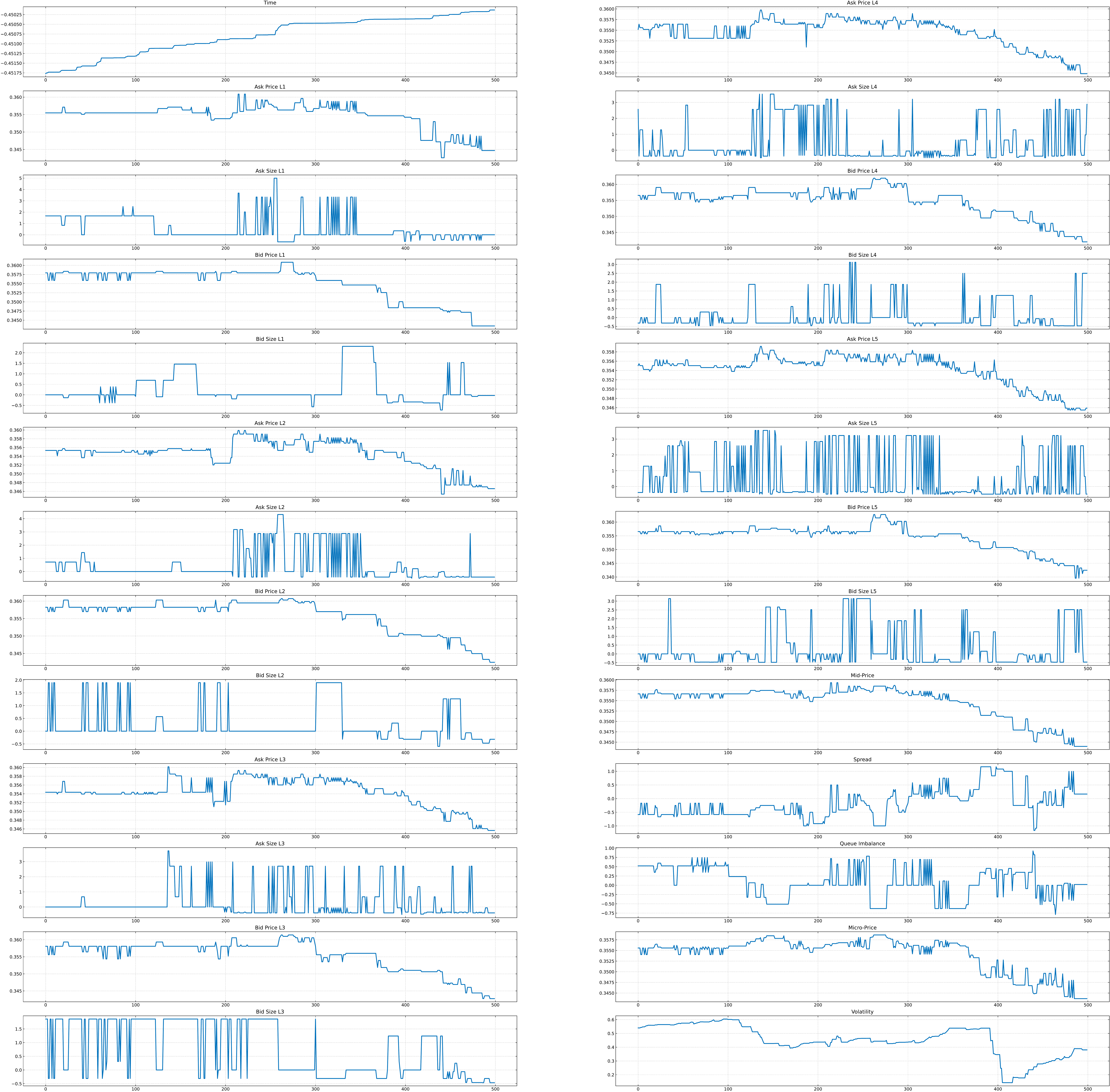}}
		\caption{Evolution of order features over 500 trades.} 
		\label{fig:order}		
\end{figure}

The evolution over 500 trades of the features considered to produce the attention heatmaps are shown in Figure \ref{fig:order}.

\newpage

\section{Kernel Sizes}
\label{app:h}

Table \ref{table:cnn_kernel} explores different kernel sizes, $s \in \left\{ 1,2,3,5,10,25,50\right\}$, for the convolutional operation of the MN-Conv-Trans model for the estimation of AAPL's survival function with a lookback window of $T=500$ trades. Recall that a convolutional network with kernel size $s=1$ results on canonical self-attention, in which case, there is an evident decline in the negative RCLL in comparison to larger kernel sizes. There are no substantial variations in the performance among the other kernel sizes. Therefore, a value of $s=3$ seems reasonable to avoid an unnecessary increase in parametric complexity.

\begin{table}[H]
	\fontsize{11.0}{14.0}\selectfont
	\centering
	\caption{\footnotesize{Performance variation while when different kernel's sizes for the MN-Conv-Trans DCC network. AAPL stock with a lookback window of 500 trades.}}
	\begin{tabular}{cc}
		\specialrule{.1em}{.1em}{.1em}
		\textbf{Kernel Size} & \textbf{Mean $\boldsymbol{\pm}$ STD Negative RCLL}                             \\ \specialrule{.1em}{.1em}{.1em}
		\textbf{$s=1$}         & 3.245 $\pm$ 0.207 \\
		\textbf{$s=2$}         & 3.184 $\pm$ 0.343  \\
		\textbf{$s=3$}         & 3.171 $\pm$ 0.157  \\
		\textbf{$s=5$}         & 3.189 $\pm$ 0.433   \\
		\textbf{$s=10$}        & 3.179 $\pm$ 0.367 \\
		\textbf{$s=25$}        & 3.196 $\pm$ 0.383  \\
		\textbf{$s=50$}        & 3.170 $\pm$ 0.370   \\ \specialrule{.1em}{.1em}{.1em}
	\end{tabular}
	\label{table:cnn_kernel}
\end{table}

\section{Survival Functions in Transaction Time}
\label{app:transaction}

\begin{figure}[H]
\includegraphics[height =6.75cm,width=0.49\textwidth]{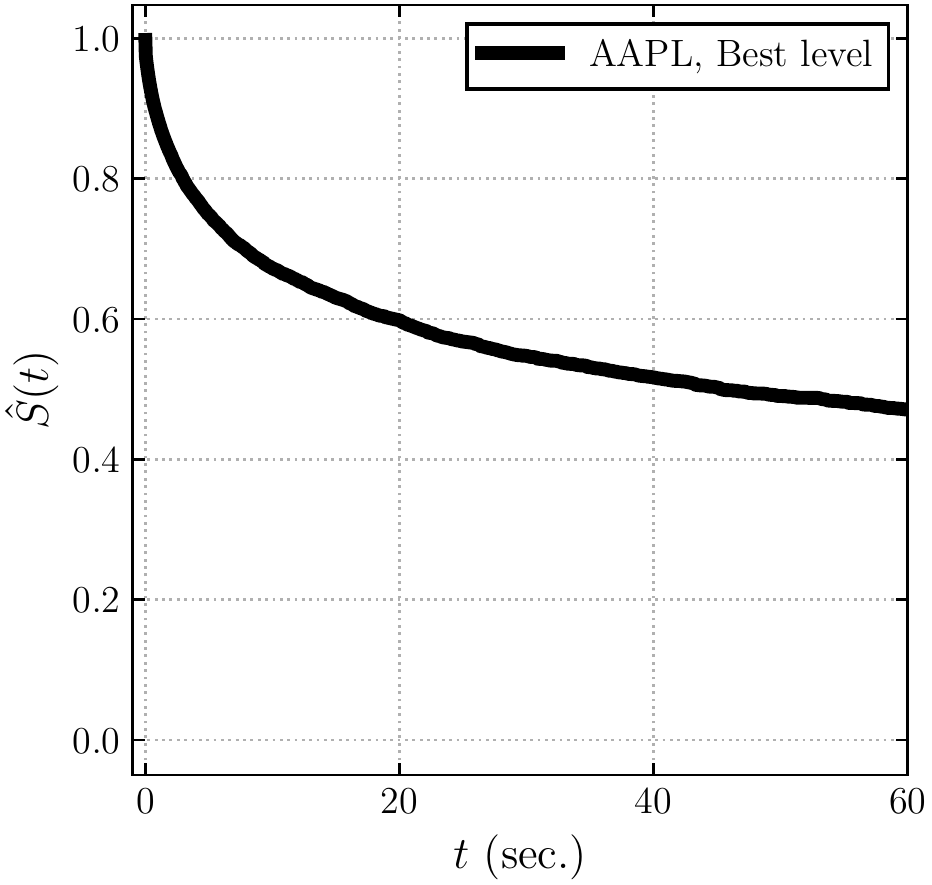}
\hspace*{\fill}
\includegraphics[height = 6.75cm,width=0.49\textwidth]{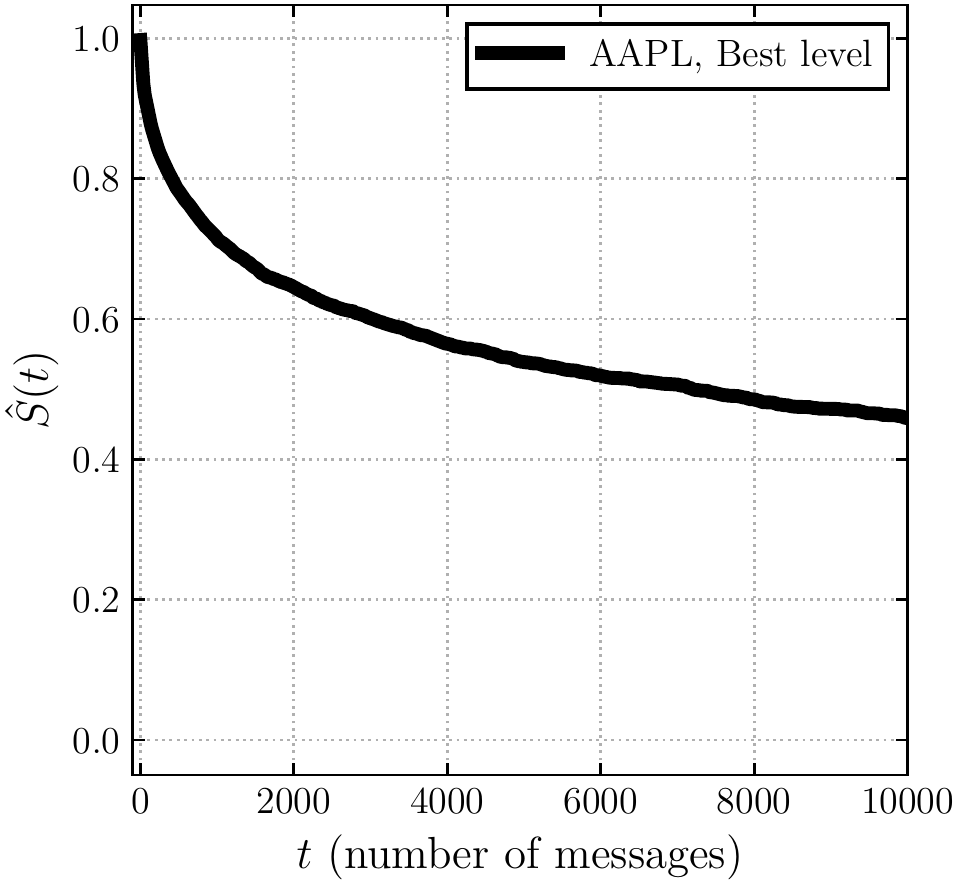}
\vspace*{\fill}
\includegraphics[height =6.75cm,width=0.49\textwidth]{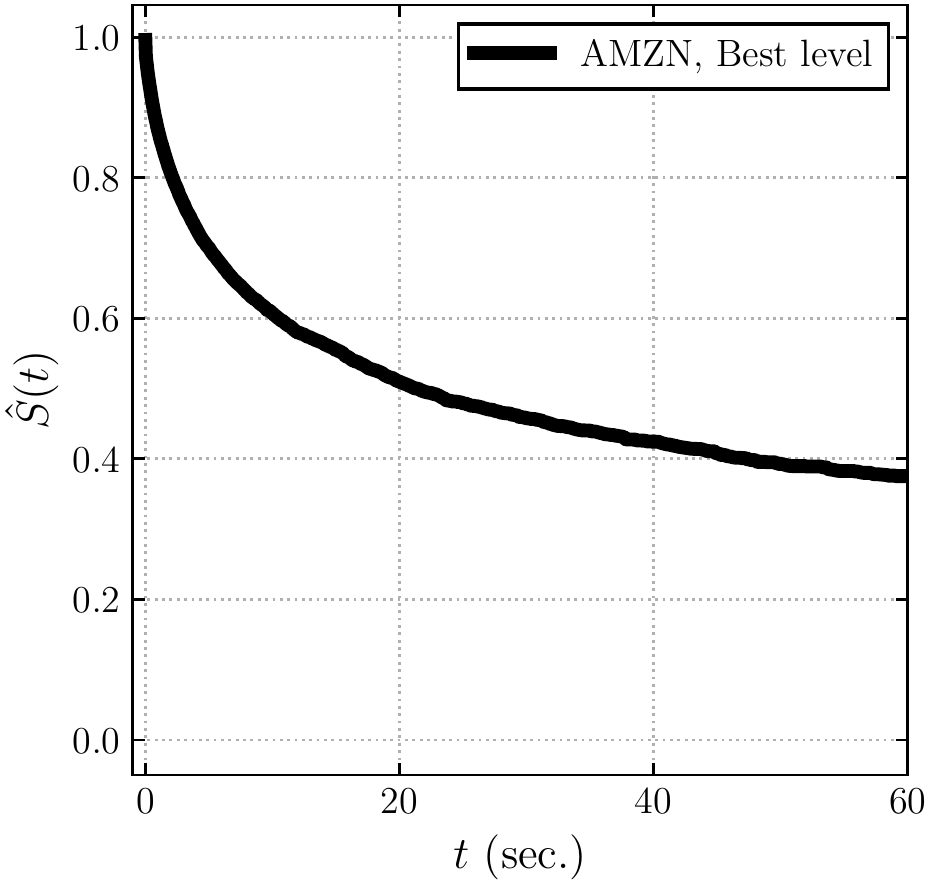}
\hspace*{\fill}
\includegraphics[height = 6.75cm,width=0.49\textwidth]{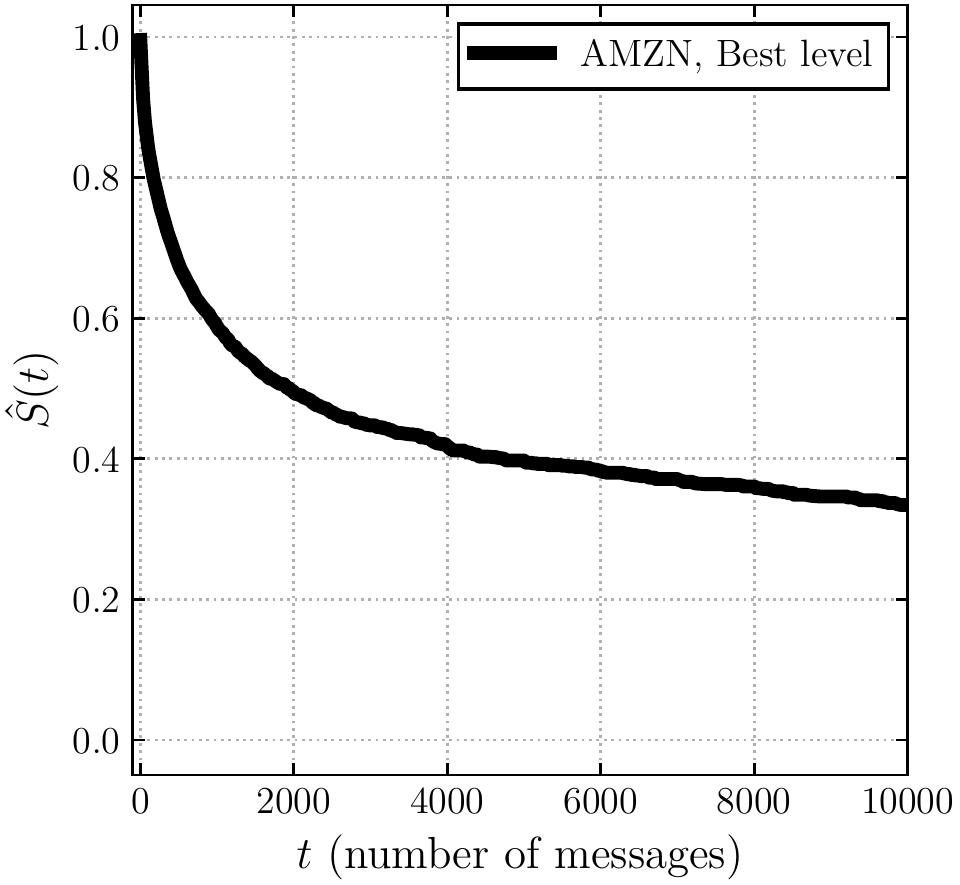}
\vspace*{\fill}
\includegraphics[height =6.75cm,width=0.49\textwidth]{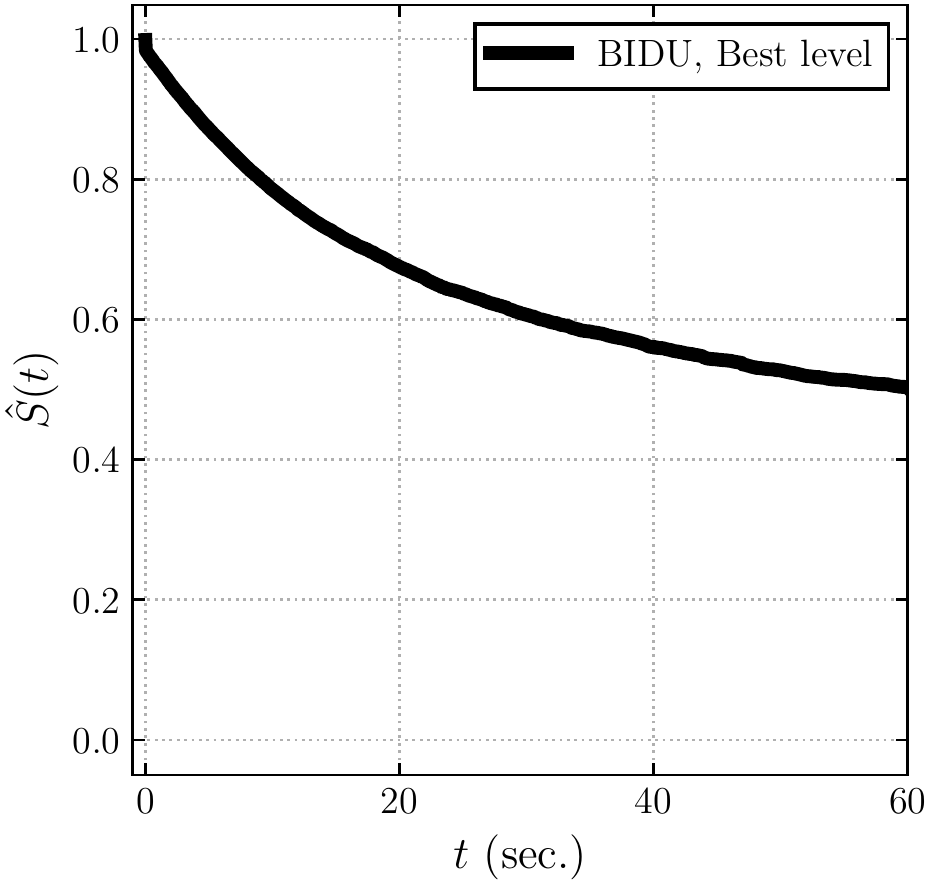}
\hspace*{\fill}
\includegraphics[height = 6.75cm,width=0.49\textwidth]{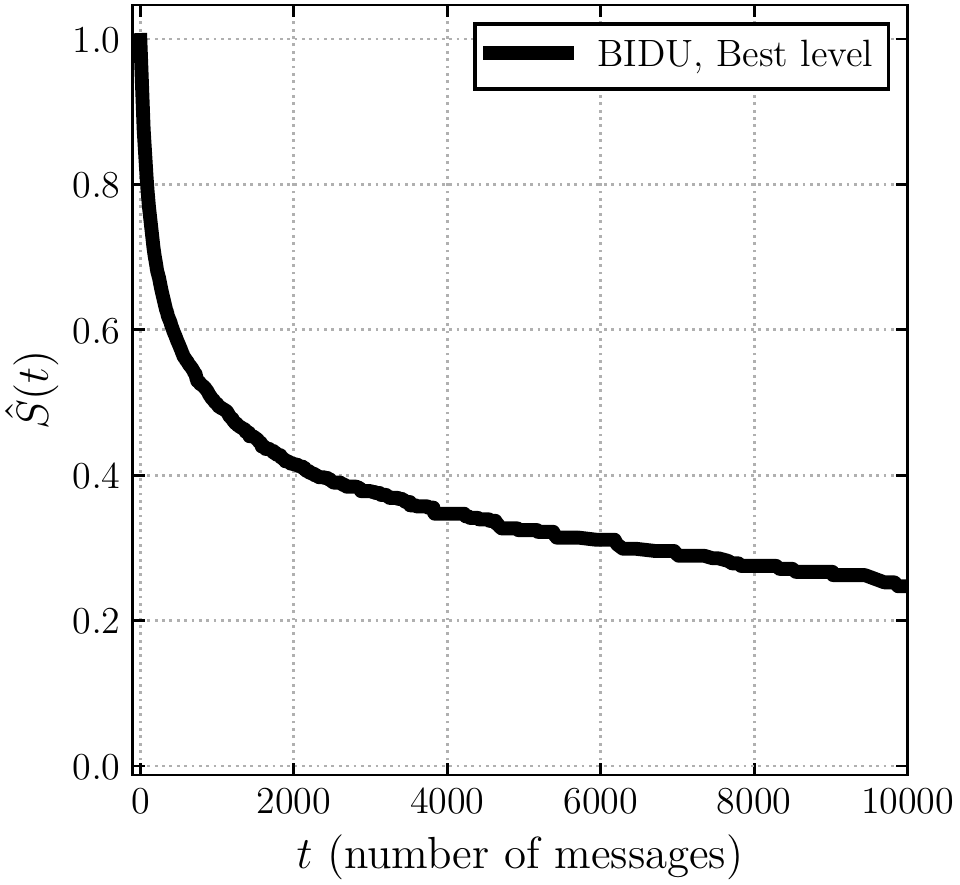}
\label{fig:}
\end{figure} 

\begin{figure}[H]
\includegraphics[height =6.75cm,width=0.49\textwidth]{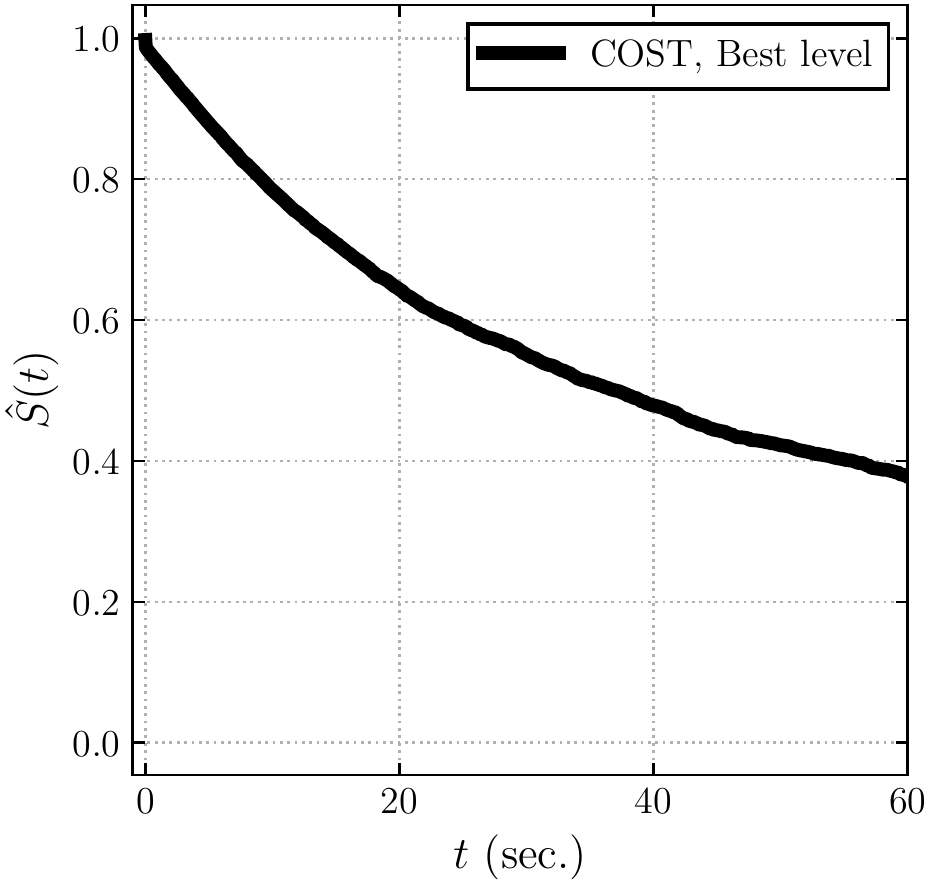}
\hspace*{\fill}
\includegraphics[height = 6.75cm,width=0.49\textwidth]{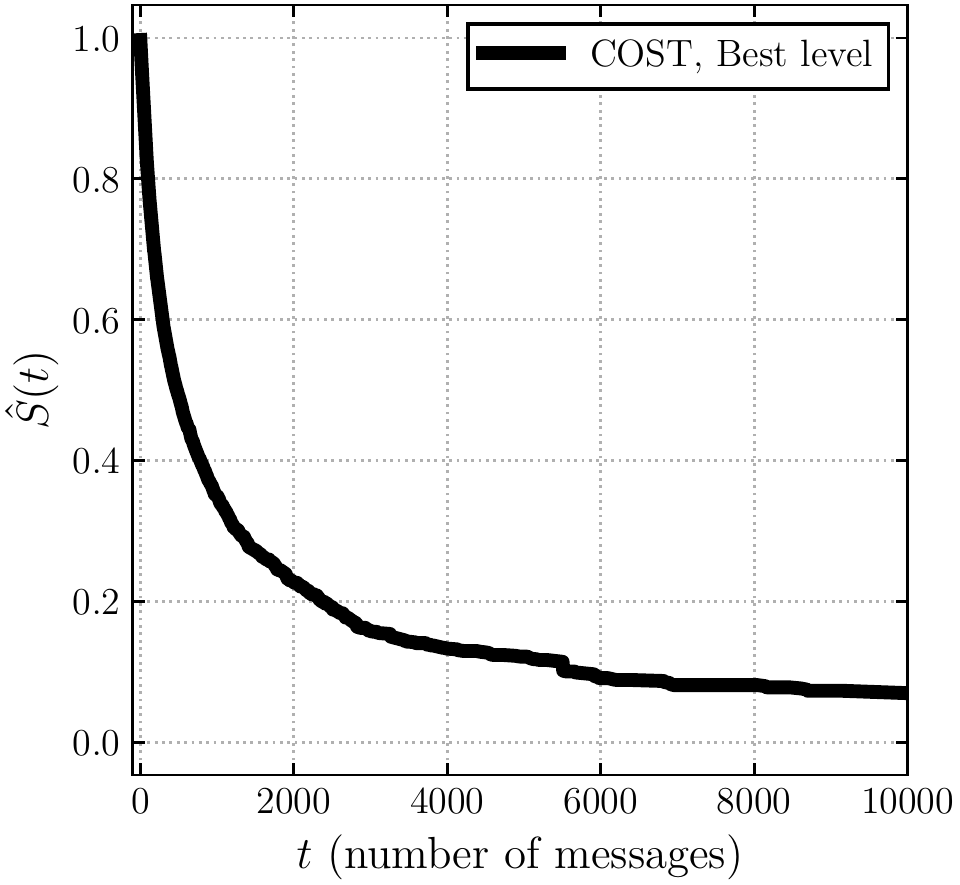}
\vspace*{\fill}
\includegraphics[height =6.75cm,width=0.49\textwidth]{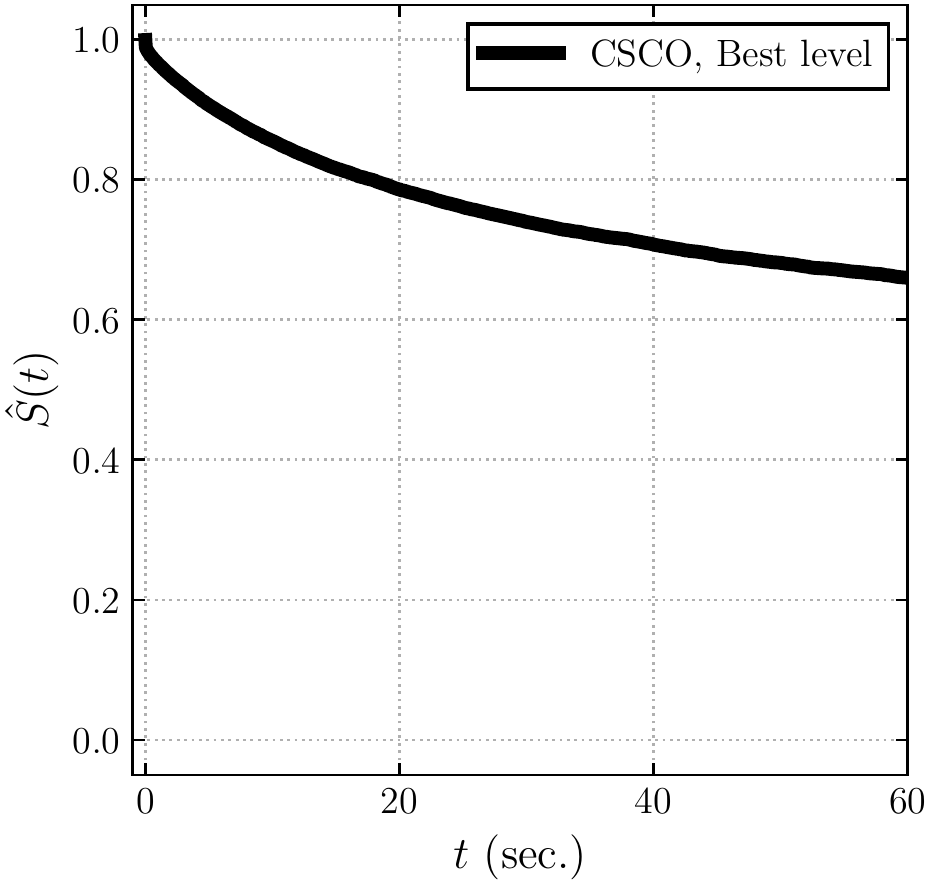}
\hspace*{\fill}
\includegraphics[height = 6.75cm,width=0.49\textwidth]{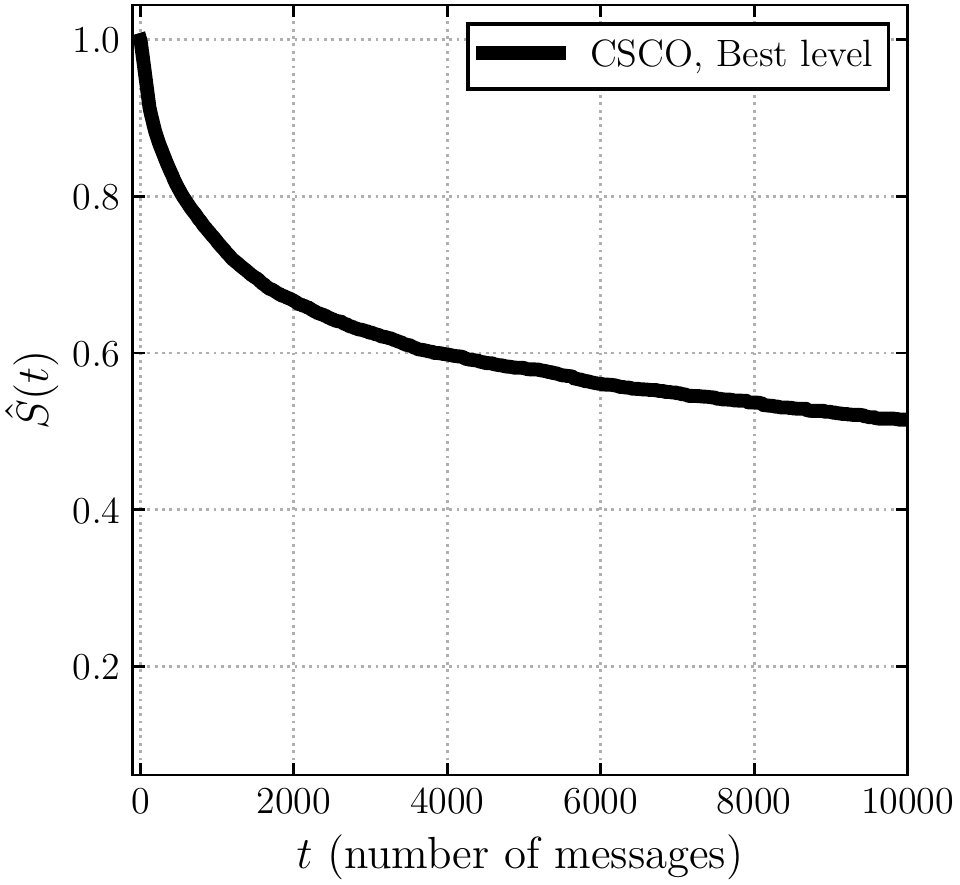}
\vspace*{\fill}
\includegraphics[height =6.75cm,width=0.49\textwidth]{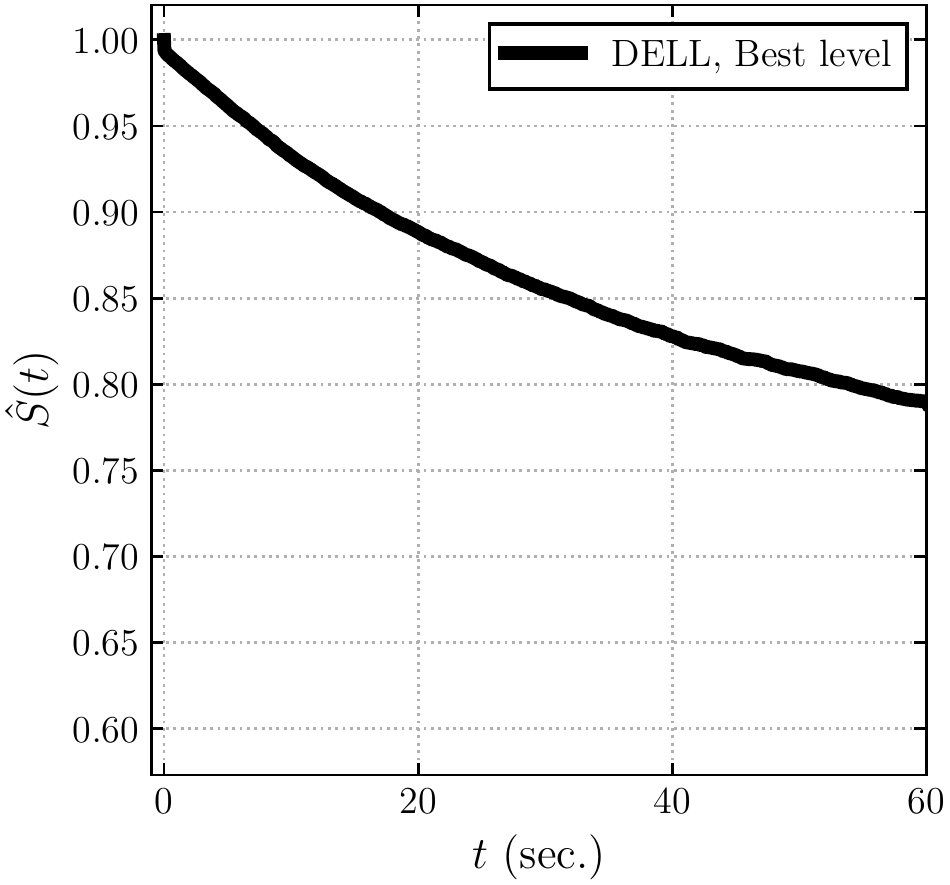}
\hspace*{\fill}
\includegraphics[height = 6.75cm,width=0.49\textwidth]{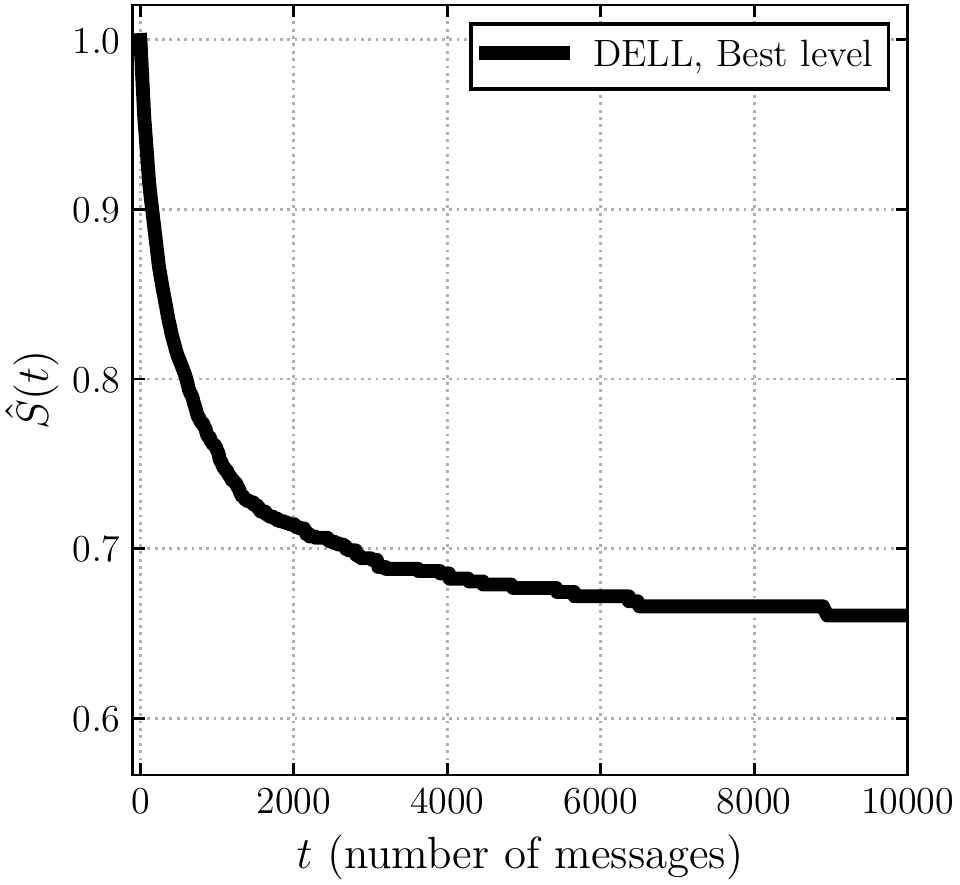}
%\caption{Kaplan-Meier estimates of survival functions when placing orders at different depths of the bid-ask spread. \textbf{Top left: }CSCO, \textbf{Top left: }INTC, \textbf{Middle left: }AAPL, \textbf{Middle right: }AMZN, \textbf{Bottom left: }BIDU, \textbf{Bottom right: }DELL.}
\label{fig:}
\end{figure} 

\begin{figure}[H]
\includegraphics[height =6.75cm,width=0.49\textwidth]{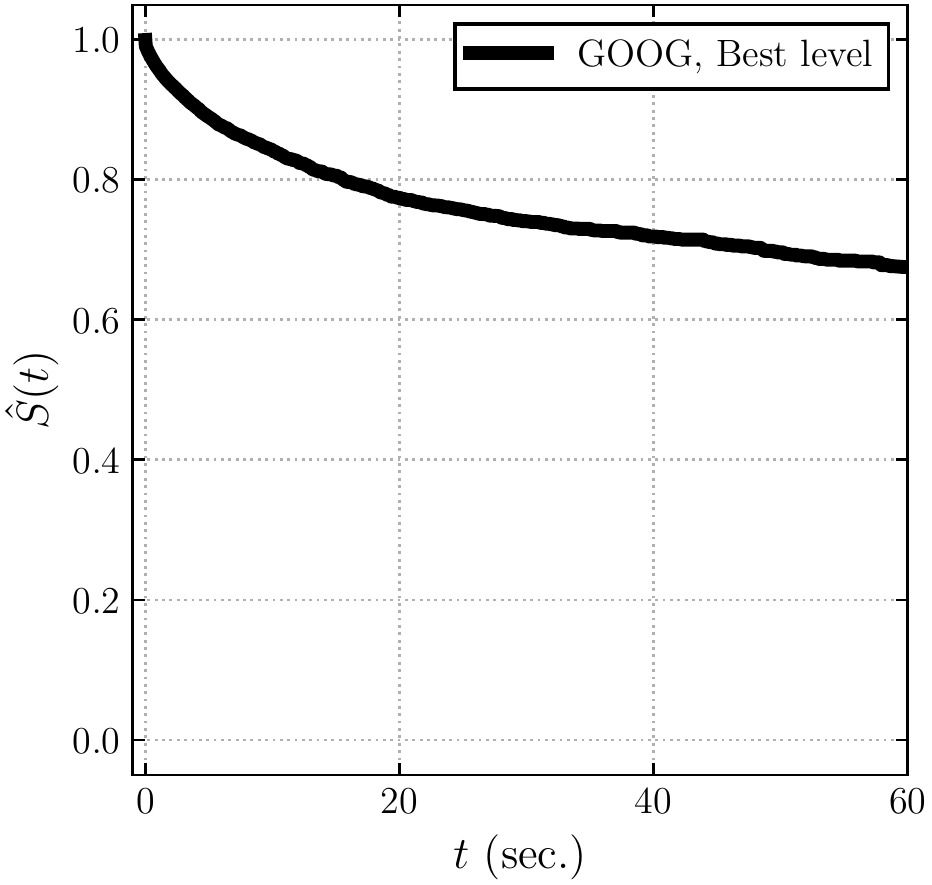}
\hspace*{\fill}
\includegraphics[height = 6.75cm,width=0.49\textwidth]{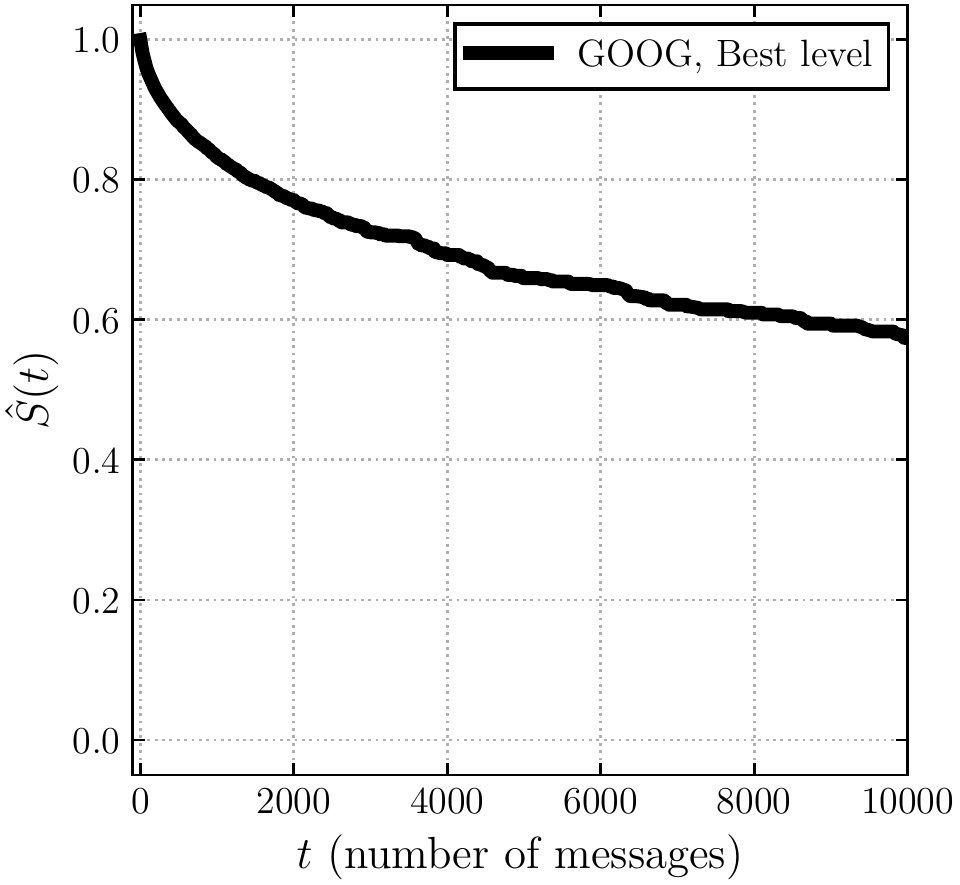}
\vspace*{\fill}
\includegraphics[height =6.75cm,width=0.49\textwidth]{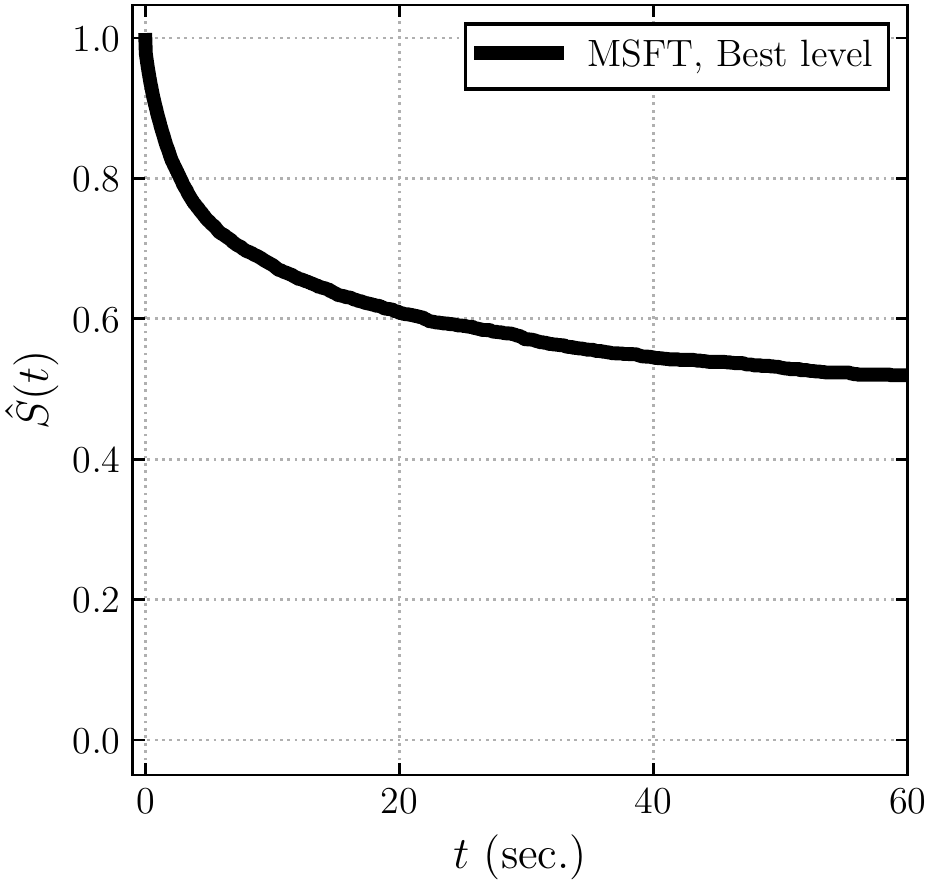}
\hspace*{\fill}
\includegraphics[height = 6.75cm,width=0.49\textwidth]{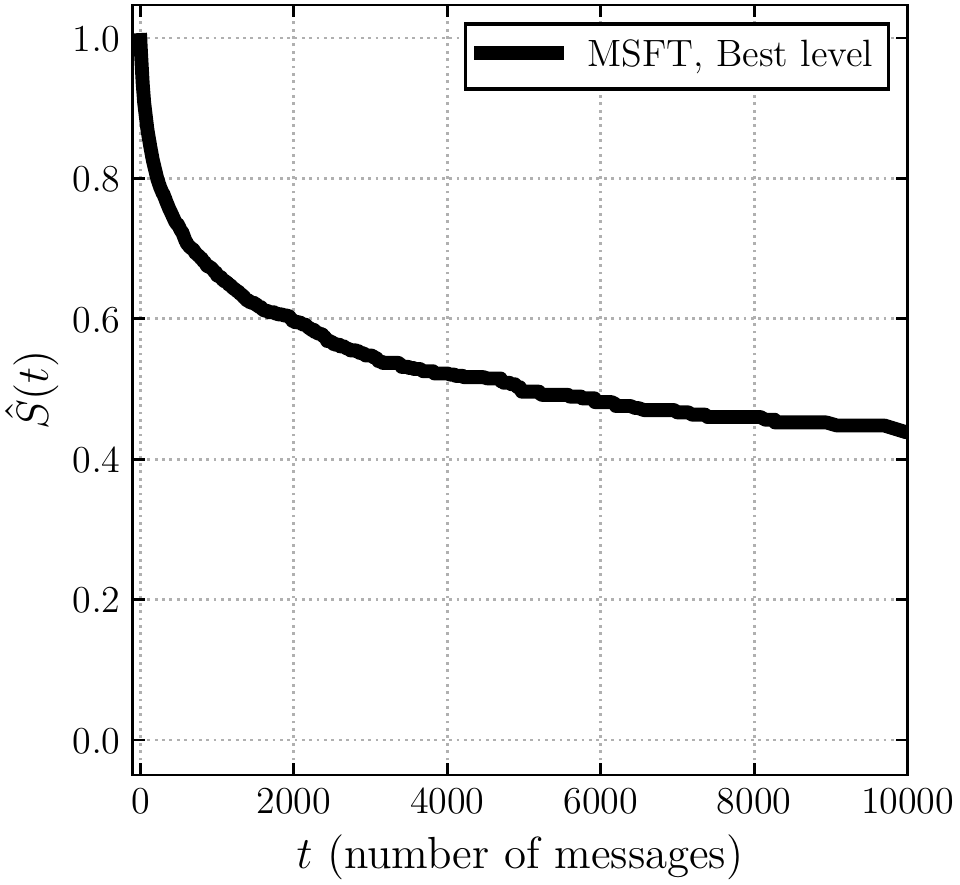}
\vspace*{\fill}
\includegraphics[height =6.75cm,width=0.49\textwidth]{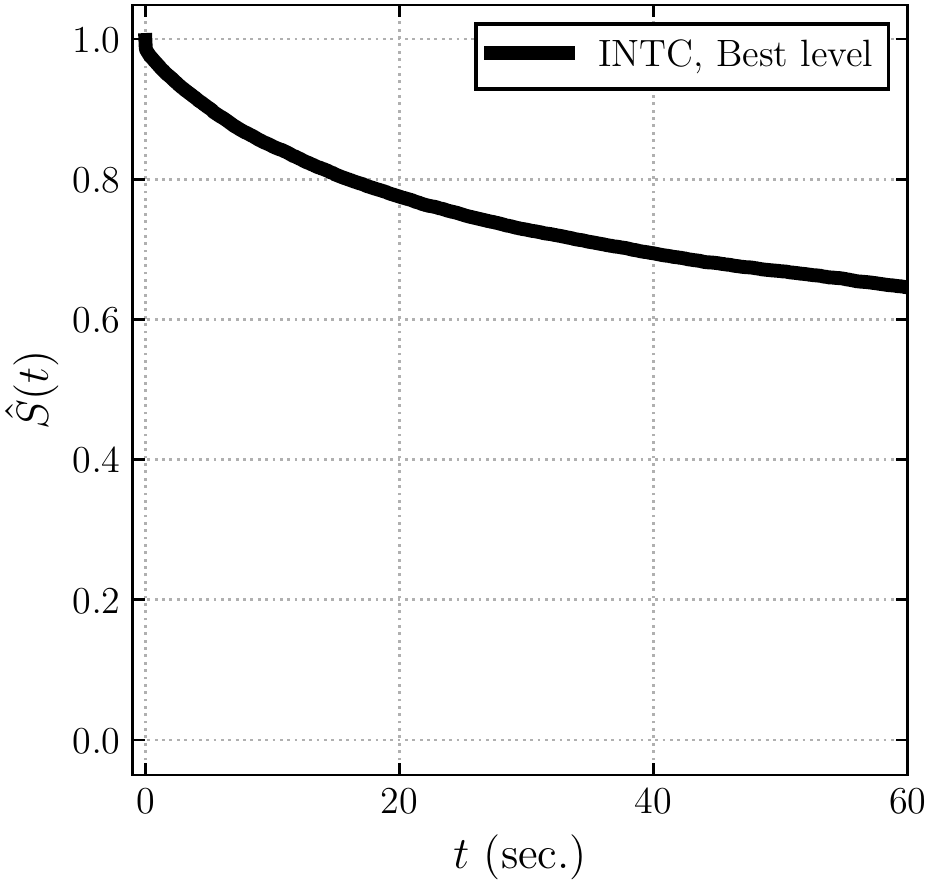}
\hspace*{\fill}
\includegraphics[height = 6.75cm,width=0.49\textwidth]{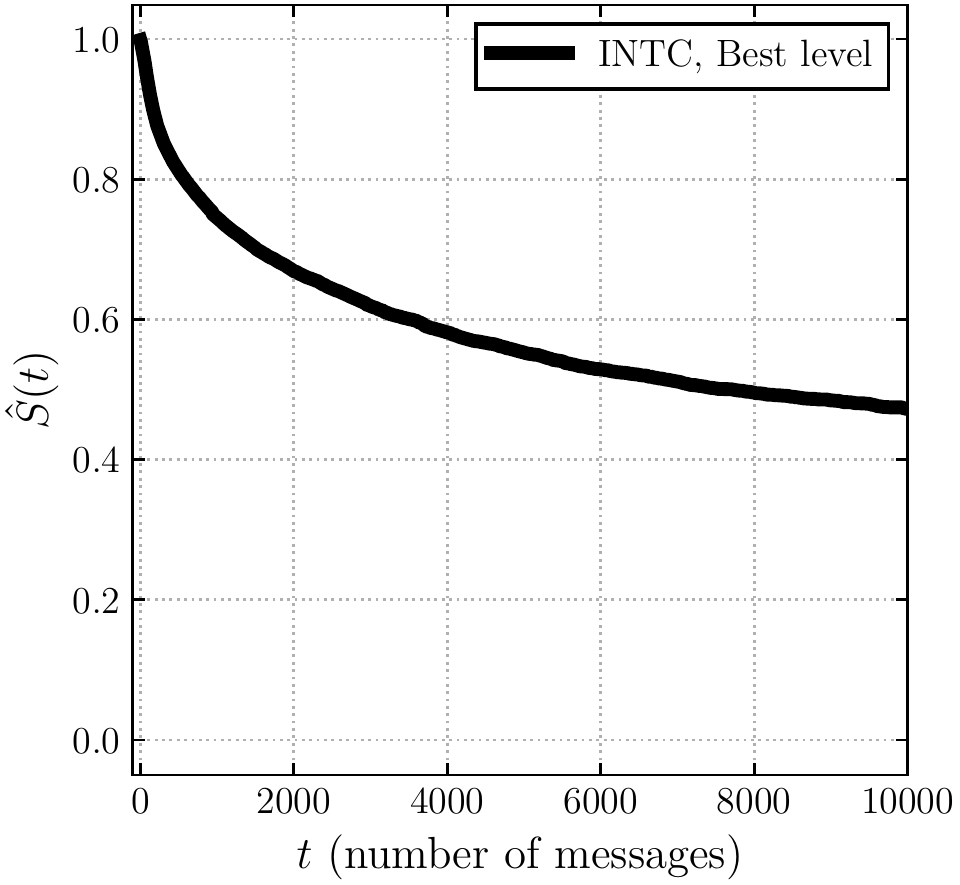}
\label{fig:}
\end{figure} 

\section{Performance of Order Flow Representations}
\label{app:i}

\begin{table}[H]
	\fontsize{10.0}{14.0}\selectfont
	\centering
	\caption{\footnotesize{Models evaluated using the negative right-censored log-likelihood for a lookback window of $T = 500$ on the order flow data, and percentage improvement, over the MN-MLP, for each of the evaluated models.}}
	\makebox[\textwidth]{\begin{tabular}{cccccccccc}
			\specialrule{.1em}{.1em}{.1em} \multicolumn{10}{c}{\textbf{Mean $\boldsymbol{\pm}$ STD Negative RCLL}}                                                                                       \\ \specialrule{.1em}{.1em}{.1em}
			& \textbf{AAPL}   & \textbf{AMZN}   & \textbf{BIDU}   & \textbf{COST}   & \textbf{CSCO}   & \textbf{DELL}   & \textbf{GOOG}   & \textbf{INTC}   & \textbf{MSFT}   \\ \specialrule{.1em}{.1em}{.1em}
			\multirow{3}{*}{\textbf{DeepSurv}}      & 8.053  & 8.346  & 7.236  & 9.712  & 8.441  & 8.441  & 8.242  & 6.211  & 9.117  \\
			& $\pm$     & $\pm$     & $\pm$     & $\pm$     & $\pm$     & $\pm$     & $\pm$     & $\pm$     & $\pm$     \\
			& 0.019  & 0.011  & 0.051  & 0.047  & 0.022  & 0.022  & 0.004  & 0.057  & 0.043  \\ \hline
			\multirow{3}{*}{\textbf{DeepHit}}       & 10.015 & 10.102 & 9.627  & 10.036 & 9.843  & 9.843  & 10.142 & 9.795  & 10.112 \\
			& $\pm$     & $\pm$     & $\pm$     & $\pm$     & $\pm$     & $\pm$     & $\pm$     & $\pm$     & $\pm$     \\
			& 0.091  & 0.024  & 0.064  & 0.043  & 0.083  & 0.083  & 0.042  & 0.157  & 0.062  \\ \hline
			\multirow{3}{*}{\textbf{MN-MLP}}        & 10.535 & 10.801 & 13.518 & 12.502 & 13.158 & 12.160 & 10.800 & 12.938 & 10.963 \\
			& $\pm$     & $\pm$     & $\pm$     & $\pm$     & $\pm$     & $\pm$     & $\pm$     & $\pm$     & $\pm$     \\
			& 0.339  & 0.347  & 0.497  & 0.431  & 0.495  & 0.451  & 0.311  & 0.422  & 0.326  \\ \hline
			\multirow{3}{*}{\textbf{MN-CNN}}        & 5.176  & 5.427  & 7.921  & 6.741  & 7.434  & 6.754  & 5.469  & 7.331  & 5.646  \\
			& $\pm$     & $\pm$     & $\pm$     & $\pm$     & $\pm$     & $\pm$     & $\pm$     & $\pm$     & $\pm$     \\
			& 0.343  & 0.173  & 0.162  & 0.224  & 0.212  & 0.291  & 0.621  & - 0.0  & 0.163  \\ \hline
			\multirow{3}{*}{\textbf{MN-LSTM}}       & 3.746  & 4.838  & 7.32   & 5.941  & 7.193  & 6.487  & 4.890  & 8.583  & 4.352  \\
			& $\pm$     & $\pm$     & $\pm$     & $\pm$     & $\pm$     & $\pm$     & $\pm$     & $\pm$     & $\pm$     \\
			& 0.136  & 0.248  & .613   & 0.195  & 0.129  & 0.198  & 0.538  & 0.387  & 0.231  \\ \hline
			\multirow{3}{*}{\textbf{MN-Conv-Trans}} & \textbf{3.158}  & \textbf{3.546}  & \textbf{6.107}  & \textbf{5.220}  & \textbf{6.211}  & \textbf{5.245}  & \textbf{3.595}  & \textbf{5.997}  & \textbf{3.772}  \\
			& $\boldsymbol{\pm}$     & $\boldsymbol{\pm}$     & $\boldsymbol{\pm}$     & $\boldsymbol{\pm}$     & $\boldsymbol{\pm}$     & $\boldsymbol{\pm}$     & $\boldsymbol{\pm}$     & $\boldsymbol{\pm}$     & $\boldsymbol{\pm}$     \\
			& \textbf{1.138}  & \textbf{1.002}  & \textbf{1.008}  & \textbf{1.013}  & \textbf{1.103}  & \textbf{1.162}  & \textbf{1.001}  & \textbf{1.173}  & \textbf{0.918}  \\ \specialrule{.1em}{.1em}{.1em}
			\multicolumn{10}{c}{\textbf{Improvement over MN-MLP (\%)}}                                                                                \\ \specialrule{.1em}{.1em}{.1em}
			& \textbf{AAPL}   & \textbf{AMZN}   & \textbf{BIDU}   & \textbf{COST}   & \textbf{CSCO}   & \textbf{DELL}   & \textbf{GOOG}   & \textbf{INTC}   & \textbf{MSFT}   \\ \specialrule{.1em}{.1em}{.1em}
			\textbf{DeepSurv}                       & 23.56  & 22.73  & 46.47  & 22.32  & 35.85  & 30.58  & 23.69  & 51.99  & 16.84  \\
			\textbf{DeepHit}                        & 4.94   & 6.47   & 28.78  & 19.72  & 25.19  & 19.05  & 6.09   & 24.29  & 7.76   \\
			\textbf{MN-MLP}                         & -      & -      & -      & -      & -      & -      & -      & -      & -      \\
			\textbf{MN-CNN}                         & 50.87  & 49.75  & 41.40  & 46.08  & 43.50  & 44.46  & 49.36  & 43.34  & 48.50  \\
			\textbf{MN-LSTM}                        & 64.44  & 55.21  & 45.85  & 52.48  & 45.33  & 46.65  & 54.72  & 33.66  & 60.30  \\
			\textbf{MN-Conv-Trans}                  & \textbf{70.02}  & \textbf{67.17}  & \textbf{54.82}  & \textbf{58.25}  & \textbf{52.80}  & \textbf{56.87}  & \textbf{66.71}  & \textbf{53.65}  & \textbf{65.59}  \\ \specialrule{.1em}{.1em}{.1em}
	\end{tabular}}
	\label{table:OF_experiments}
\end{table}

\newpage

\bibliographystyle{elsarticle-harv}
\bibliography{references}

\end{document}